\documentclass[useAMS,usenatbib,usegraphicx,]{mn2e}
  
\usepackage{aas_macros}
\usepackage{threeparttable}
\usepackage{grffile}
\usepackage{wallpaper}
\usepackage{mathbbol}
\usepackage{rotating}

\usepackage{tocloft}
\newcommand{\vek}{\underline}
\newcommand{\fat}{\textbf}
\newcommand{\ita}{\textit}

\newcommand{\beq}{\begin{equation}}
\newcommand{\eeq}{\end{equation}}  
\newcommand{\RNum}[1]{\uppercase\expandafter{\romannumeral #1\relax}} 
         
\title[SN Feedback in MCs]{Supernova Feedback in Molecular Clouds: Global Evolution and Dynamics}
  \author[B. K\"ortgen et al.]
  {Bastian~K\"ortgen~$^1$\thanks{bkoertgen@hs.uni-hamburg.de}, Daniel~Seifried$^2$, Robi~Banerjee$^1$,  
  Enrique~V\'{a}zquez--Semadeni$^3$,\newauthor{and Manuel~Zamora--Avil\'{e}s$^3$} \\
  $^1$ Hamburger Sternwarte, Universit\"at Hamburg, Gojenbergsweg 112, 21029 Hamburg, Germany\\
  $^2$ I. Physikalisches Institut, Universit\"at zu K\"oln, Z\"ulpicherstra\ss{}e 77, 50937 K\"oln, Germany\\
  $^3$ Instituto de Radioastronom\'{i}a y Astrof\'{i}sica, Universidad Nacional Aut\'{o}noma de M\'{e}xico, Campus Morelia, \\
  \ \ Apdo. Postal 3-72, Morelia, 58089, M\'{e}xico\\
  }

\date{Released 2015}

\pagerange{\pageref{firstpage}--\pageref{lastpage}} \pubyear{2015}

\begin{document}

\label{firstpage}

\maketitle

\begin{abstract}
We use magnetohydrodynamical simulations of converging warm neutral medium flows to analyse the formation and global evolution of magnetised and turbulent molecular clouds subject to supernova feedback from massive stars.
We show that supernova feedback alone fails to disrupt entire, gravitationally bound, molecular clouds, but is able to disperse small--sized ($\sim$10\,pc) regions on timescales of less than 1\,Myr. 
Efficient radiative cooling of the 
supernova remnant as well as strong compression of the surrounding gas result in non--persistent energy and momentum input from the supernovae. However, if the time between subsequent supernovae is short and 
they are clustered, large hot bubbles form that disperse larger regions of the parental cloud. On longer timescales, supernova feedback increases the amount of gas with moderate temperatures 
($T\approx300-3000\,\mathrm{K}$).
Despite its inability to disrupt molecular clouds, supernova feedback leaves a strong imprint on the star formation process. We find an overall reduction of the star formation efficiency by a factor of 2 and of the star formation rate by 
roughly factors of 2--4.
\end{abstract}

%  \begin{multicols}{2}
%%%%%%%%%%%%%%%%%%%%%%%%%%%%%%%%%%%%%%%%%%%%%%%%%%%%%%%%%%%%%%%%%%%%%%%%%%%%%%%%%%%%%%%%%%%%%%%%%%%%%%%%%%%%%%%%%%%%%%%%%%%%%%%%%%%%%%%%%%%%%%%%%%%%%%%%%%%%%%%%%%%%%%%%%%%%%%%%%%%%%%%%%%%%%%%%%%%%%%%%%%%

\section{Introduction}
\label{sec1}
The formation of molecular clouds, dense clumps, and finally stars is regulated by the interplay of gravity, magnetic fields, stellar feedback, and turbulence.\\
%Magnetic fields provide support against gravity in addition to thermal and turbulent pressure.
%This support depends on the mass--to--flux ratio, and if the cloud becomes supercritical, collapse is being initiated. In the contrary case of subcritical clouds, magnetic pressure is able to balance 
%gravity. However, Zeeman observations by \citet[][see also \citet{Crutcher12}]{Crutcher10} show that \ita{molecular clouds are supercritical} by a factor of 2--3 and should thus be in a phase of gravitational collapse.
%The low--density regime, i.e. diffuse HI clouds, is described by constant magnitude of the magnetic field as function of (column--)density, highlighting accretion \ita{along the magnetic field lines} up to a certain 
%critical density, where the field becomes supercritical.\\
%But, it is highly debated whether molecular clouds \ita{start out} from sub-- or super--critical conditions \citep[e.g.][]{Crutcher10,Hennebelle13,Li14,Koertgen15}. 
%The former case would underline the growing evidence for dynamically important magnetic fields and hence different influence on the cloud's internal dynamics.\\
The effect of turbulence is two--fold. Firstly, in the cold neutral medium (CNM) turbulent fluctuations are primarily supersonic. Thus, shocks occur, which compress the gas 
and hence provide the seeds for gravitationally unstable regions \citep[e.g.][]{MacLow04}. On the other hand, these supersonic motions constitute an effective pressure. This turbulent pressure 
acts as further support against gravity beside thermal and magnetic pressure. If the turbulence is also superalfv\'{e}nic, it is the major support in molecular 
clouds \citep[][]{Padoan99,Padoan99b,Federrath12,Federrath13}.\\
Additionally the internal cloud dynamics are mediated by stellar feedback via jets/outflows, winds, ionising radiation, and supernovae. 
The role of jets and outflows is still being subject to strong debate. On the one hand, they are able to drive turbulence in the intra--clump medium \citep[][]{Nakamura14,HxLi15} and hence maintain the 
level of energy counterbalancing gravity. On the other hand, \citet[][]{Banerjee07b} argue that the turbulent fluctuations, driven by a single source, are damped too fast as primarily compressive modes are excited. However, the combined 
effect of \ita{multiple} outflows seems to be able to disperse (not disrupt) the parental clump \citep[][]{Banerjee07b,Wang10,Nakamura14}.\\
Stellar winds are believed to have a stronger impact on the massive star's environment and hence the parental cloud. As \citet[][]{Dale13} point out, winds are most efficient in dispersing dense, massive cores in which the stars are embedded. Their 
longrange impact, however, is not sufficient. \citet[][]{Dale14} compared simulations of idealised molecular clouds including stellar winds or ionisation feedback. The main driver of cloud dispersion is the 
massive star's ionising radiation, consistent with studies by \citet[][]{Vazquez10}. Stellar winds, in contrast, only help to shape the emerging HII regions. In detail, the emerging HII regions are more spherical and stable against shell 
instabilities.
%winds yield more spherical HII regions, which are more stable against shell instabilities due to the smoothing influence of the expelled winds. 
However, the efficiency of dispersing entire (giant) molecular clouds by these two feedback mechanisms strongly depends on the cloud's mass and escape velocity. Concerning the impact of these mechanisms on the star formation 
process, \citet[][]{Dale14} and \citet[][]{Vazquez10} come to similar conclusions: ionisation feedback is most efficient in dispersing small regions. In addition,  
\citet[][]{Colin13} give a timescale for the dispersion of such regions (of size $\sim\,$10\,pc) of $t\approx 10-15\,$Myr.\\ On scales of entire molecular clouds ionisation feedback may also help to trigger the 
formation of new stars \citep[][]{Gritschneder09,Walch12,Walch13}. However, the star formation efficiency is \ita{globally} still being reduced by a factor of 10--20\,\% but not halted \citep[][]{Dale14}. The degree of 
turbulence within the dense gas, in contrast, is only essential for the inhomogeneity of the cloud and hence the ability of the hot, ionised gas to escape through low--density channels.\\
Finally, massive stars explode in a violent supernova event, thereby releasing $E_\mathrm{SN}=10^{51}\,\mathrm{erg}$ in a short period of time. 
A number of studies exist that either focus on Galactic scales, \citep[][]{Rosen95,Korpi99,Avillez00,Avillez02,Avillez04,Joung06,Joung07,Shetty08,Joung09,Ostriker11,Hill12,Shetty12,Gent13a,Gent13b,Walch14b,Hennebelle14,Gatto15}, or on 
scales of small clouds or even clumps with radii of a few pc \citep[][]{Pittard12,Rogers13,Walch15,Iffrig15,Geen15,Kim15}.\\
Recently, \citet[][]{Walch15} have reported on supernova feedback in small--sized (radius $r\,=\,16\,\mathrm{pc}$), massive ($M\approx10^5\,\mathrm{M}_\odot$) and non--magnetised molecular clouds. The authors injected momentum in a small 
sub--volume of the cloud in order to mimic the free--expansion phase of the supernova remnant (SNR). They resolved the different stages during the SNR evolution and analysed the influence of different physical mechanisms on this. For 
adiabatic expansion of the SNR in a homogeneous cloud, they yielded the complete dispersion of the latter on timescales of $t\,\leq\,1\,$Myr. However, the clouds -- homogeneous or fractal -- are \ita{not being destroyed} if radiative cooling is 
included. The hot and shock--compressed gas cools too fast. Hence, the thermal energy supply, which can be converted into kinetic energy, shrinks on the same timescales. The net energy and momentum input are thus not sufficient to 
accelerate the gas to velocities greater than the cloud's escape velocity. 
Similar results were obtained by \citet[][]{Iffrig15}, who analysed the impact of supernova explosions within or near molecular clouds. The 
authors deduced that the impact of supernova feedback is primarily determined by the position of the progenitor star. Supernovae at the border of or near to a molecular cloud do not have a significant impact on a possible cloud dispersal 
as well as on the dynamics of the dense gas which is due to the lack of momentum transfer to the latter. 
The major part of the cloud is compressed and some regions are ablated. In the case of a supernova going off within a molecular cloud, the 
momentum transfer to the dense gas is much higher and hence the fraction of gas escaping the cloud. The results indicate a reduction of the cloud mass due to single supernova explosions of up to 50\,\% for clouds with masses of $M\,\approx\,10^4\,M_\odot$ and sizes 
of approximately 20--30\,pc. However, the authors report no complete cloud dispersion.\\ 
\citet[][]{Kim15} analysed supernova explosions in a two--phase ISM \citep[see also a similar study by][]{Martizzi15}. The authors found that the 
net momentum input from supernovae is nearly independent of the morphology of the environment and may only  
change if further feedback prior to the supernova is considered \citep[see also][]{Walch15}.\\
In studies of Galactic scale simulations supernova feedback is usually taken into account since it is the main driver of
 Galactic fountain flows \citep[e.g.][but see \cite{Girichidis15} for details about the efficiency in 
launching these outflows]{Hill12,Gent13b,Walch14b}.  Usually, $E_\mathrm{SN}=10^{51}\,\mathrm{erg}$ are injected during each individual supernova event. However, some approaches 
inject $E_\mathrm{SN}=(2-3)\times10^{51}\,\mathrm{erg}$ in order to resemble additional energy input from ionisation and winds in one single event (P.Col\'{i}n, priv. communication, 2012). Studies implementing more than one supernova are 
restricted to a certain supernova rate. For example, \citet[][]{Joung06} use the \ita{observed Galactic rate} of $\nu_\mathrm{SN,gal}\,=\,1/44\,\mathrm{yr}^{-1}$ from \citet[][]{Tammann94}. More recent studies by \citet[][]{Walch14b} and 
\citet[][]{Gatto15} use a Kennicut--Schmidt relation in order to extract the star formation rate surface density, $\Sigma_\mathrm{SFR}$, and transform it to a supernova rate by convolution with an IMF. \\%Varying the slope of the 
%Kennicut--Schmidt relation\footnote{These variations are then no KS--relation in the original sense.} then gives different supernova rates.\\
\citet[][]{Gatto15} conducted a large parameter study of supernova feedback on Galactic scales. They investigated the influence of different supernova driving mechanisms on the thermal and dynamical state of the interstellar medium (ISM). 
%Specifically, they analysed random, peak, and mixed driving, i.e. whether the supernovae are placed randomly throughout the volume, are correlated with high--density regions, or a mixture of both. 
%In addition, the authors varied the supernova rate by varying the gas surface density. 
Most relevant are their results from 'peak driven' supernovae -- i.e. the supernovae exploded in regions of significantly enhanced density -- which state that this driving mechanism fails to explain the large fraction of molecular gas as well as the volume filling fraction of hot, ionised gas. The
former is most likely due to disruption of dense, cold branches by the interaction of the SNR with the densest gas. The latter originates in very efficient cooling of hot gas in the shock--compressed regions within the dense clumps. The gas 
temperatures are cooled efficiently to $T\,<\,10^6\,\mathrm{K}$. This is supported by \citet[][]{Walch14b}, who yield realistic disc structure and volume filling fractions of the hot gas for non--peak driven supernovae.  Both studies underline the 
importance of feedback mechanisms prior to supernova feedback. In addition, recent investigations by \citet[][]{MLi15} 
have shown that the volume filling fraction of the hot component of the ISM needs to be $f_\mathrm{V}\sim0.6$ in 
order to induce thermal runaway. In this case, the bubbles created by multiple supernovae connect to each other. In 
the contrary case of $f_\mathrm{V}<0.6$, the bubbles of supernova remnants do not connect and cooling dominates.\\
Our study bridges the gap between small--scale, i.e. 1 to a few 10 pc, simulations \citep[][]{Pittard12,Walch15,Iffrig15} and large--scale (kpc) disc simulations \citep[][]{Korpi99,Ostriker11,Walch14b,Tasker15} by performing a set of simulations on intermediate scales of a few hundred pc.
Section \ref{sec2} introduces the numerical model as well as the initial conditions. The following section \ref{sec3} 
gives the results of our study on supernova feedback, thereby focussing on the global evolution of the (dense) gas. In section \ref{secdisc} we briefly discuss the limitations of our model. The study closes with a summary in section \ref{secsum}.

\section{Numerical setup and initial conditions}
\label{sec2}
\subsection{Details of the numerics}
For this study we use the finite volume adaptive mesh--refinement (AMR) code FLASH (version 2.5) \citep{FLASH00,Dubey08}. 
%During each timestep a Riemann problem is solved at the cell interfaces, yielding the respective fluxes for the hyperbolic partial differential equations. The MHD fluxes are computed by a multiwave 
%Riemann solver developed by \citet[][]{Bouchut07,Bouchut09} and implemented in FLASH by \citet[][]{Waagan11}, which preserves positive states for density and internal energy. 
The code solves the ideal MHD equations, the Poisson equation for self--gravity of the gas, as well as heating and cooling. The MHD fluxes are computed by a multiwave 
Riemann solver developed by \citet[][]{Bouchut07,Bouchut09} and implemented in FLASH by \citet[][]{Waagan11}, which preserves positive states for density and internal energy.
 Since we are interested in the 
process of star formation, we also include sink particles to follow collapsing regions \citep{Federrath10}.
%and the local Jeans length is resolved with at least ten grid cells 
%in order to fulfill the Truelove criterion \citep{Truelove97}. 
In order to form a sink particle, the gas has to pass several checks, which are described in detail in \citet[][]{Federrath10}. Beside these checks, the density has to exceed a threshold density of 
$n_\mathrm{sink}=3\times10^5\,\mathrm{cm}^{-3}$.
Periodic boundary conditions (BCs) are applied for the hydrodynamics and self--gravity is treated with isolated BCs.
Refinement of certain regions is achieved by resolving the local Jeans length with 10 grid cells, hence a factor of 2.5 larger than the usual Truelove criterion \citep[][]{Truelove97}, except at the maximum refinement level where it 
is still resolved with four grid cells.

%Once the maximum refinement level is reached, the local Jeans length becomes less resolved due to increasing density. However, the Jeans length of the threshold density is still resolved with four grid cells.}

%\subsubsection{Notes on the Truelove criterion}
%The Truelove criterion \citep[][]{Truelove97} states that the local Jeans length has to be resolved with at least four grid 
%cells. This is necessary in order to prevent artificial fragmentation. However, we resolve the local Jeans length with only 
%\ita{two} grid cells. Hence, the diameter of the control volume in the sink particle creation algorithm is resolved with a 
%total of four grid cells. Thus, the Truelove criterion is fulfilled to a certain extent, which is enough for our purposes. 
%Treating the Jeans criterion in such a way leads to high threshold densities, which are more comparable to observational 
%data \citep[see e.g.][]{Vazquez11}, and still reliable fragmentation properties (e.g. the formation of truely collapsing 
%objects). We further do not need to resolve the Jeans length with at least ten grid cells, since we are not interested in 
%resolving turbulent structures within the control volume \citep[][]{Federrath11a}.

\subsection{Initial Conditions}
%\begin{figure}
 %\includegraphics[width=0.5\textwidth]{setup3dnorm-crop.pdf}
% \caption{Setup of the initial conditions. The black dashed lines denote the flow axes and the position of the collision layer, respectively. Adapted from \citet[][]{Vazquez07}.}
% \label{fig1}
%\end{figure}
Our numerical setup is adapted from \citet[][see also
\citet{Banerjee09a}, \citet{Vazquez11}, and \citet{Koertgen15}]{Vazquez07}. The physical size of the numerical box is $L_\mathrm{box}=256\,\mathrm{pc}$. Two cylindrical 
flows of warm neutral medium (WNM) collide at the centre of the domain. Each flow has a linear dimension of $l=112\,\mathrm{pc}$ and a radius of $r=64\,\mathrm{pc}$.
A schematic picture of the initial setup is shown in \citet[][see their figure 1]{Koertgen15}. We have chosen the initial density to be 
$n=1\,\mathrm{cm}^{-3}$ and the temperature as $T=5000\,\mathrm{K}$, typical for the WNM. The total mass in the 
flows is $M_\mathrm{flows}=90,000\,\mathrm{M}_\odot$. The respective column density is 
$N_\mathrm{flows}=6.9\times10^{20}\,\mathrm{cm}^{-2}$.
The sound speed at the given temperature is $c_\mathrm{s}=5.7\,\mathrm{km/s}$, which we take as a reference speed. The mean molecular weight is $\mu\,=\,1.27$. 
We choose the speed of the colliding flows such that the \ita{isothermal} Mach number is $\mathcal{M}_\mathrm{flow}=2$. The dynamical time of each flow is thus 
$t_\mathrm{flow}=L_\mathrm{flow}/v_\mathrm{flow}=9.6\,\mathrm{Myr}$.
%With this temperature, we can define a sound speed of the warm neutral medium and we choose the velocity of every single flow in such 
%a way that the resulting \ita{isothermal} sonic Mach number is $\mathcal{M}_\mathrm{f}=2$. 
In addition we add a turbulent velocity field to the flows to mimic the general turbulent behavior of the ISM \citep{MacLow04}. 
The turbulent fluctuations are calculated in Fourier space with a Burgers type spectrum, i.e. $E(k)\propto k^{-2}$ for $k\geq k_\mathrm{int}$, where $k_\mathrm{int}$ is the wave number of the 
integral scale ($\lambda_\mathrm{int}=64\,\mathrm{pc}$).
Furthermore, these turbulent fluctuations trigger the onset of dynamical instabilities such as the non-linear thin-shell instability \citep[NTSI,][]{Vishniac94} or the Kelvin--Helmholtz instability 
\citep[e.g.][]{Heitsch05,Heitsch08b}.%\far{Contact Manuel for his simulation with the FLAT cloud in hydro case and turb Mach number eq. 0.8. What is the flow Mach number? Higher Mach number flows lead to NO clouds because of instabilities?} 
The initially uniform magnetic field has a strength of $\left|\vek{B}\right|=3\,\mu\mathrm{G}$ and is aligned with the flows, that is, 
$\vek{B}\propto\hat{\vek{x}}$ where 
$\hat{\vek{x}}$ is the unit vector in the x direction. 
Using the description for the critical mass--to--flux ratio by \citet{Nakano78}, i.e. $\mu_{\mathrm{crit}}=0.16/\sqrt{G}$, the two streams are in total initially 
subcritical ($\mu/\mu_\mathrm{crit}\,=\,0.97$), but can become supercritical very fast due to accretion of mass along the field lines. We use a maximum refinement level of 
$\mathcal{L}_{\mathrm{max}}=11$, which corresponds to a maximum physical resolution of $\Delta x_{\mathrm{max}}\approx0.03\,\mathrm{pc}$. A detailled overview of the simulation runs is given in table \ref{tab1}.
The main parameters are summarised as follows:
\begin{itemize}
\item Flow mass: $M_\mathrm{flows}=90,000\,\mathrm{M}_\odot$
\item Column density: $N_\mathrm{flows}=6.9\times10^{20}\,\mathrm{cm}^{-2}$
\item Sound speed: $c_\mathrm{s}=5.7\,\mathrm{km/s}$
\item Box size: $L_\mathrm{box}=256\,\mathrm{pc}$
\item Maximum resolution: $\Delta x_\mathrm{max}\approx0.03\,\mathrm{pc}$.
\end{itemize}

\begin{table}
 \caption{List of performed simulations, showing the simulation name in column one, the \ita{isothermal} sonic Mach number in column two. In the third column we list the isothermal sonic RMS Mach number. The last two columns 
show the initial magnetic field strength and the peak resolution, respectively.}
  %\begin{center}
   \begin{tabular}{p{1.7cm} p{0.7cm} p{0.7cm} p{1.4cm} p{0.5cm} p{1.3cm}}
\hline
\hline
   \fat{Run Name}	&\fat{$\mathcal{M}_\mathrm{flow}$}	&\fat{$\mathcal{M}_\mathrm{turb}$}	&\fat{Feedback?}	&\fat{$\left|\vek{\mathrm{B}}\right|$}	&\fat{Min. $\Delta$}\\
				&							&							&				&($\mu\mathrm{G}$)		&(pc)\\
  \hline
 % LR0.0N			&2							&0.0							&No				&3					&0.125\\
  %LR0.0Y			&2							&0.0							&Yes				&3					&0.125\\
%  LR0.4N			&2							&0.4							&No				&3					&0.125\\
%  LR0.8N			&2							&0.8							&No				&3					&0.125\\
%  LR0.8Y			&2							&0.8							&Yes				&3					&0.125\\
%  LR1.2N			&2							&1.2							&No				&3					&0.125\\
%  LR1.2Y			&2							&1.2							&Yes				&3					&0.125\\
%  LR1.6N			&2							&1.6							&No				&3					&0.125\\
%  LR2.0N			&2							&2.0							&No				&3					&0.125\\
 % LR2.0Y			&2							&2.0							&Yes				&3					&0.125\\
%\hline
  HR0.8N			&2							&0.8							&No				&3					&0.03\\
  HR0.8Y			&2							&0.8							&Yes				&3					&0.03\\
  HR1.0N			&2							&1.0							&No				&3					&0.03\\
  HR1.0Y			&2							&1.0							&Yes				&3					&0.03\\ 
  HR1.2N			&2							&1.2							&No				&3					&0.03\\
  HR1.2Y			&2							&1.2							&Yes				&3					&0.03\\
\hline
  HR0.8HN		&2							&0.8							&No				&0					&0.03\\
  HR0.8HY		&2							&0.8							&Yes				&0					&0.03\\
\hline
\hline
   \end{tabular}
%  \end{center}
 \label{tab1}
\end{table}

\subsection{The supernova subgrid model}
\label{secsubgrid}
The supernova (SN) feedback is directly coupled to the sink particles\footnote{For a resolution study we refer the reader to the appendix.}. 
A Kroupa--IMF \citep[][]{Kroupa01} is fitted to the \ita{total sink particle mass} each timestep in order to evaluate the number of 
massive stars within the cloud. The analytical evaluation of the IMF results in a minimum (or critical) mass $M_\mathrm{Kroupa}=160\,\mathrm{M}_\odot$ in order to form a massive star of 
mass $M_*\geq10\,\mathrm{M}_\odot$.
The SN model is only being activated if the total mass in all sink particles exceeds $M_\mathrm{Kroupa}$. We 
then let the \ita{most massive} sink particle explode in case:
\begin{itemize}
\item The particle has accreted at least 
\item[] $M_\mathrm{sink}\geq30\mathrm{M}_\odot$.
\item The sink particle's life time is $t_\mathrm{lifetime}\geq2\,\mathrm{Myr}$ 
\item[] according to standard mass--lifetime relations 
\item[]\citep[][]{WWW09}.
\end{itemize}
These two criteria ensure that feedback is not being initiated before the mass and lifetime correspond to these of an O--type star. Although it is more likely to form a B--star with such a small mass reservoir, the respective lifetime would be too 
long ($50\,\mathrm{Myr}$) to affect molecular clouds via supernova feedback during their lifetime \citep[20--30\,Myr,][]{Blitz07}.
%Such a feedback description would lead to inconsistent results regarding the cloud characteristics (e.g. gravitational binding). O--stars instead feed back onto the cloud in a reasonable period of time and deviations due to missing physics become smaller. 
Once, these two criteria are fullfilled, $E_\mathrm{SN}=10^{51}\,\mathrm{erg}$ -- with $E_\mathrm{th}=0.65\,E_\mathrm{SN}$ 
and $E_\mathrm{kin}=0.35\,E_\mathrm{SN}$%\footnote{\far{Please note that the canonical fractions are 72\,\% thermal and 
%28\,\% kinetic energy, but the difference to the used fractions here is negligible.}} 
-- are injected into a spherical control volume (CV, with a radius of 
$\mathcal{R}_\mathrm{CV}=0.06\,\mathrm{pc}$), similar to the original solution by \citet[][]{Sedov59}. %\far{The CV is hence over--pressured due to the increased thermal energy.} %The kinetic energy part is injected \far{with the shock--velocity calculated according to} the Sedov--Taylor (ST) solution \citep[][]{Sedov59}, with the ambient density (which occurs in the ST solution) being the average density within the control volume. 
The thermal energy is adjusted by increasing the pressure in the cells within the CV and the respective velocity 
increases radially with distance from the centre of the CV. We point out that in case of a SN going off the cooling time of the dense gas is still 6--7 times longer than the sound crossing time of the hot ($T\sim10^6\,\mathrm{K}$)
 gas. Right after the SN energy is injected, the timestep of the simulation is automatically adjusted according to the 
Courant condition so that the fast pressure waves and the momentum input are properly resolved. 
The mass of the exploding star is uniformly mapped back onto the numerical grid. Sink particles, which have gone off as a SN become passive particles, that is, 
accretion is switched off and they do not feed back onto the gas a second time (i.e. if they still possess more than $30\,\mathrm{M}_\odot$). Passive particles are not considered anymore for the 
$M_\mathrm{Kroupa}$--criterion.\\
%\far{After each SN the simulation timestep is automatically adjusted during runtime according to the Courant condition.}\\
If more than one massive star exists (i.e. $M_\mathrm{Sinks}/M_\mathrm{Kroupa}>1$), the supernovae explode according to the criterion: one SN per 44 M$_\odot$ of stars. This criterion is deduced from the SN type II rate 
of $1/44\,\mathrm{yr}^{-1}$ \citep[][]{Tammann94} and the average star formation rate of 
$\sim 1\,\mathrm{M}_\odot\mathrm{yr}^{-1}$ in the Milky Way \citep[see e.g.][and references therein]{Robitaille10}. 
%SN--rate, which is $\nu_\mathrm{SN}=\left(44\,\mathrm{M}_\odot\right)^{-1}$ for type--II SN. 
In such cases it is \ita{the most massive} particle that explodes. Multiple
 massive stars are then not allowed to go off as a SN during subsequent timesteps until the total mass in sink particles has grown by 44\,M$_\odot$ compared to the mass at the time of the last SN.
%The rate is a combination of the observed annual Galactic SN--rate \citep[][]{Tammann94} and the 
%Galactic star formation rate \citep[1\,M$_\odot$yr$^{-1}$,][]{MacLow04} and is thus independent of the chosen volume\footnote{For further explanation, please see appendix.}. 
Note that this rate is higher than typical rates estimated from IMF calculations and thus gives an upper limit on the efficiency of supernova feedback.  Note further that we do not include type I supernovae due to their low rate 
\citep[see e.g.][]{Tammann94}.

\subsection{Heating and cooling}
The ISM is subject to various heating and cooling processes, which affect the thermodynamic behaviour of the gas. Hence, heating and cooling are included as source terms in the energy equation.
They are incorporated following the recipe by \citet[][with modifications by \citet{Vazquez07}]{Koyama00}. This prescription results in a thermally unstable regime in the density range 
$1\,\mathrm{cm}^{-3}\leq n\leq10\,\mathrm{cm}^{-3}$, corresponding 
to a temperature interval of $500\,\mathrm{K}\leq T \leq5000\,\mathrm{K}$ if thermal equilibrium conditions are applied.\\
The fitting functions for heating, $\Gamma$, and cooling, $\Lambda$, give 
\beq
\Gamma\quad\qquad=\quad2.0\times10^{-26}\,\mathrm{erg\,s}^{-1}
\eeq
\begin{eqnarray}
\frac{\Lambda\left(T\right)}{\Gamma}&=&10^{7}\mathrm{exp}\left(\frac{-1.184\times10^{5}}{T+1000}\right)
\end{eqnarray}
\begin{eqnarray*}
\qquad&\qquad&+1.4\times10^{-2}\sqrt{T}\mathrm{exp}\left(\frac{-92}{T}\right)\,\mathrm{cm}^{3}.
\end{eqnarray*}
We remark that the cooling function stays almost constant for typical temperatures within supernova remnants ($T\geq10^{6}\,\mathrm{K}$).
%\begin{figure}
 %\includegraphics[height=0.45\textwidth,angle=-90]{cooling_function-crop.pdf}
% \caption{Cooling function in $\mathrm{erg\,cm^{3}s^{-1}}$ as function of gas temperature, T.}
% \label{figcooling}
%\end{figure}

\section{Results}
\label{sec3}
In the following we will present the results of this study. The focus will be on the global evolution of the molecular cloud and its resulting dynamics. The impact of supernova feedback on the process of star formation is being 
investigated in one of the last results sections.
\subsection{Evolution of cloud masses}
\begin{figure*}
\includegraphics[height=0.45\textwidth,angle=-90]{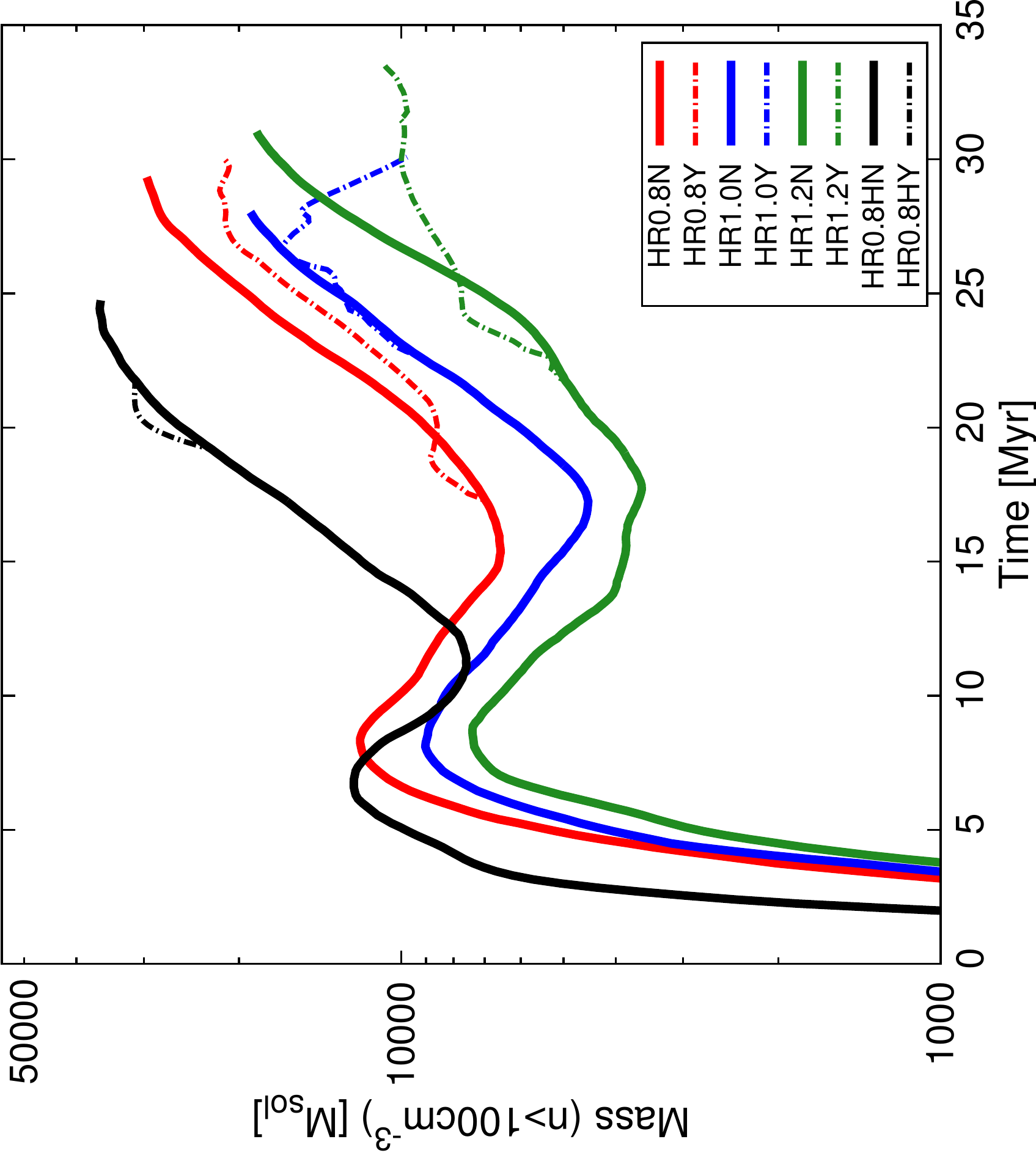}\quad\includegraphics[height=0.45\textwidth,angle=-90]{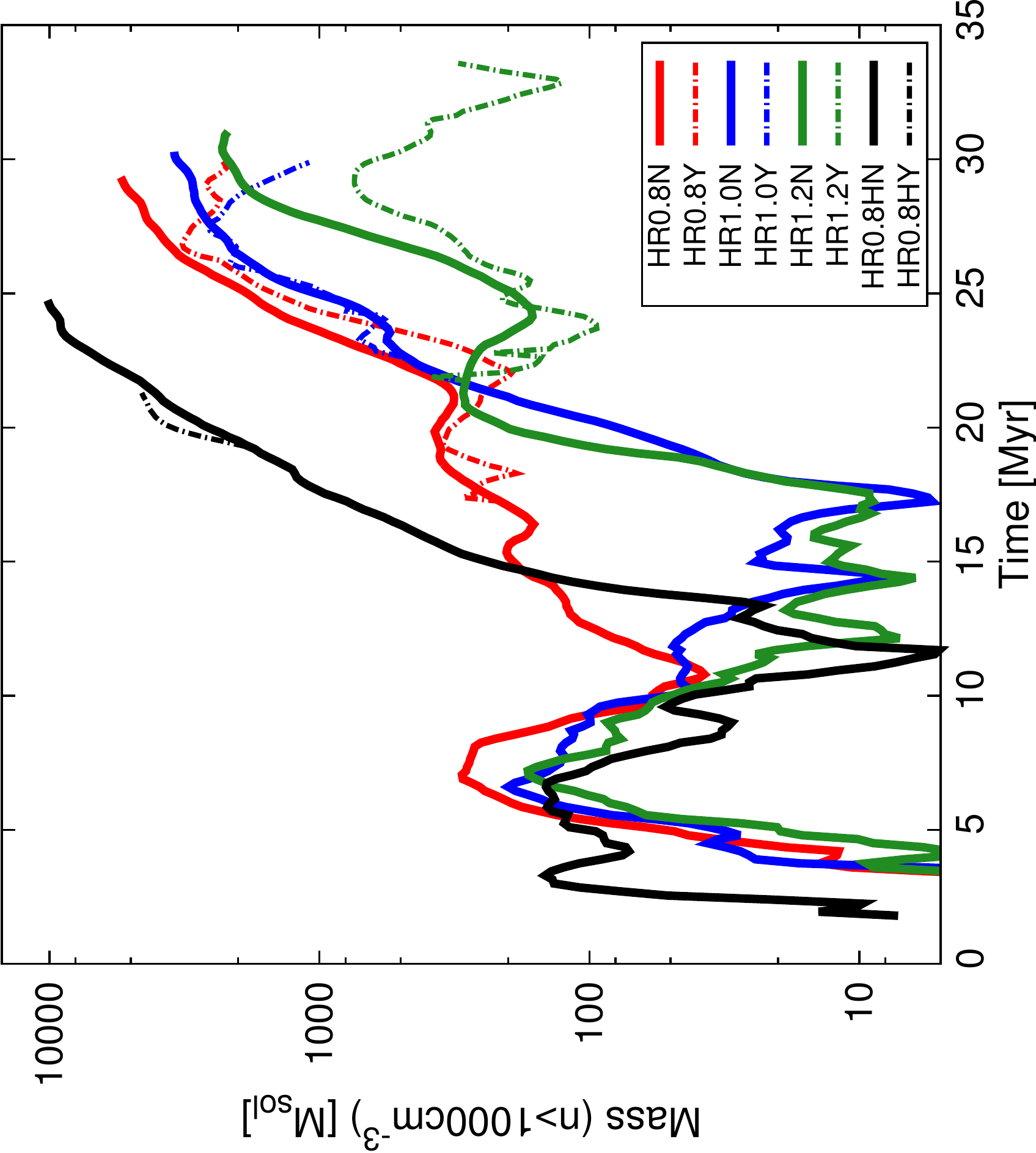}
\caption{Temporal evolution of the mass of the cloud (left, $n\geq100\,\mathrm{cm}^{-3}$) and the densest parts in the cloud interior (right,$n\geq1000\,\mathrm{cm}^{-3}$) . The data are evaluated within a cylindrical volume of radius 
$r=50\,$pc and height $h=40\,$pc. Solid lines denote the runs without feedback, dash--dotted lines those with feedback.
After the converging flows have vanished (at $t\approx 10\,$Myr) global collapse is initiated and the clouds become 
more massive. Note the different evolution of the cloud and the densest parts for HR0.8, respectively. Supernova feedback has a two--fold impact on the global evolution of the cloud mass (see text).}
\label{fig1}
\end{figure*}
The left panel in figure \ref{fig1} shows the evolution of the mass of the \ita{clouds}. 
As a \ita{cloud} in the simulations we define the regions with density of $n\geq100\,\mathrm{cm}^{-3}$, which
 must not necessarily be spatially connected. These clouds have a mean temperature of $\sim 30\,$K. The decrease of dense gas mass between 8 and 17 Myr in the 
MHD runs is a consequence of the re--expansion of the compressed material \citep[see e.g.][and references therein]{Koertgen15}. 
For the hydrodynamic runs this stage is from 5 to 12 Myr and is thus faster. This is because of the lack of a magnetic field, which would decelerate the gas. After this stage (from 12 Myr in the hydrodynamic case and from 17 Myr 
on in the MHD case), global contraction of the cloud leads 
to an increase of mass \citep[see also][]{Banerjee09a}. Generally the clouds are more massive for the hydrodynamic runs due to the 
lack of magnetic support. However, the accretion properties of the clouds seem very similar in the later stages from 20 Myr on, as can be inferred from the increase of their masses. These stages are independent of the initial conditions since the flows have vanished 
and the initial turbulence has fully decayed. In the end, the clouds have masses between $M_\mathrm{cloud}=2\times10^{4}-4\times10^{4}\,\mathrm{M}_\odot$. The difference of the cloud masses for the MHD runs is due to 
the fact that initially the stronger turbulence disperses the gas more efficient. This also prevents the build up of a massive cloud.\\
The impact of supernova feedback on the clouds is two--fold. On the one hand, the first supernova explosion results in a compression of the surrounding gas, thereby \ita{increasing} the total mass of the dense gas. 
On the other hand, this period of compression lasts only until the point, where parts of the clouds are heated up and dispersed by the transmitted shocks. The 
increase of the mass depends on where the supernova goes off,that is, either in a compact or a structured environment \citep[see also][]{Iffrig15}. For example, \citet[][]{MLi15} point out that supernovae going off in a 
structured environment do interact with the high--density regions, but the long--term impact of the remnant is primarily
 onto the low--density medium. A supernova going off in a structured environment sweeps up less mass as its 
counterpart going off in a compact/homogeneous medium of same average density.  For runs HR0.8Y and HR0.8HY, the cloud is compact enough to provide a significant obstacle to the 
emerging shock wave. An increase of mass is also seen in the cloud of run HR1.2Y, although the increase takes a longer time. % due to the more inhomogeneous cloud. 
In contrast, the supernova explosion in run HR1.0Y results in only a 
small increase of a few hundred solar masses. The denser regions are not significantly compressed. From the  stage of dispersion of parts of the cloud on, the cloud mass stays lower in comparison to the clouds without feedback. If more and more supernovae go off, the growth in 
mass is either stopped or turned into a stage of decreasing mass (as in case HR1.0Y). The total decrease in cloud mass is between a factor of 1.5--2, in agreement with a previous study by 
\citet[][]{Iffrig15}. However, the efficiency of supernova feedback, that is, its impact onto the dense gas (e.g. 
reduction of the total gas mass), depends on the initial turbulence within the flows and thus the final cloud mass and (mean) density. Please note that the 
negligible difference in the cloud mass in runs HR0.8HN and HR0.8HY is coincidental and may increase with 
continuation of the simulation.

\subsubsection{Evolution of the densest parts}
In the right panel of figure \ref{fig1} we present the evolution of the densest parts of the molecular cloud with densities of $n\geq1000\,\mathrm{cm}^{-3}$.
The evolution of the densest parts is similar to the evolution of the cloud. The strong fluctuations during the early stages indicate that these regions are diluted due to the energy injection from the 
WNM flows and turbulence. All clouds undergo a lateral expansion phase (from $\sim 7\,\mathrm{Myr}$) during which 
the amount of high--density gas decreases.
However, the densest regions in run HR0.8N do not show such variation. The more compact cloud interior is nearly 
unaffected by the re--expansion of the cloud and keeps on accreting. %This indicates that dispersion is primarily restricted to the outer edges of the cloud. 
The later evolutionary stages of all clouds -- from 
15 Myr on for the hydro case and from 17 Myr on for the MHD simulations -- are dominated by global cloud contraction. Nevertheless, some of the high--density material can still be dispersed by the momentum injection 
and turbulence delivered by the colliding streams. %\far{All clouds undergo an lateral expansion phase ($\sim 7$ to $17\,\mathrm{Myr}$), while the mass of 
%the dense gas is decreasing.
%\far{The periods of mass loss seen e.g. at around $t\sim16-17\,\mathrm{Myr}$ in run HR0.8N and at around 
%$t\sim24\,\mathrm{Myr}$ in run HR1.2N can be attributed to turbulent dispersion (compare e.g. with figure \ref{figekinegra} where an increase of the kinetic energy is observed at these times).}\\
The supernova explosions yield periods of varying total mass in the densest parts. This is due to dilatational and compressive stages and is seen in all runs. In the end, the mass of the densest parts in the MHD runs is reduced by factors of about three for HR0.8Y and HR1.0Y to of about ten for HR1.2Y. 
%In contrast, the hydro simulation only indicates a period of compression. However, at this stage almost 10\,\% of the mass are contained in the dense parts, which may affect the evolution of the SNR. 

\subsection{Cloud Dynamics}
\begin{figure*}
\includegraphics[height=0.33\textwidth]{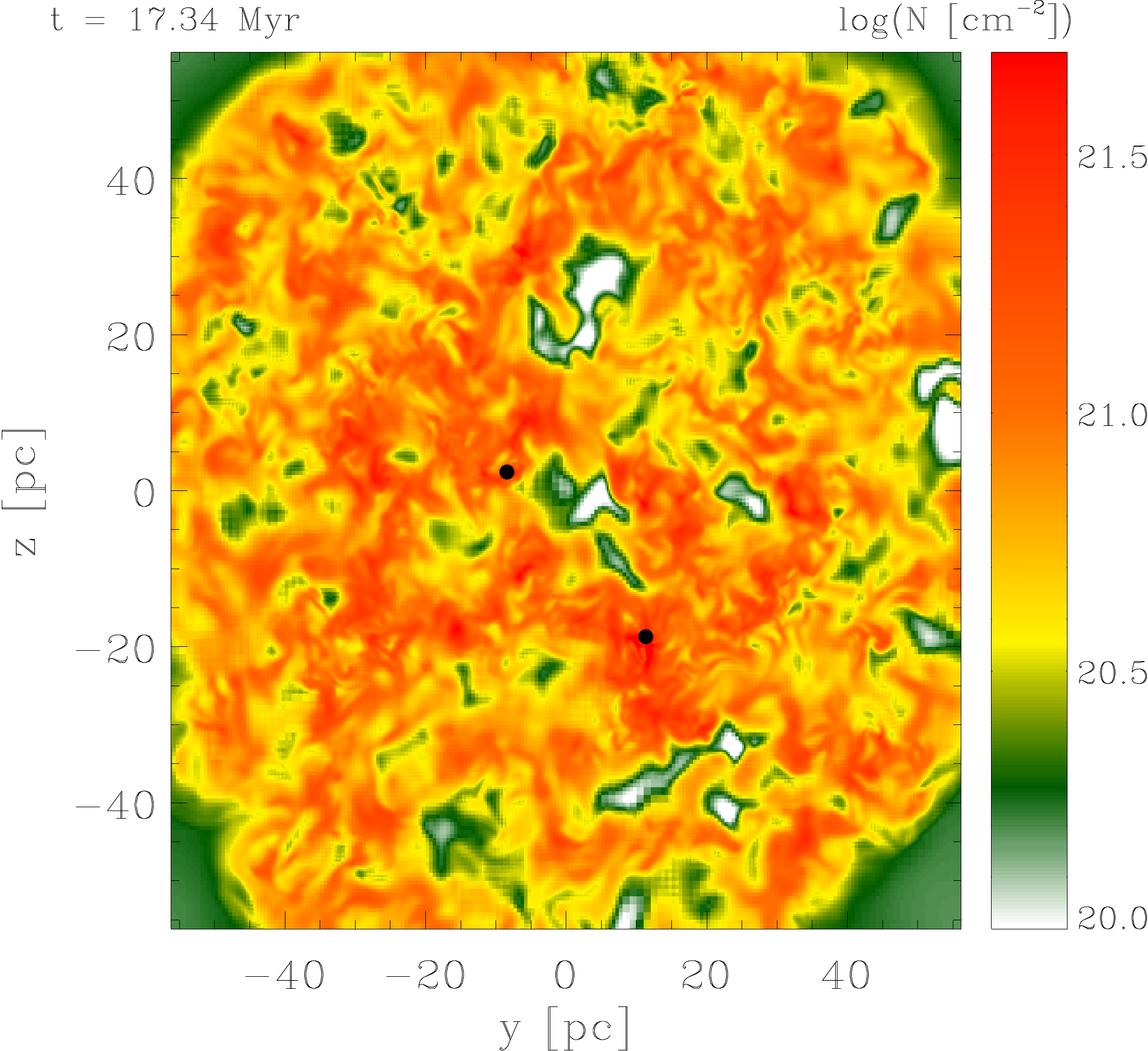}\includegraphics[height=0.33\textwidth]{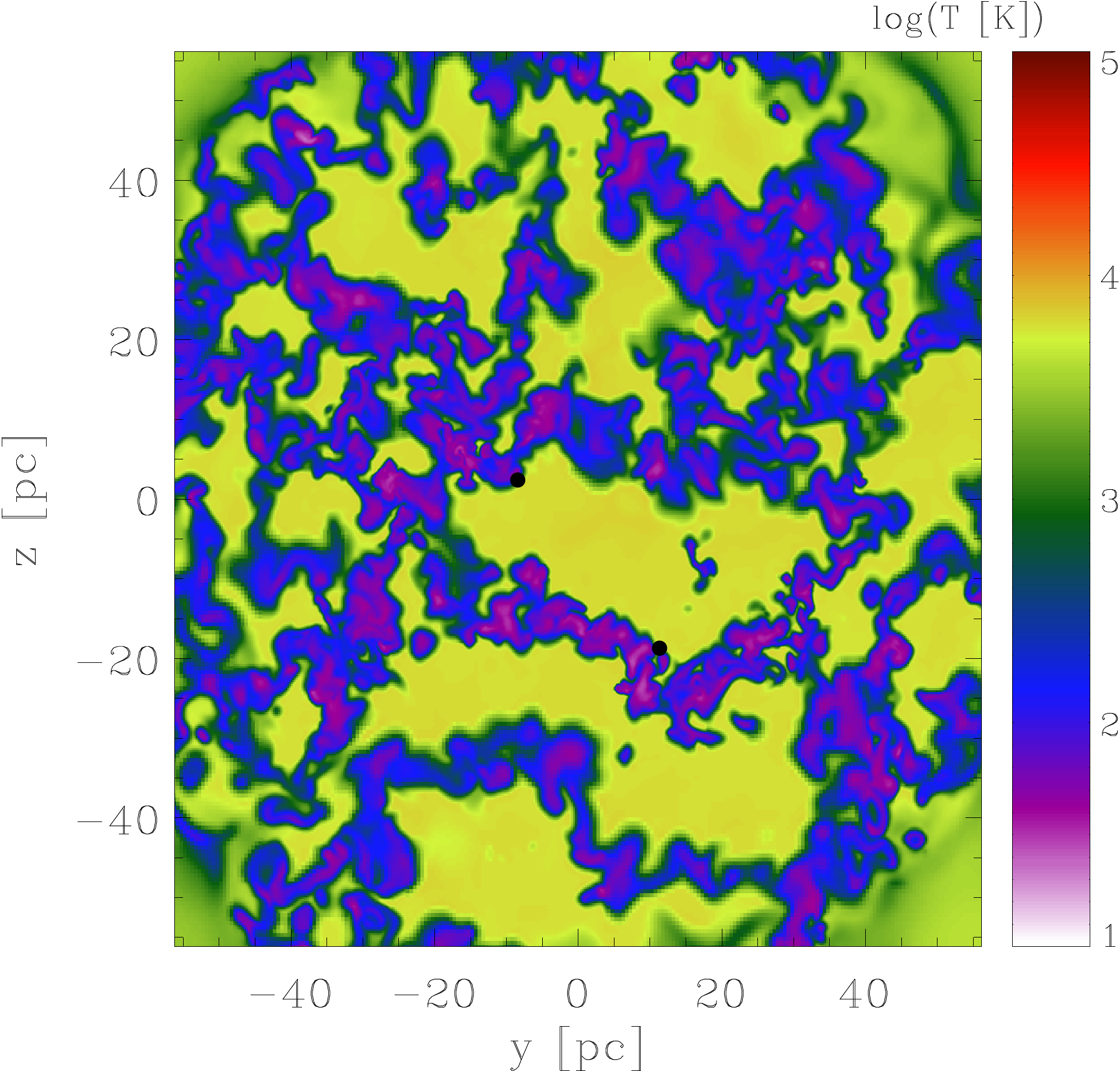}\includegraphics[height=0.33\textwidth]{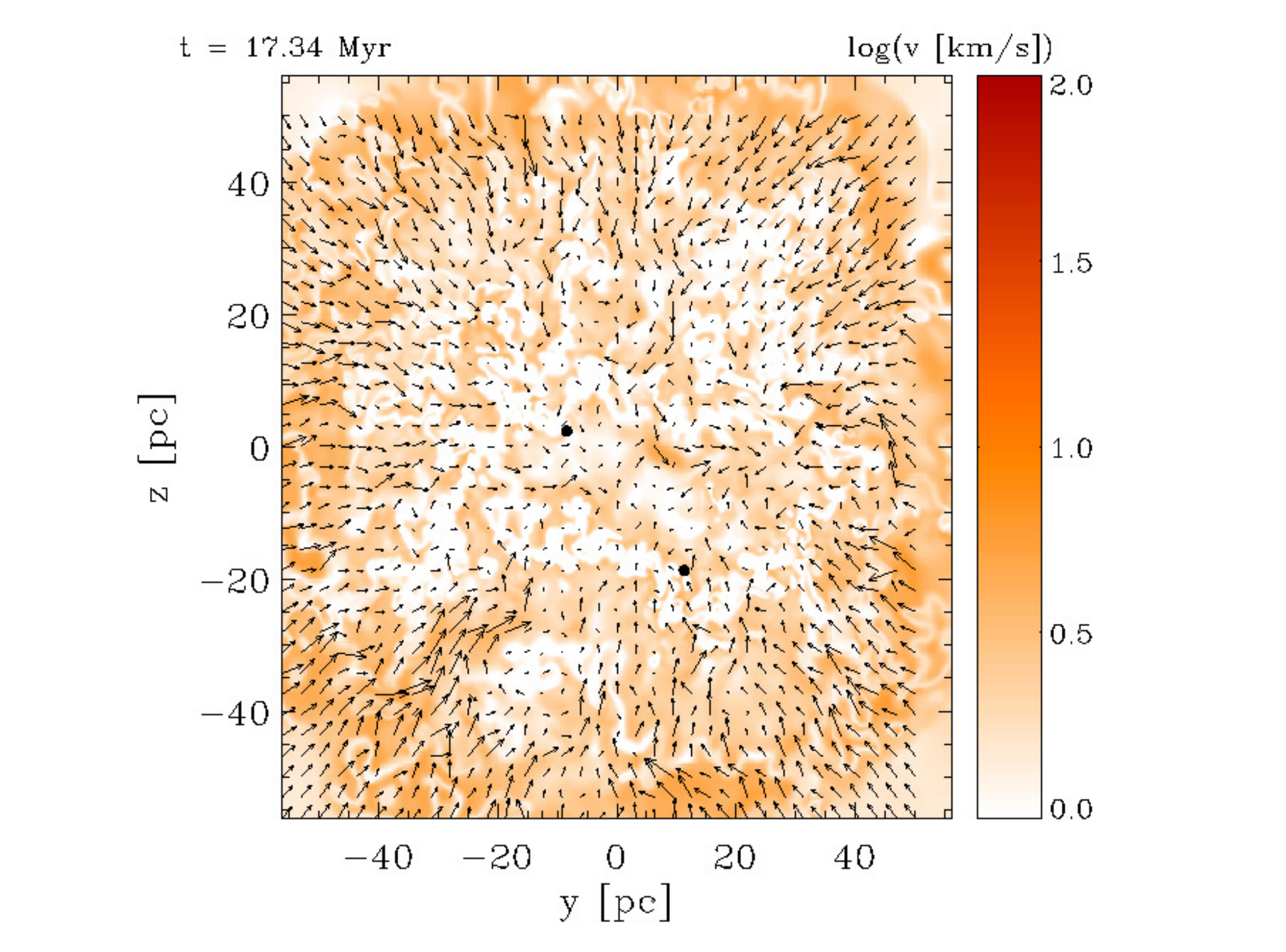}\\
\includegraphics[height=0.33\textwidth]{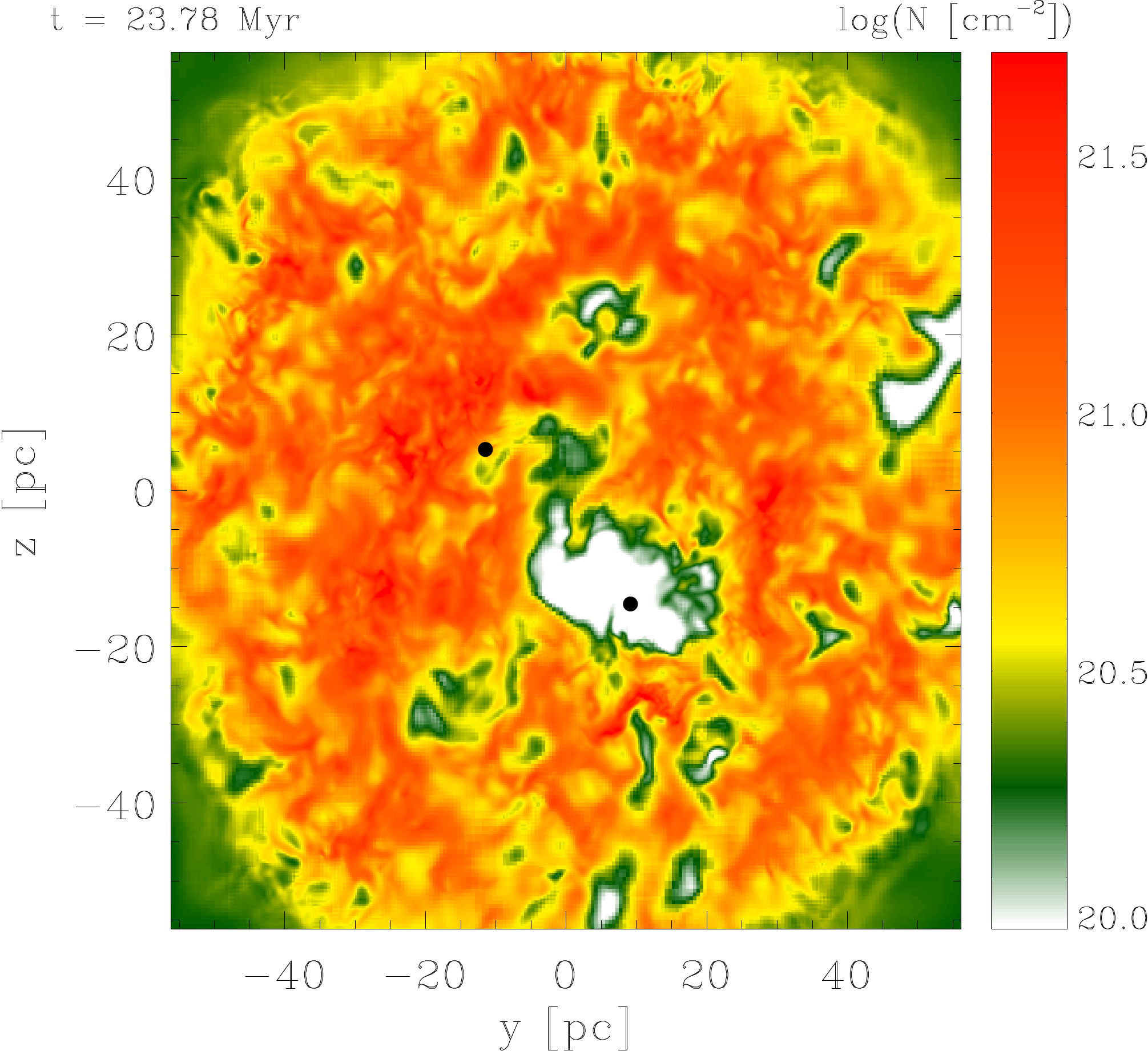}\includegraphics[height=0.33\textwidth]{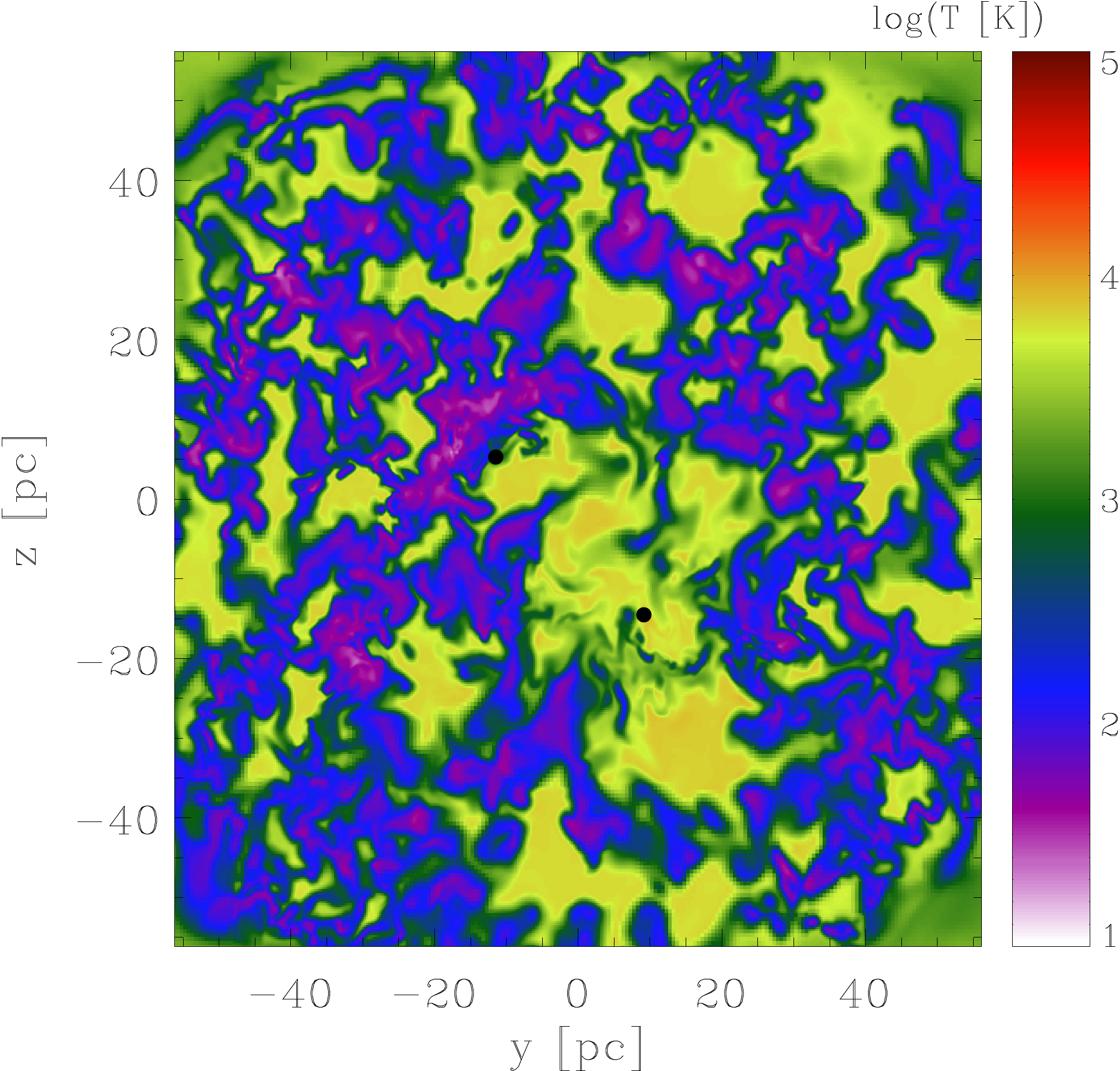}\includegraphics[height=0.33\textwidth]{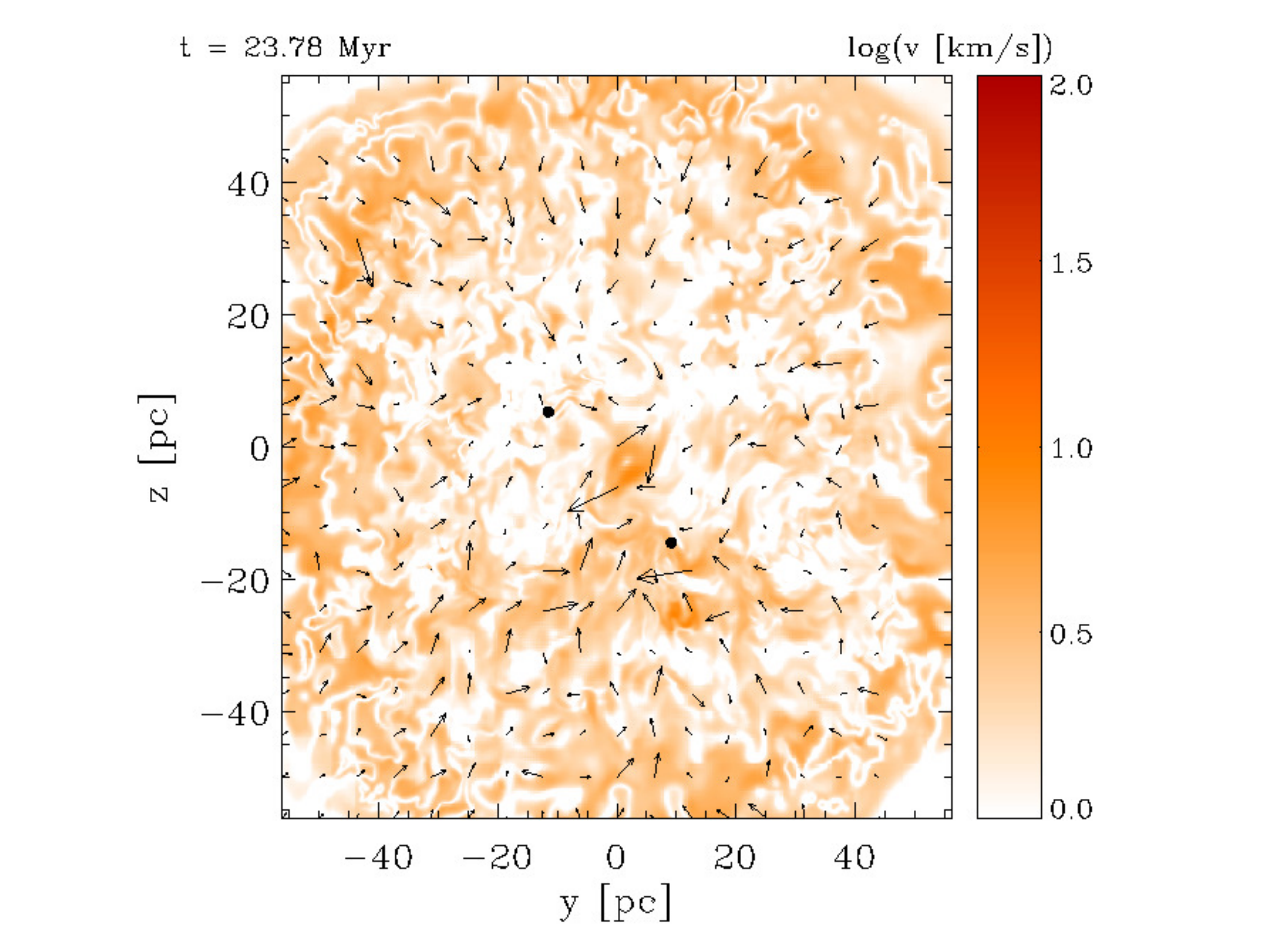}\\
\includegraphics[height=0.33\textwidth]{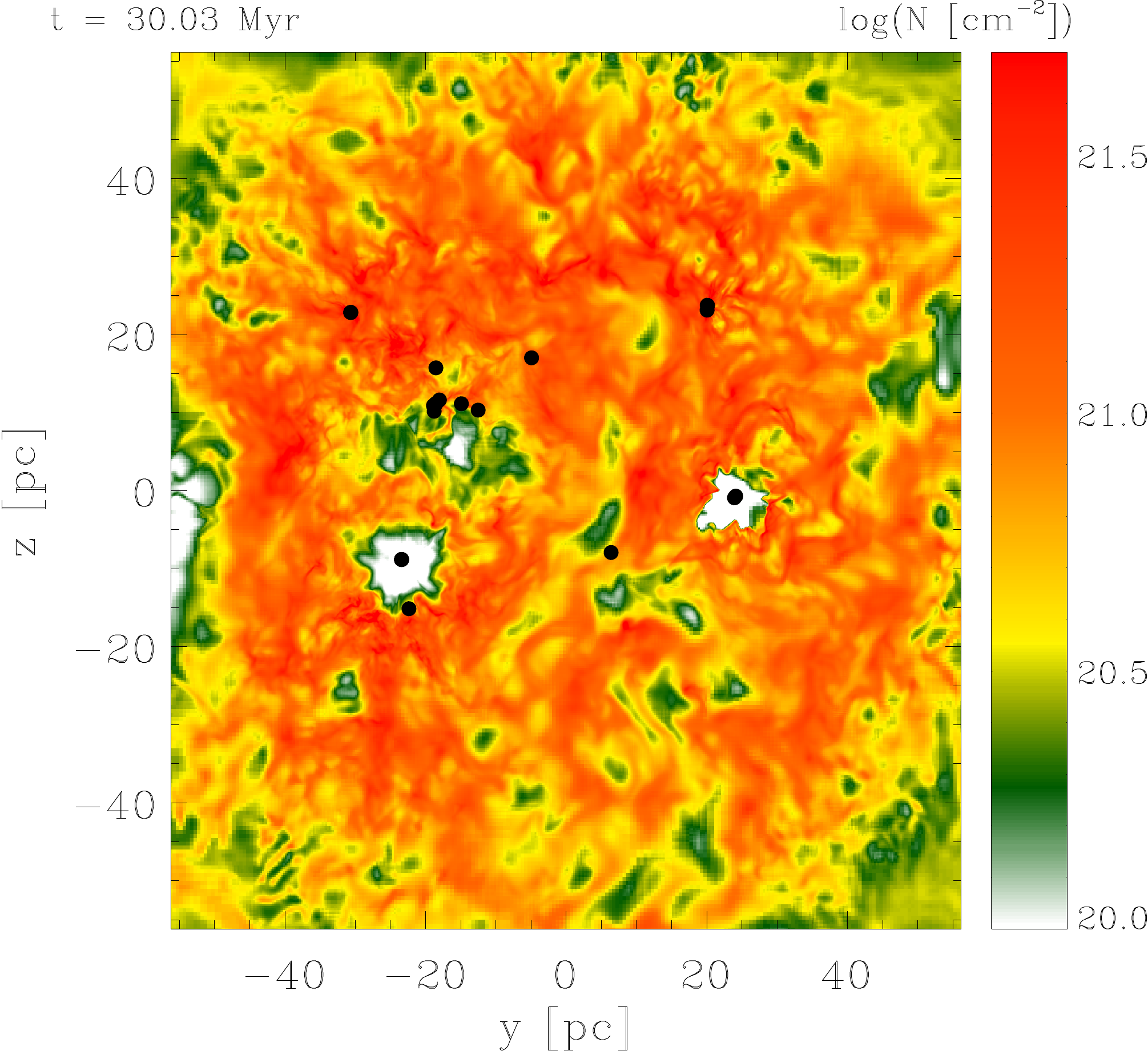}\includegraphics[height=0.33\textwidth]{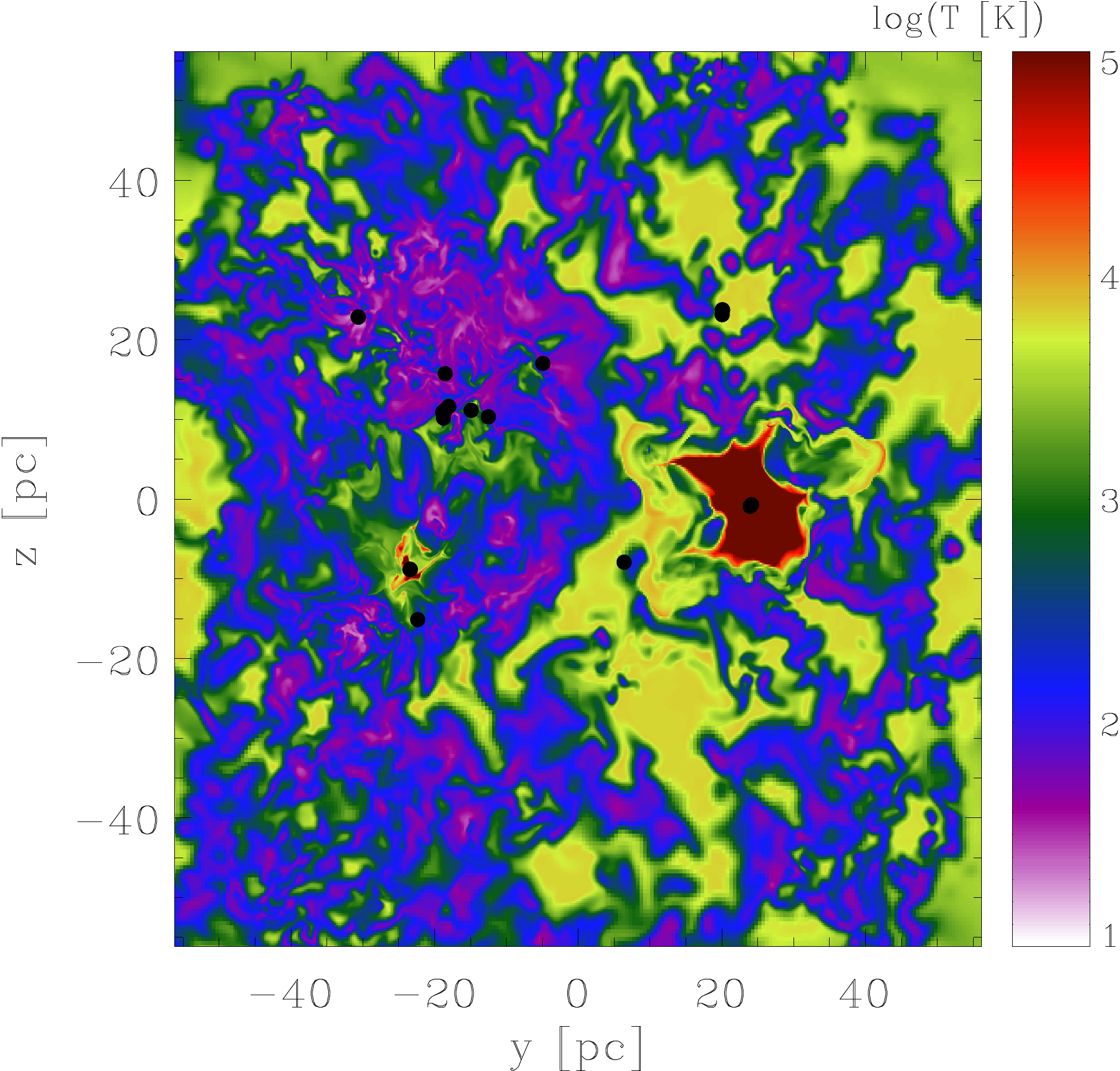}\includegraphics[height=0.33\textwidth]{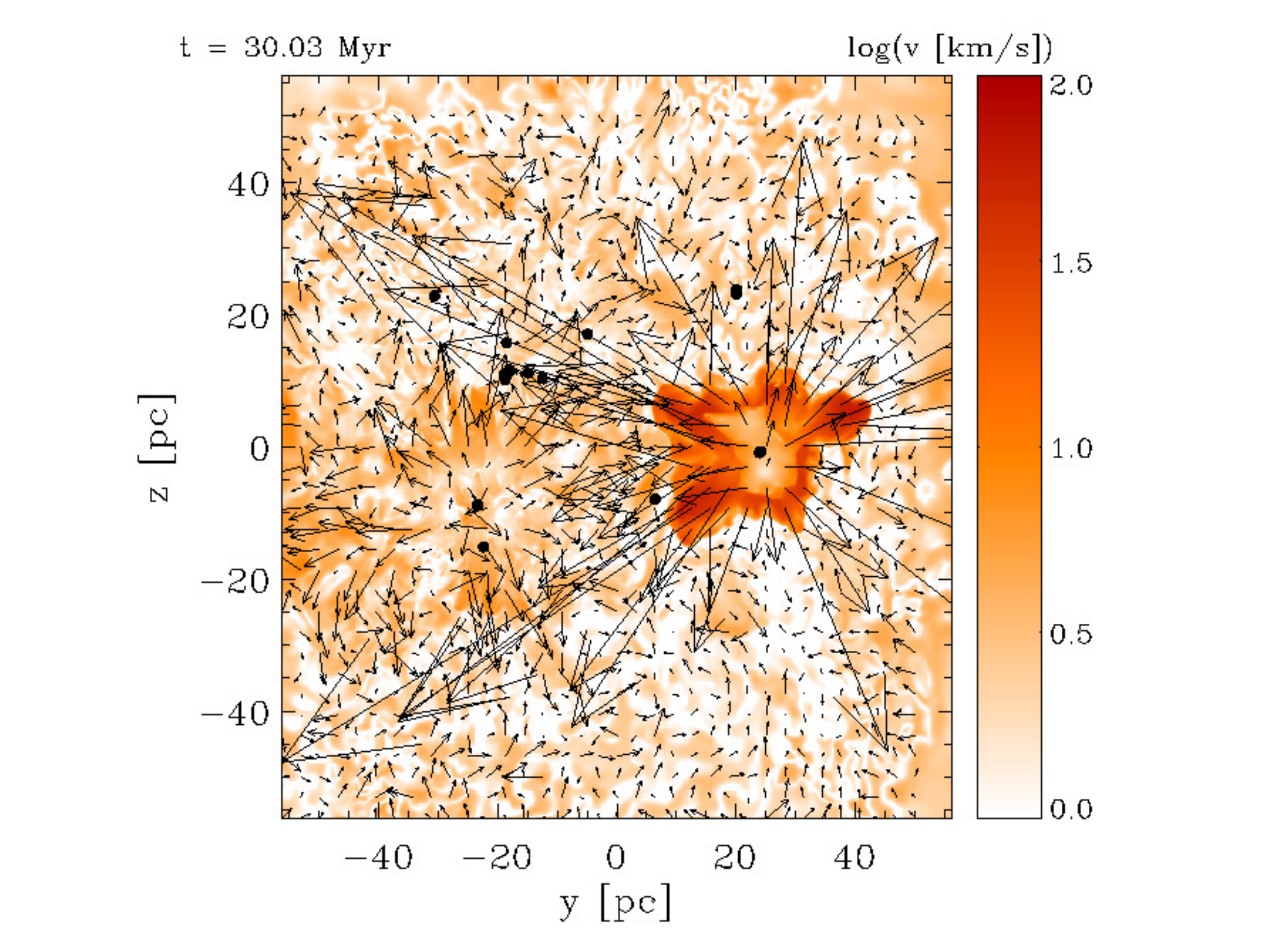}
\caption{Data of run HR0.8Y. \ita{Left to right:} Column density along the x--axis (parallel to the WNM streams, integration length is 40\,pc), temperature, 
and absolute value of the velocity in the midplane (x=0). \ita{Top to bottom:} Different evolutionary stages prior to and after supernova feedback. It is clearly seen that 
supernovae do not have a huge impact on the cloud dynamics and structure. The effects are only localised to some 
small regions of a few tens of parsec. In all cases, the typical vector in the velocity plots has a magnitude of 
$v_\mathrm{typ}=5\,\mathrm{km/s}$. Note that the vector arrows are plotted with a linear scale. Sink particles are 
represented by the black dots.}
\label{fig2d}
\end{figure*}

Figure \ref{fig2d} shows a temporal sequence of the column density, temperature and total velocity for 
run HR0.8Y. The first row shows the molecular cloud 30\,kyr \ita{before} the first supernova, the other two rows after supernovae have gone off. Prior to the first supernova, the cloud reveals a filamentary network and clumps as well as low--density 
cavities in between. This is a result of the interaction of turbulence and gravity, mediated by the ambient 
magnetic field \citep[e.g.][]{Hennebelle13,Chen14,Koertgen15}. The filaments are best being identified in the temperature slice as the cold branches with temperatures of about $T=30\,\mathrm{K}$, immersed in a warm medium with 
$T=1000-5000\,\mathrm{K}$.\\
%When supernovae go off, it is interesting to see that the \ita{global} cloud morphology does not change after four supernovae have 
%gone off (last row). The supernovae drive shock waves, which interact with the surrounding gas by compressing it.
 %Thus, most of the energy is converted into compressive work done on the gas.
The supernovae act only locally since the shock terminates after $\sim15\,\mathrm{pc}$.
The cold, dense gas is redistributed, as is best seen in the temperature slices. %However, some part of the injected energy escapes through low--density channels within the molecular cloud \far{and hence reduces the influence of the 
%supernovae}. 
In addition, the column density map before the 
first supernova and at the end of the simulation reveal different patterns at the cloud outskirts. Prior to supernova feedback, the outer border of the cloud is more spherical. After supernova feedback,  the outskirts are more structured and the
 density of the WNM halo surrounding the cloud is higher. These parts of the molecular cloud do also reveal a slower inward--directed 
 velocity, as can be identified from the length and orientation of the velocity vectors in the velocity pattern.
\begin{figure*}
%\centering
\begin{tabular}{cc|cc}
$\mathcal{M}_s=1.2$,no Feedback &$\mathcal{M}_s=1.2$,SN Feedback &$\mathcal{M}_s=1.0$,no Feedback &$\mathcal{M}_s=1.0$,SN Feedback\\
\includegraphics[height=0.23\textwidth]{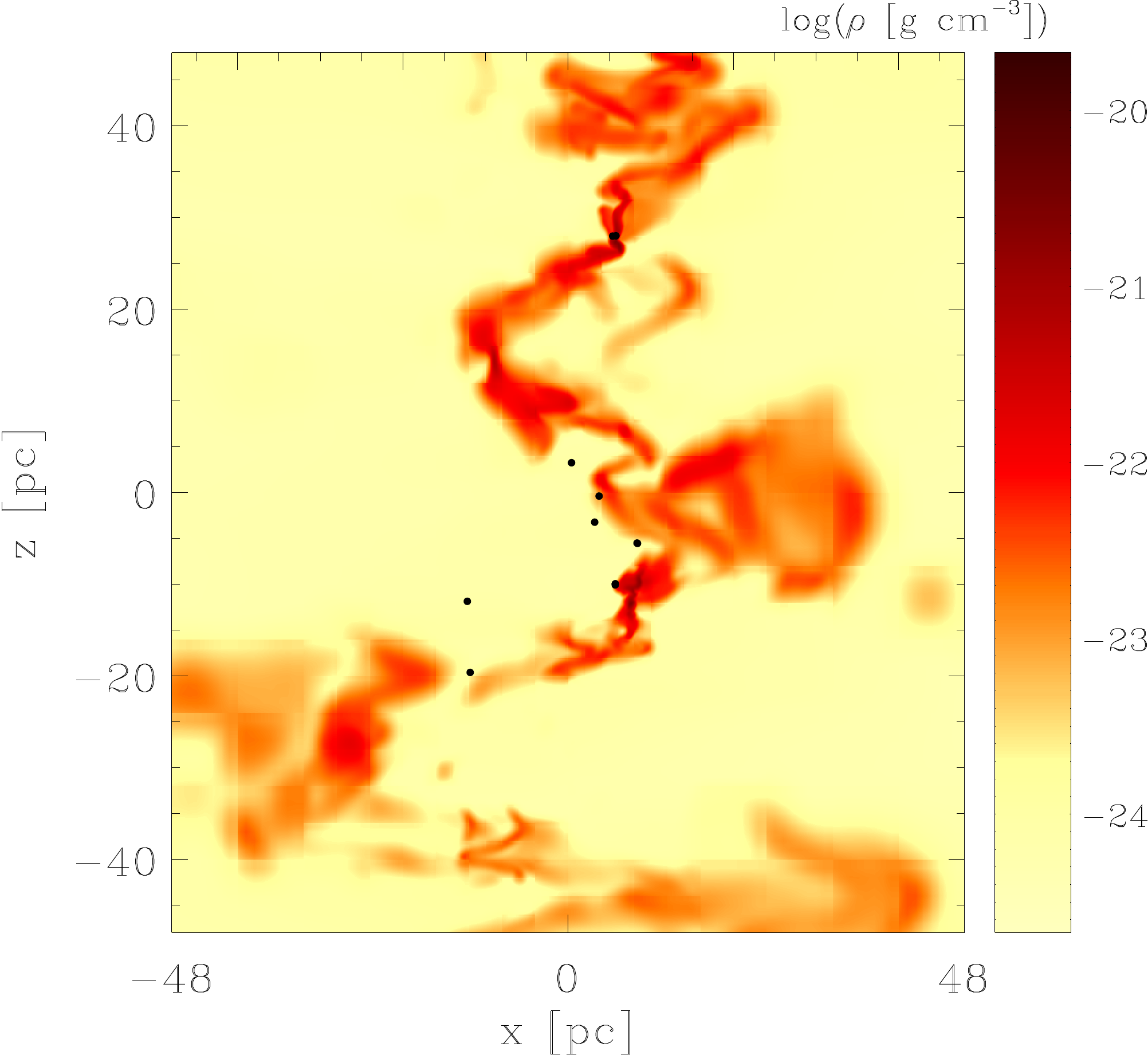}&\includegraphics[height=0.23\textwidth]{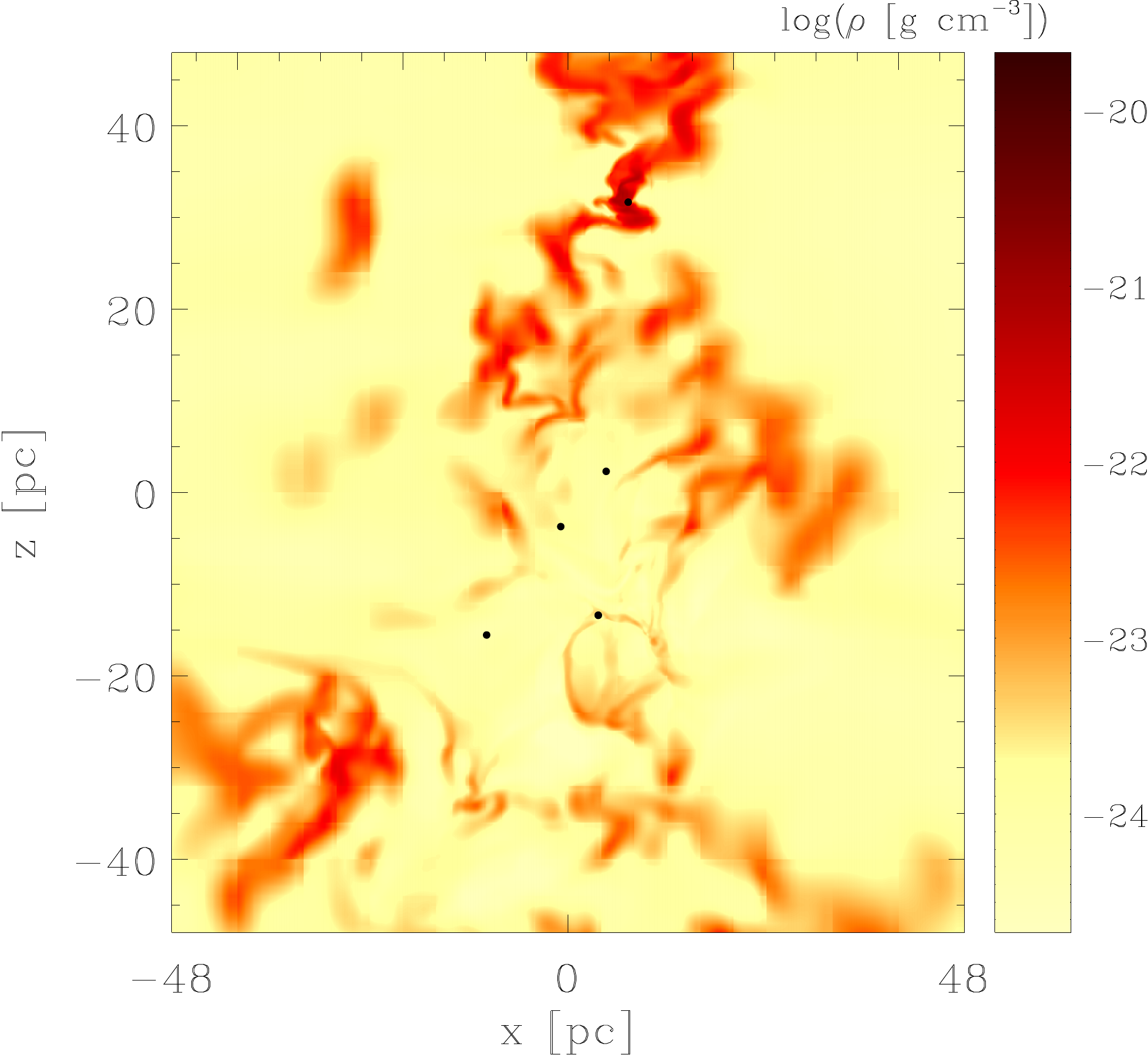}&\includegraphics[height=0.23\textwidth]{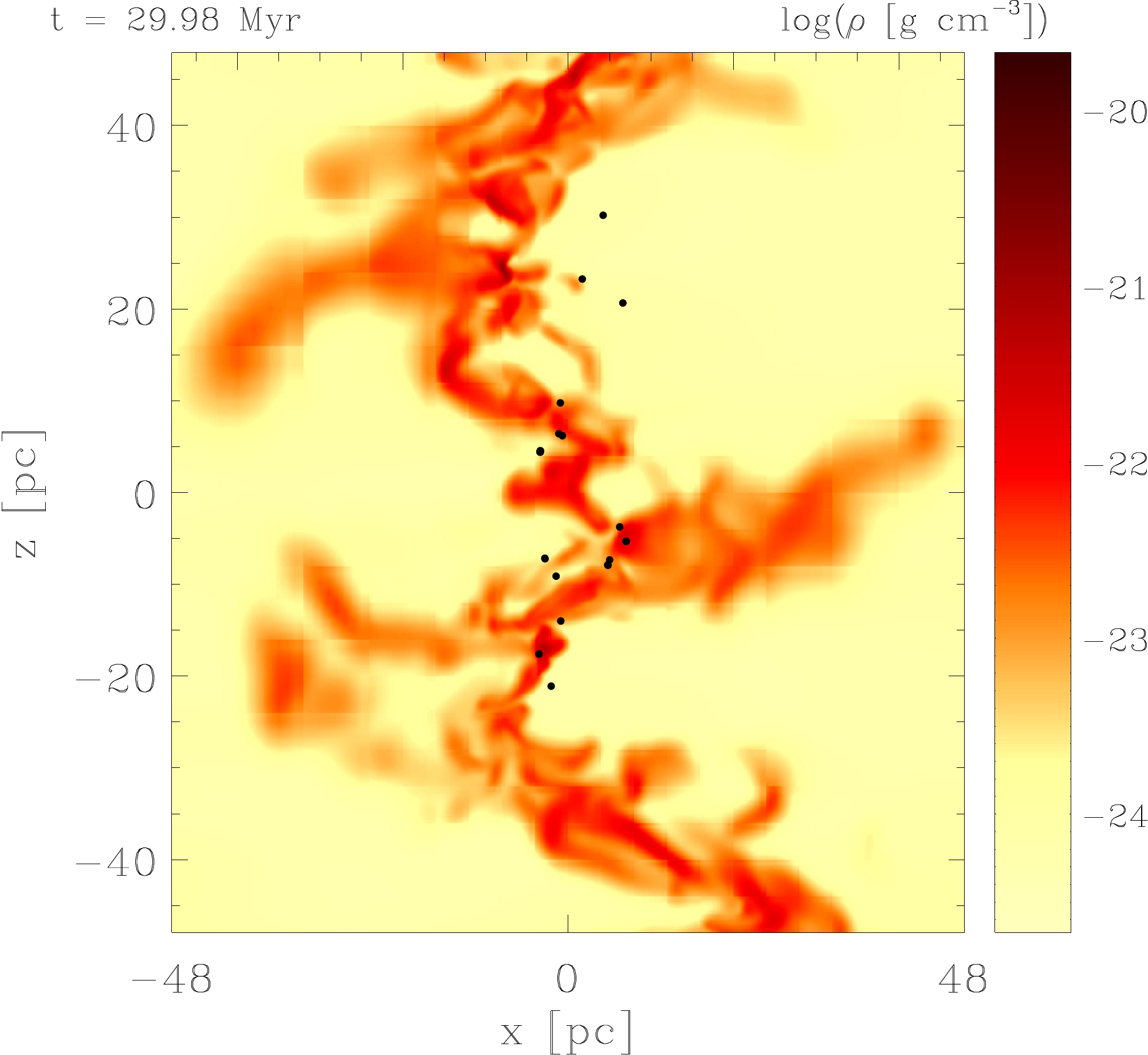}&\includegraphics[height=0.23\textwidth]{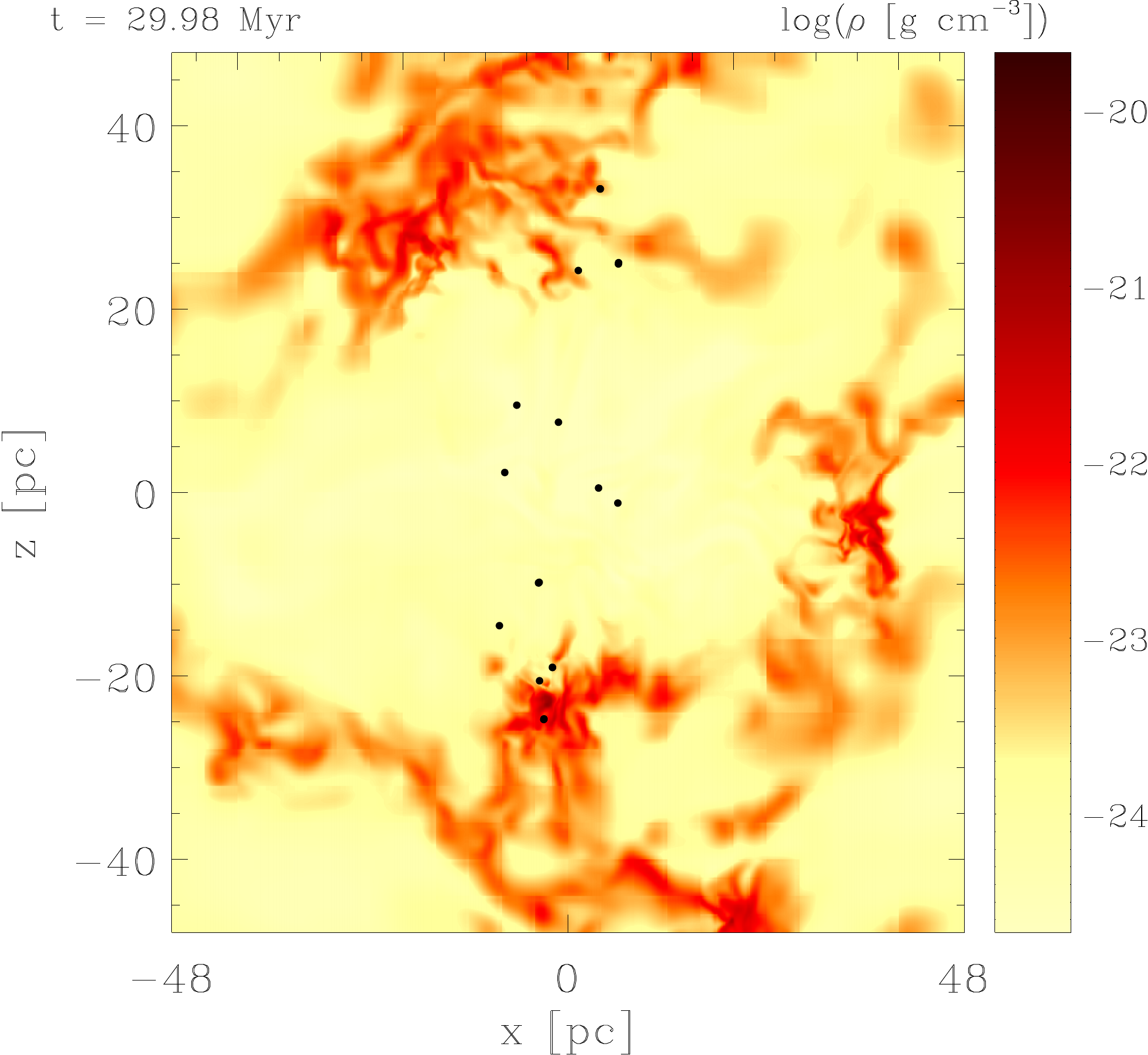}\\
\includegraphics[height=0.23\textwidth]{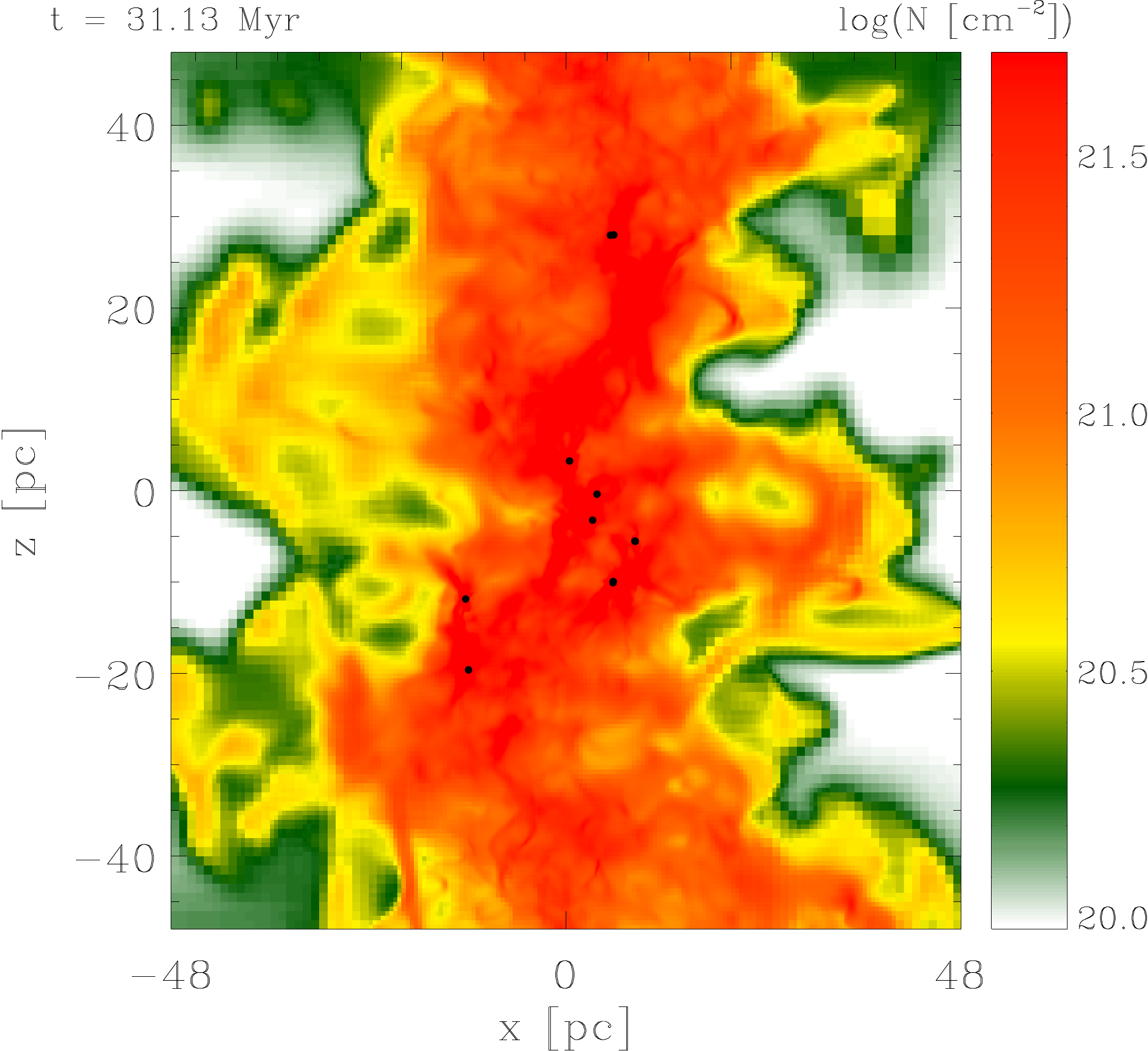}&\includegraphics[height=0.23\textwidth]{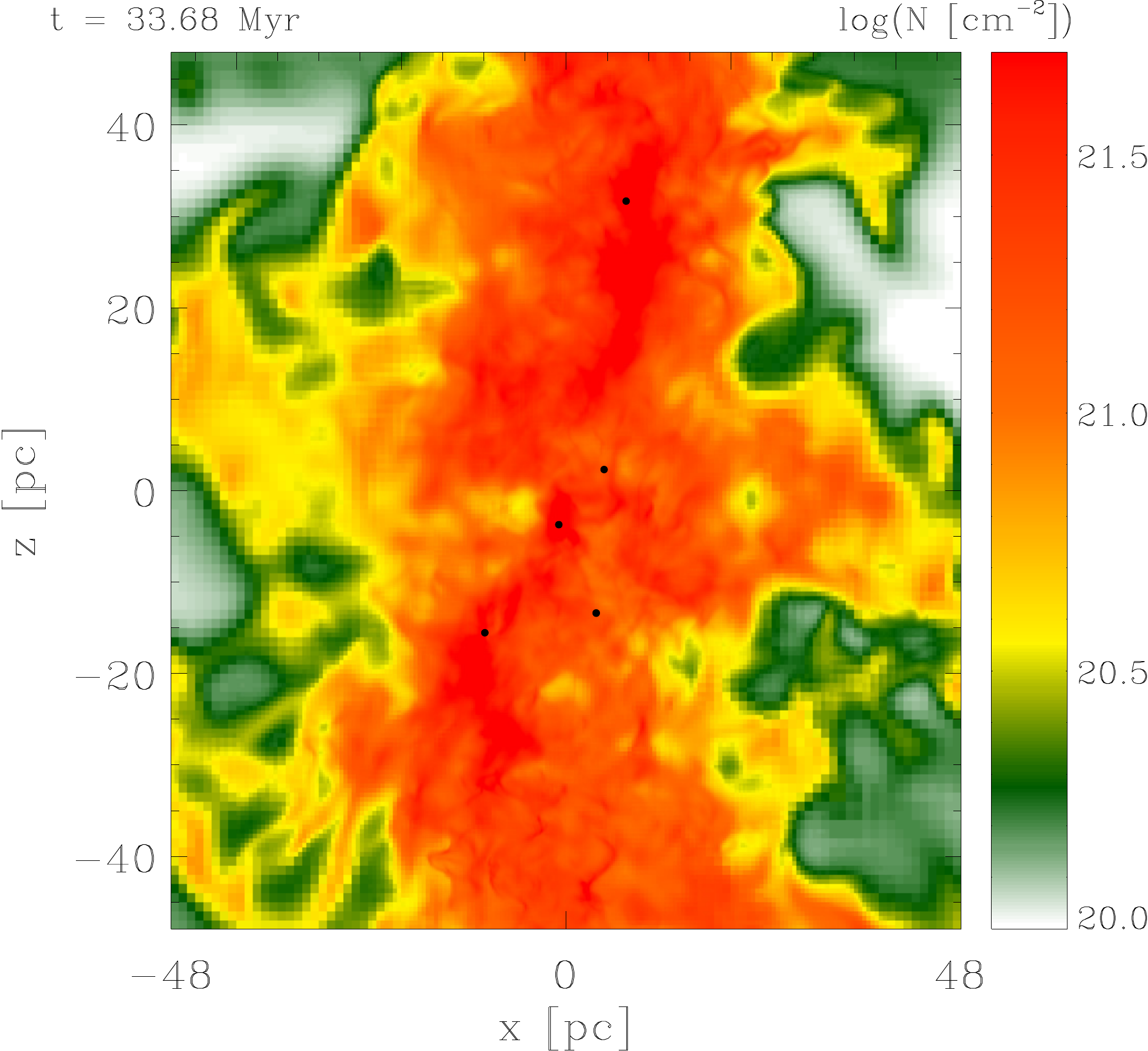}&\includegraphics[height=0.23\textwidth]{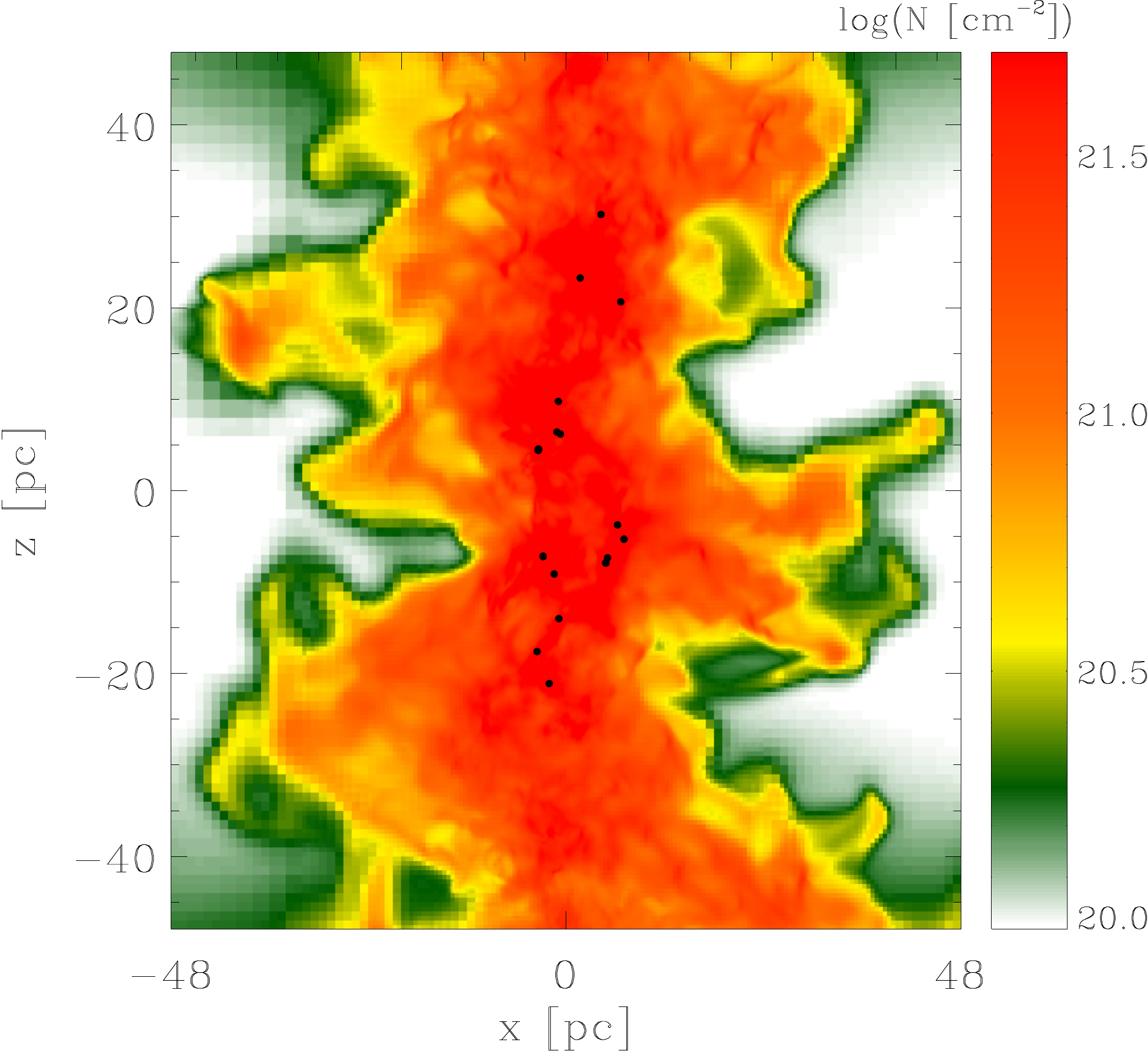}&\includegraphics[height=0.23\textwidth]{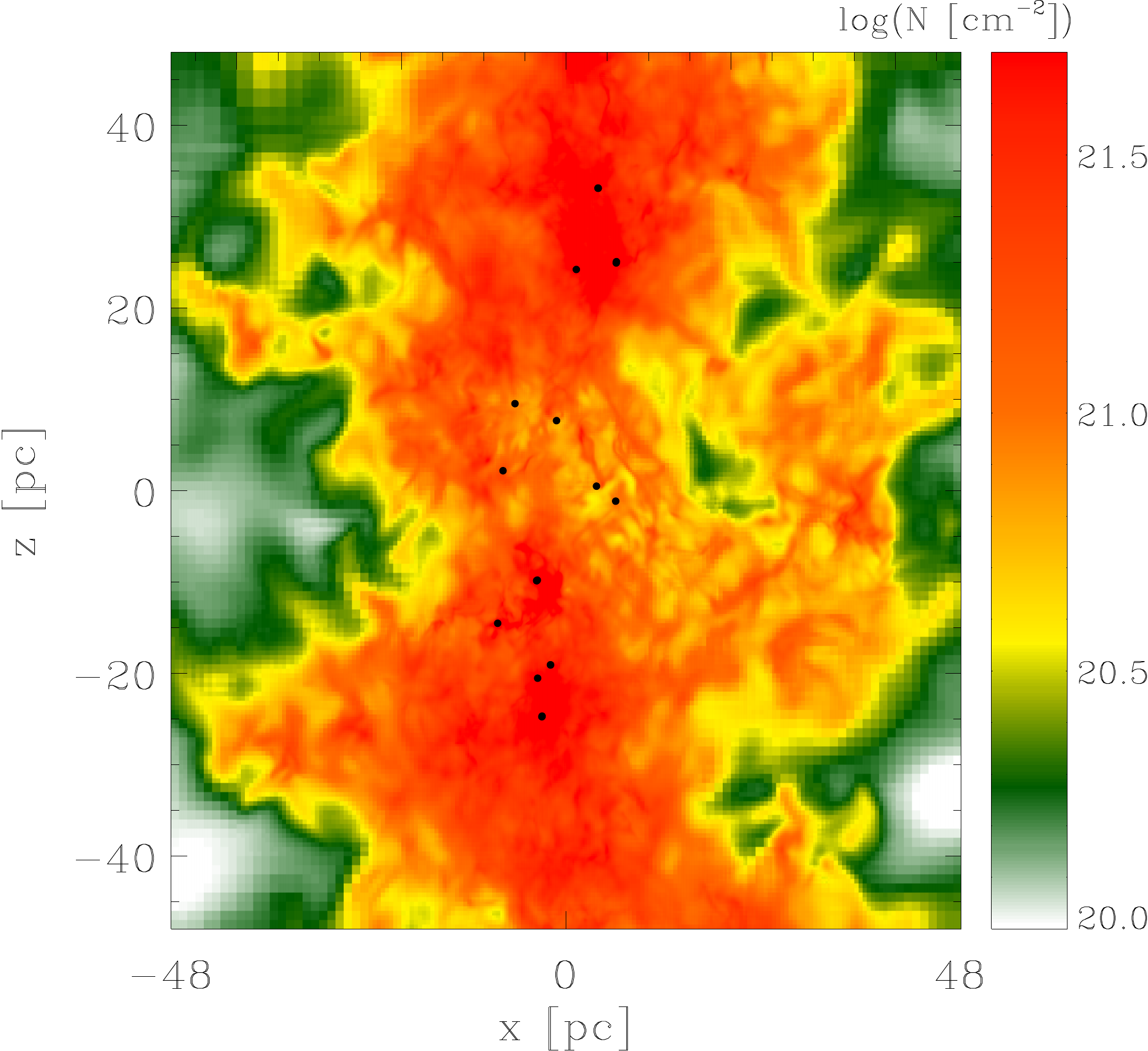}\\
\includegraphics[height=0.23\textwidth]{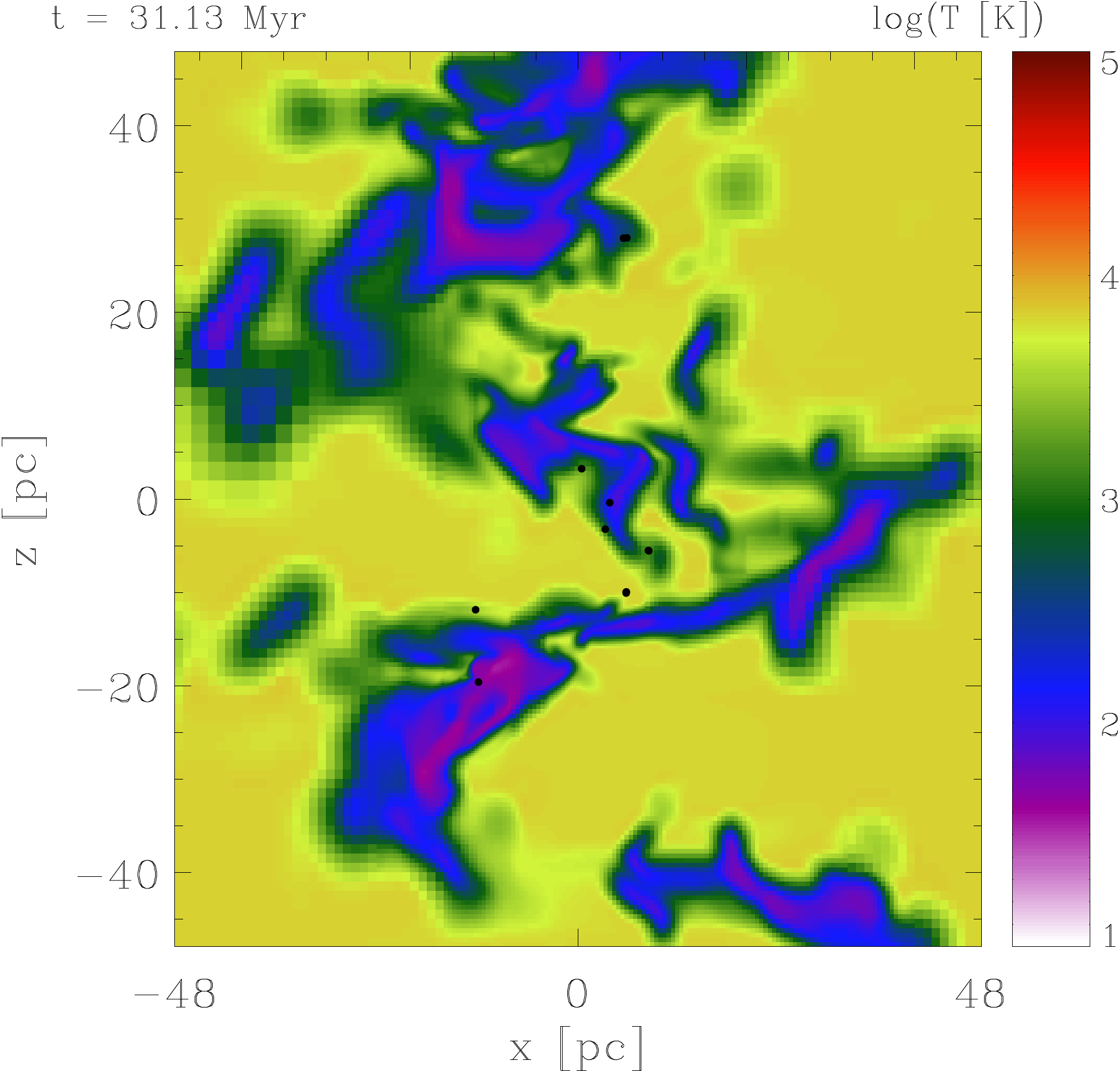}&\includegraphics[height=0.23\textwidth]{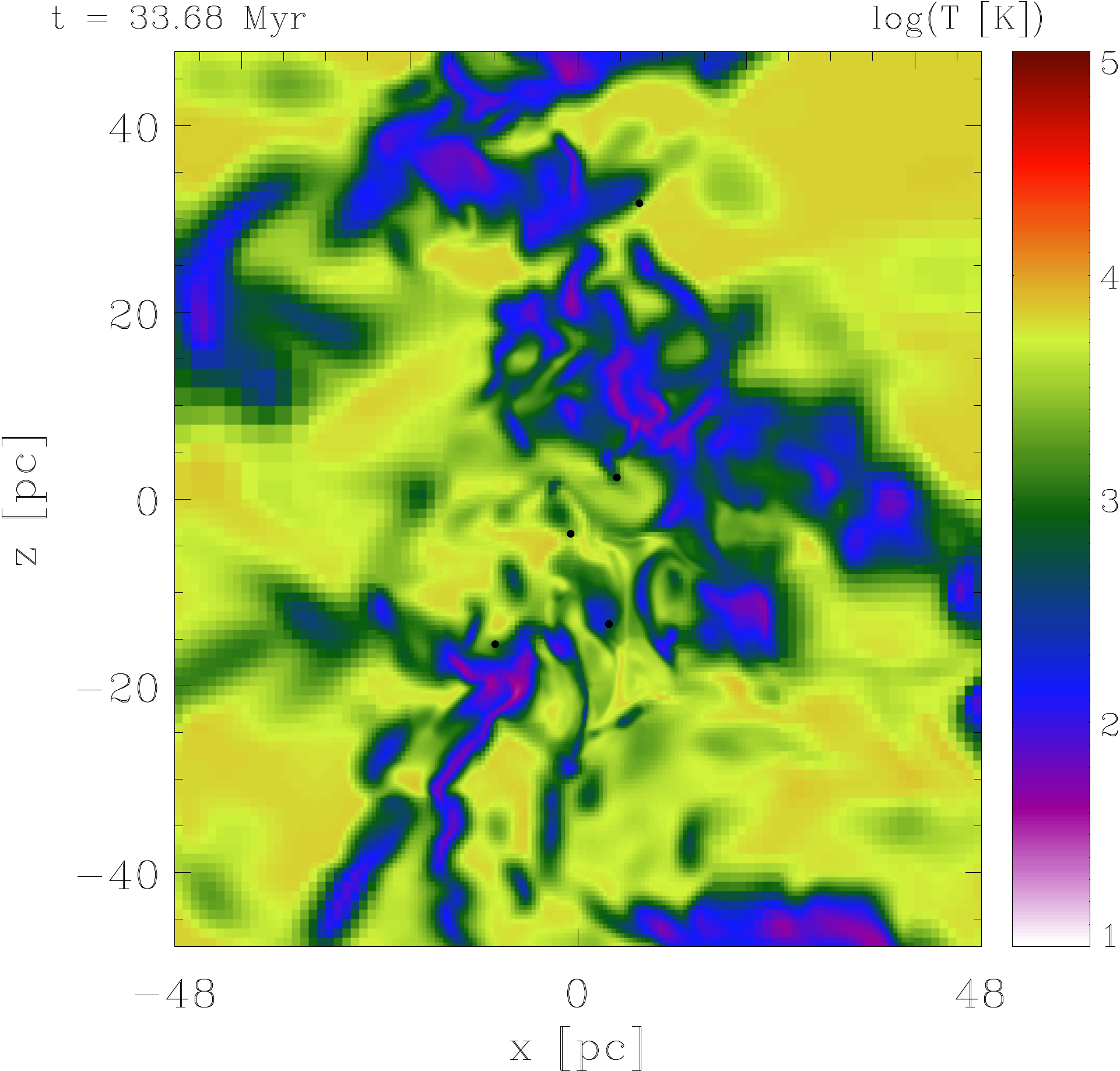}&\includegraphics[height=0.23\textwidth]{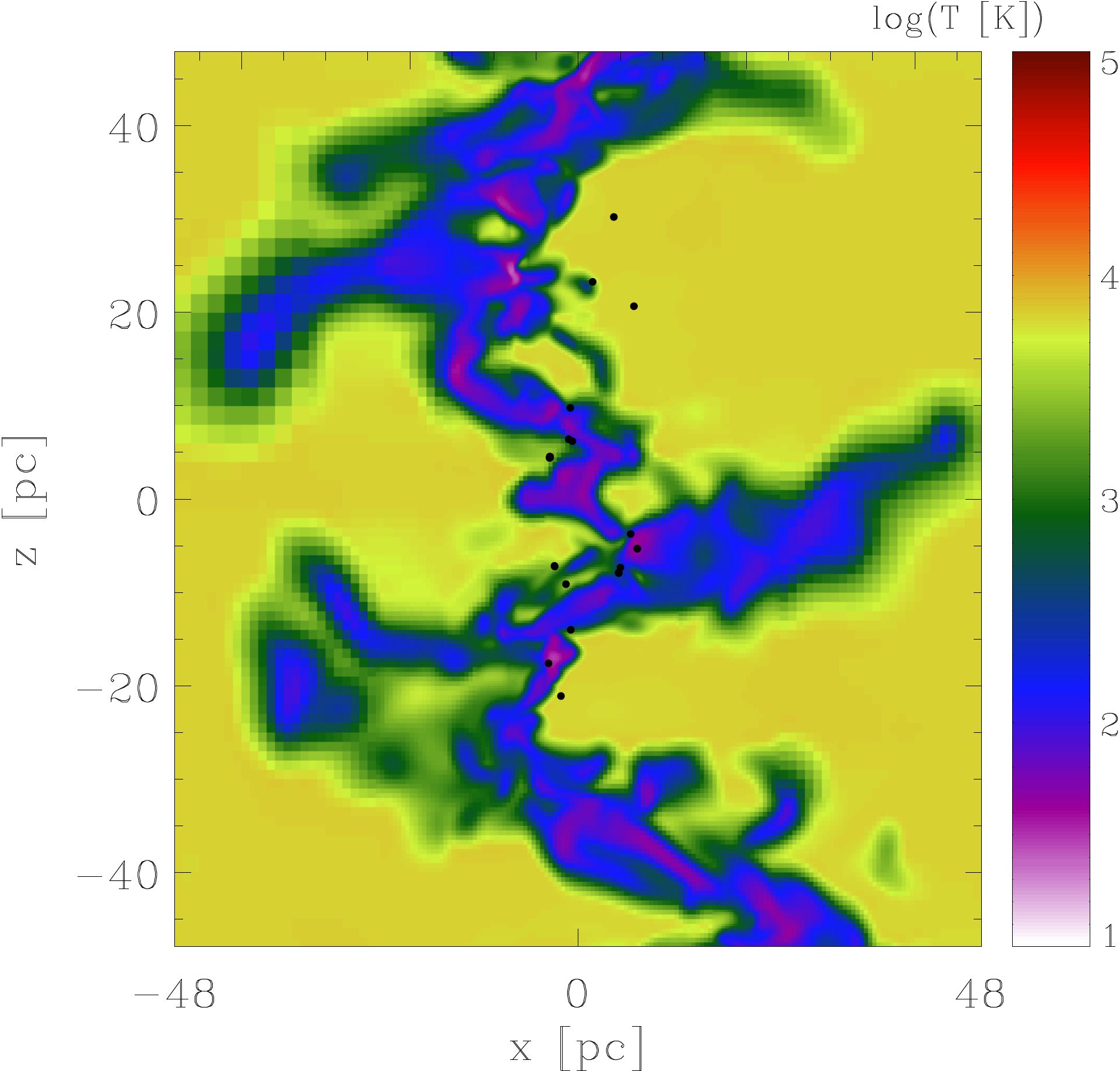}&\includegraphics[height=0.23\textwidth]{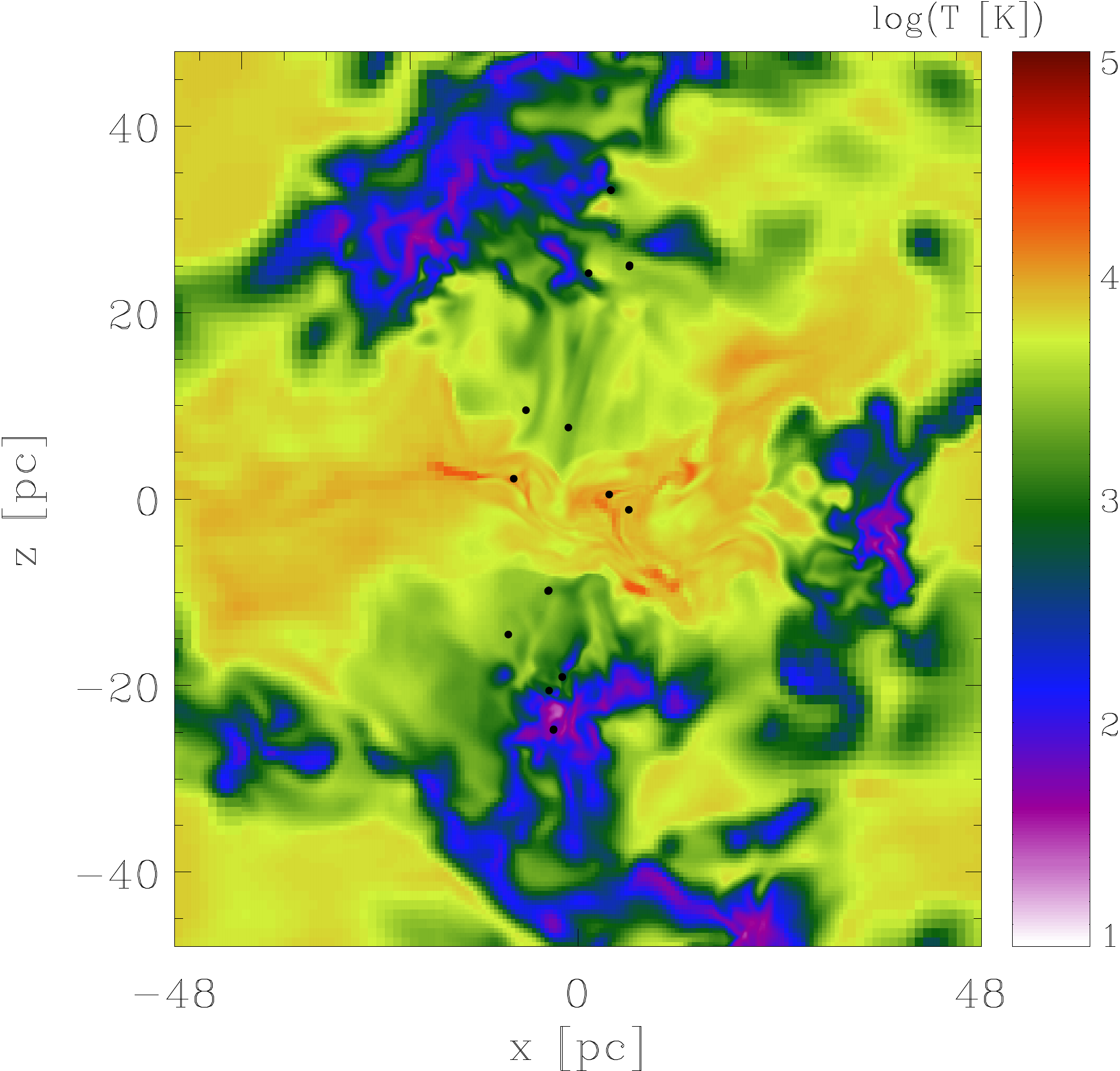}\\
\includegraphics[height=0.23\textwidth]{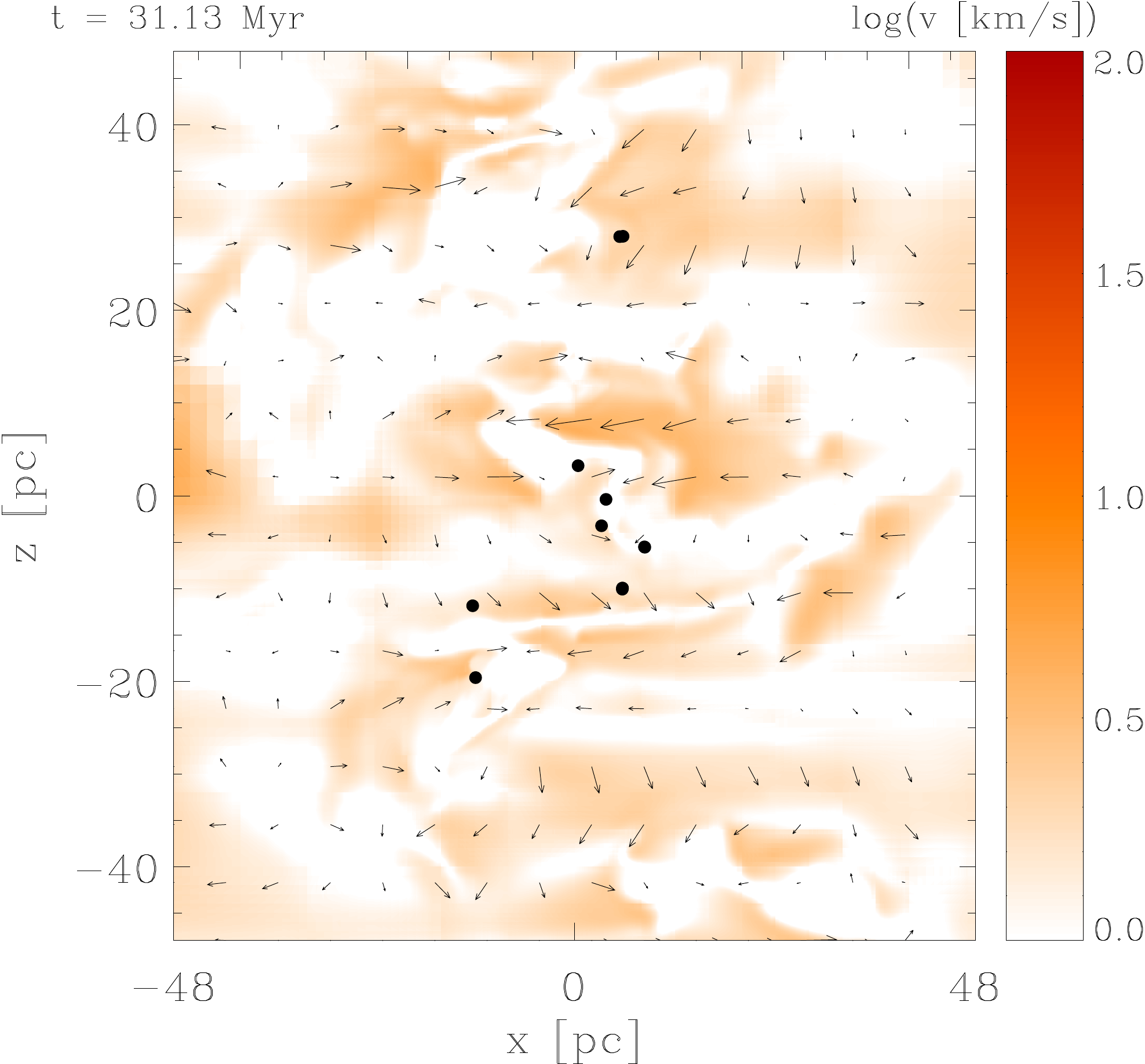}&\includegraphics[height=0.23\textwidth]{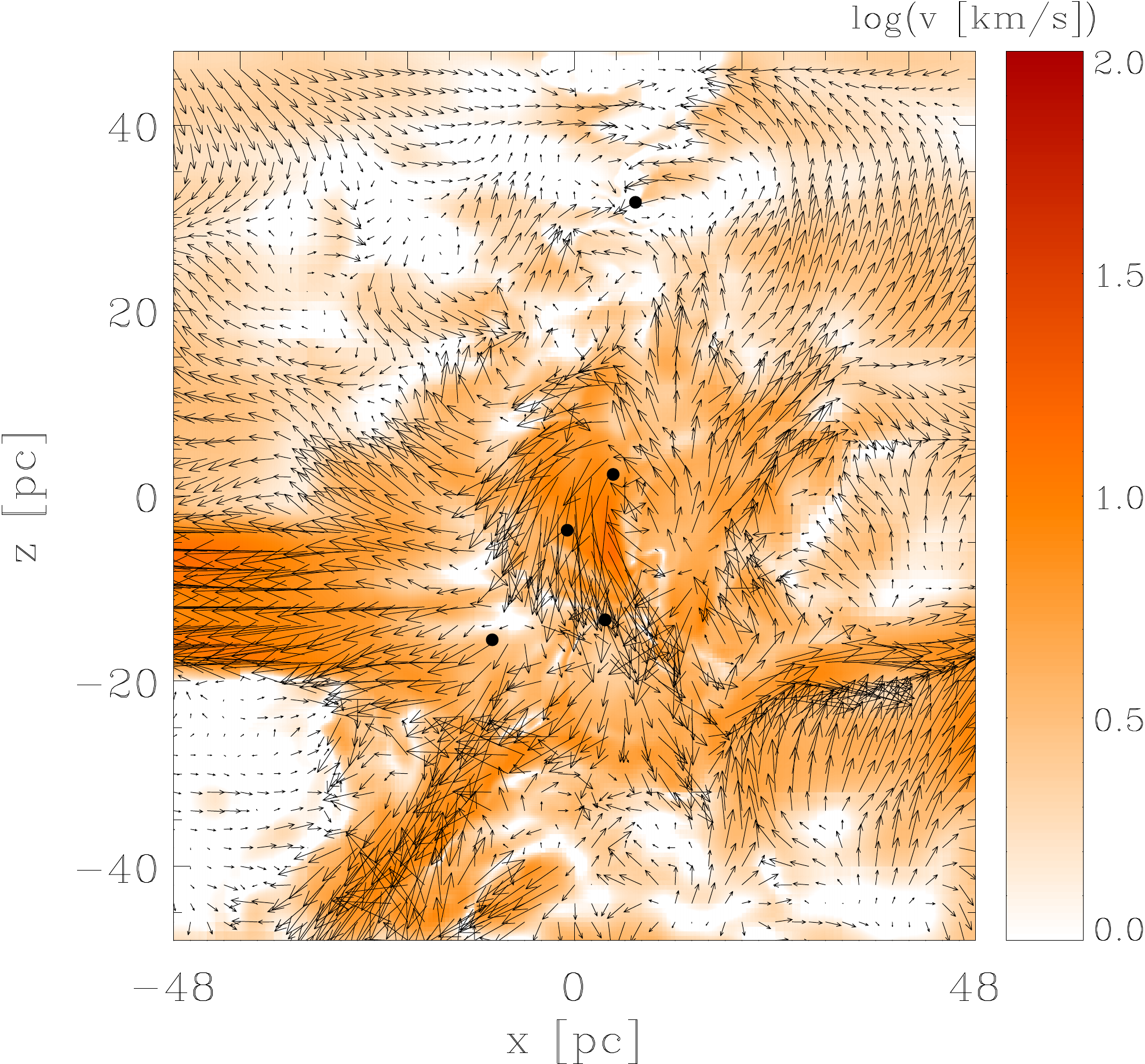}&\includegraphics[height=0.23\textwidth]{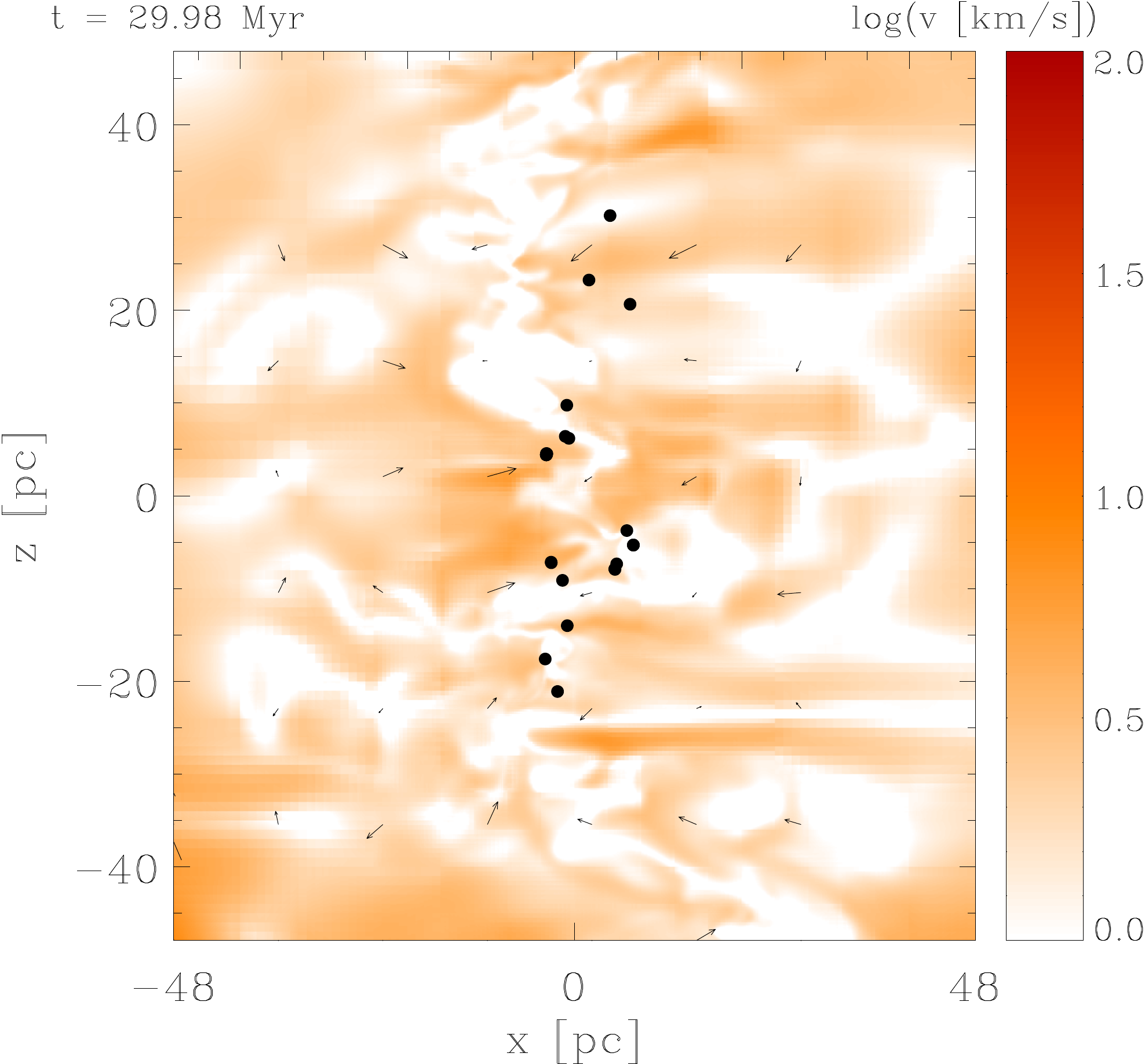}&\includegraphics[height=0.23\textwidth]{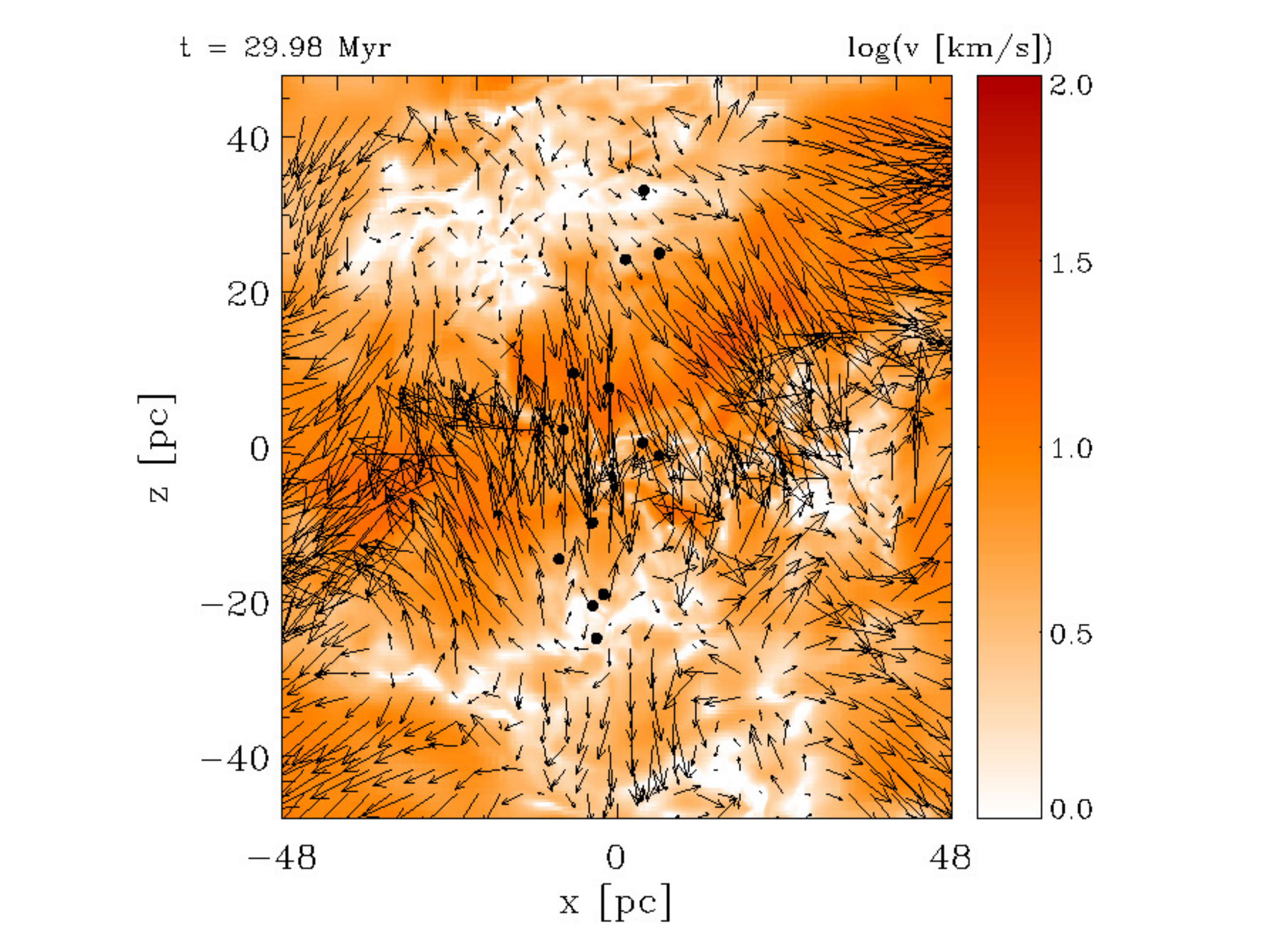}\\
\end{tabular}
\caption{From top to bottom: Density in the y=0--plane (for runs HR1.0) and in the y=+15\,pc--plane (for runs HR1.2), column density along the y--axis (perpendicular to the WNM streams), temperature, and velocity magnitude. Left two columns for runs HR1.2N and HR1.2Y, respectively.
Right two columns for runs HR1.0N and HR1.0Y, respectively. The cloud is dispersed within a localised region. However, the column density 
map reveals a cloud with a teneous region in its centre. The latter indicates that supernovae are not able to disrupt the whole cloud. The length of a typical vector is the same as in figure \ref{fig2d}.  Note that the vector arrows are plotted with a linear scale.}
\label{fig2d2}
\end{figure*}
Figure \ref{fig2d2} shows the final stage of runs HR1.0N, HR1.0Y, HR1.2N, and HR1.2Y in the xz--plane. The effect of the supernovae is more drastic for runs HR1.0N,HR1.0Y. This is because most of the massive stars have exploded within a distinct region in the centre of the molecular 
cloud. However, this effect is clearly seen only in the slices of the midplane (y=0). The column density map shows a 
compact molecular cloud with a slightly teneous region in its centre. Again, the supernovae result in a redistribution of matter, which 
is seen by the increased column density at the outer edges of the cloud. 

%\subsubsection{Phase Space Diagrams}
\subsubsection{Thermal State of the Cloud}
\begin{figure*}
\begin{tabular}{c c c}
%&\LARGE{\fat{$M_\mathrm{RMS}=0.8$}}	&\\%[0.2cm]
%\hline
\Large{\fat{19.3 Myr}}&\Large{\fat{24.3 Myr}}&\Large{\fat{29.3 Myr}}\\

\includegraphics[height=5.3cm,angle=-90]{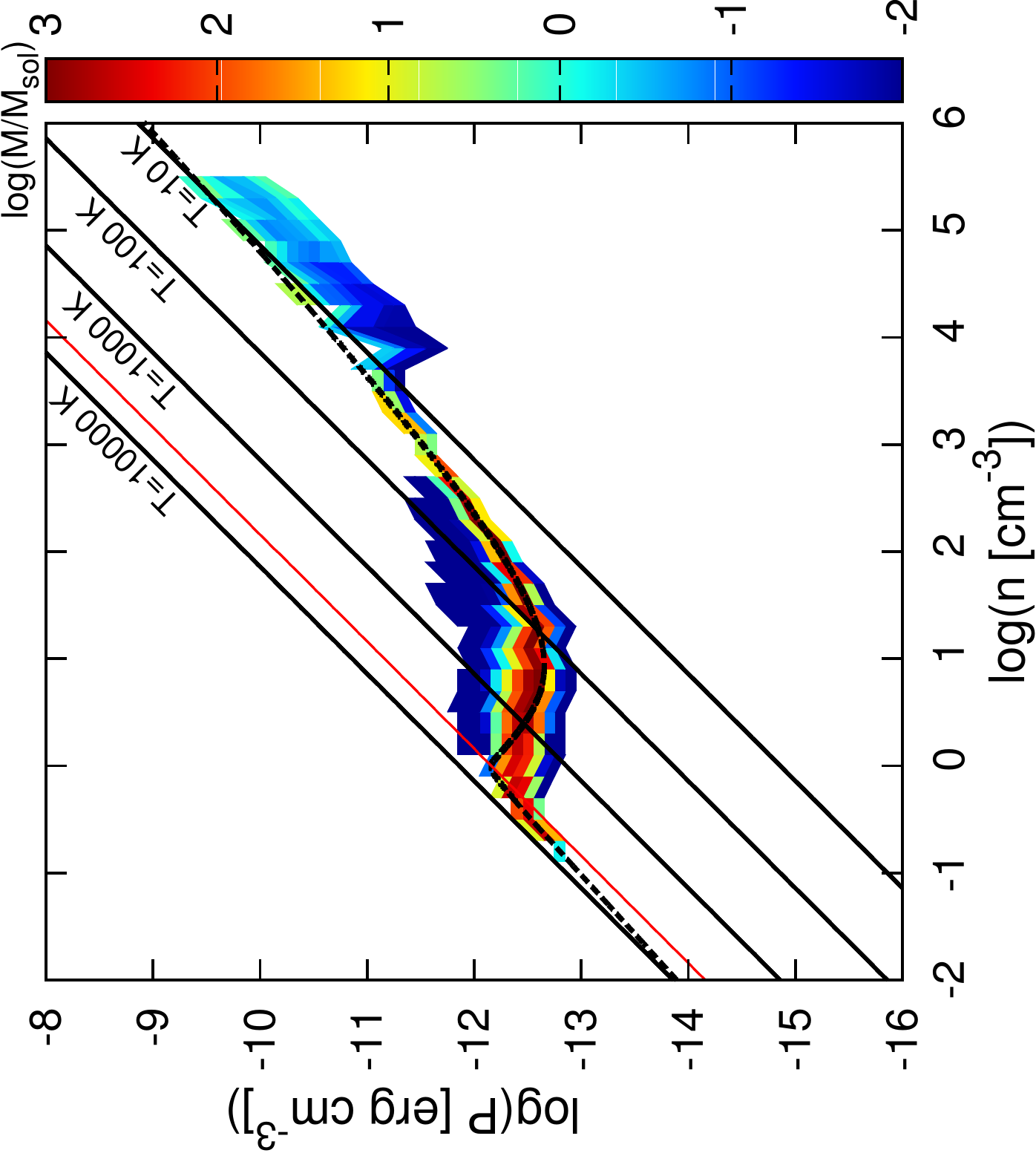}&\includegraphics[height=5.3cm,angle=-90]{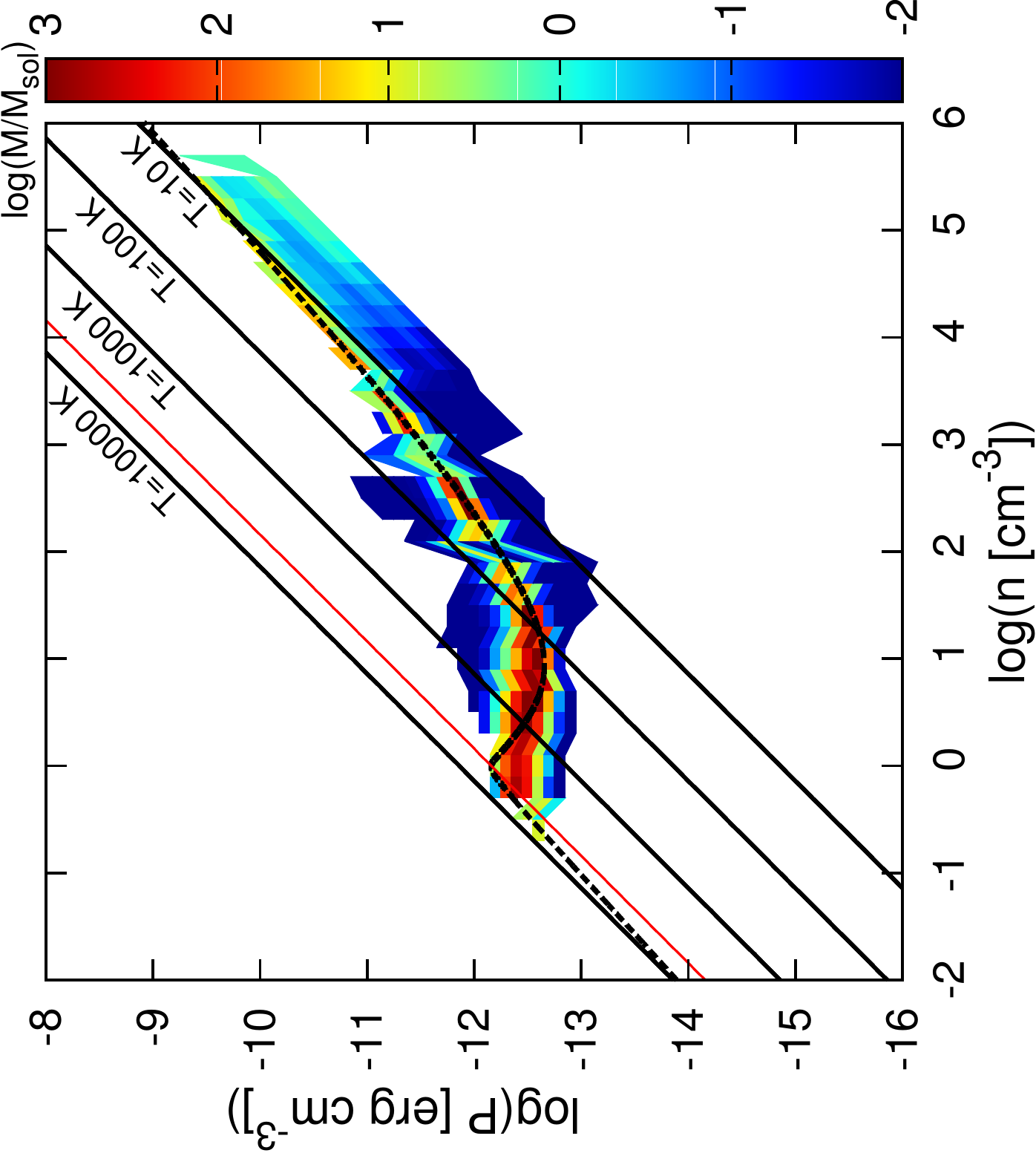}&\includegraphics[height=5.3cm,angle=-90]{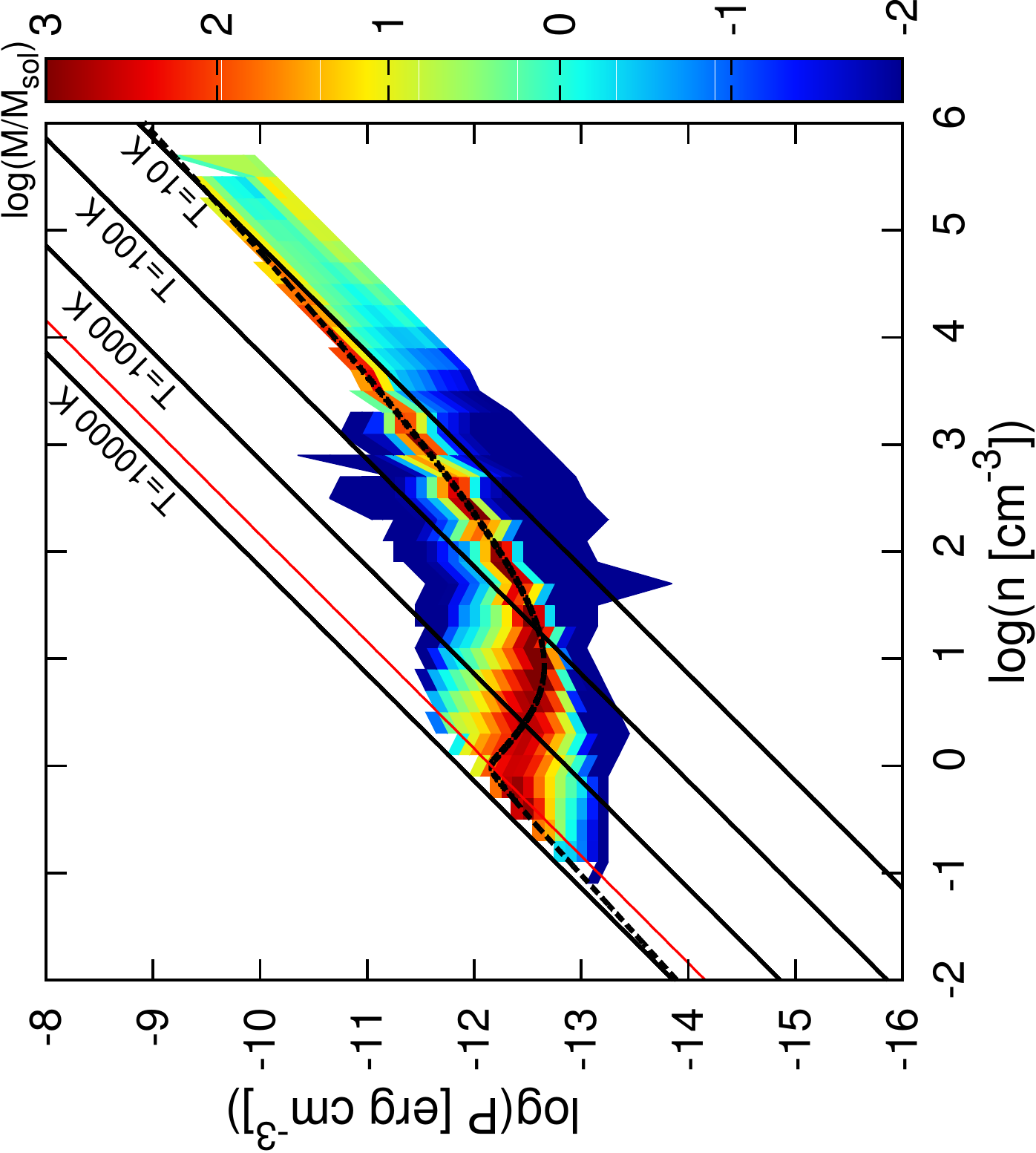}\\
\includegraphics[height=5.3cm,angle=-90]{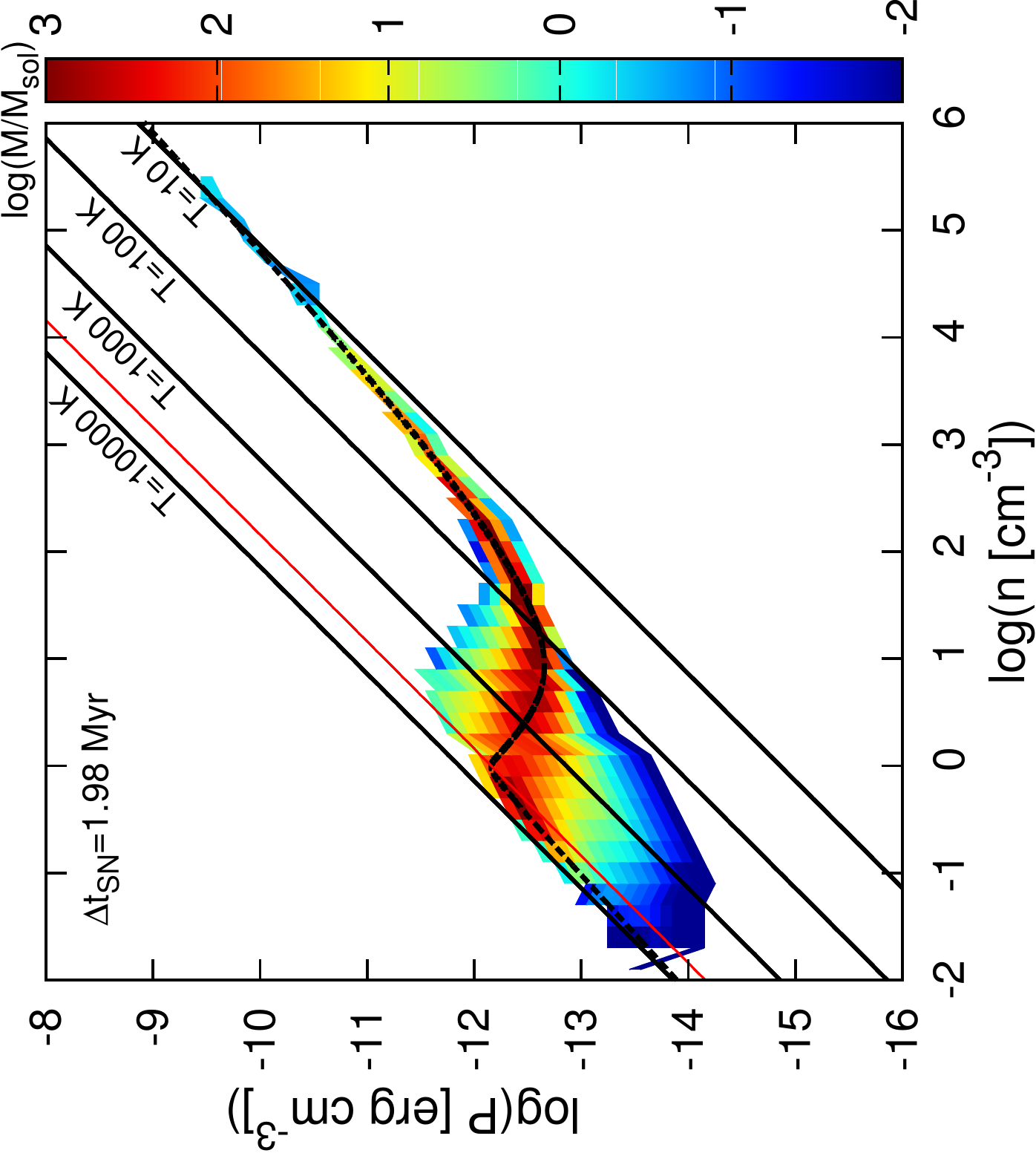}&\includegraphics[height=5.3cm,angle=-90]{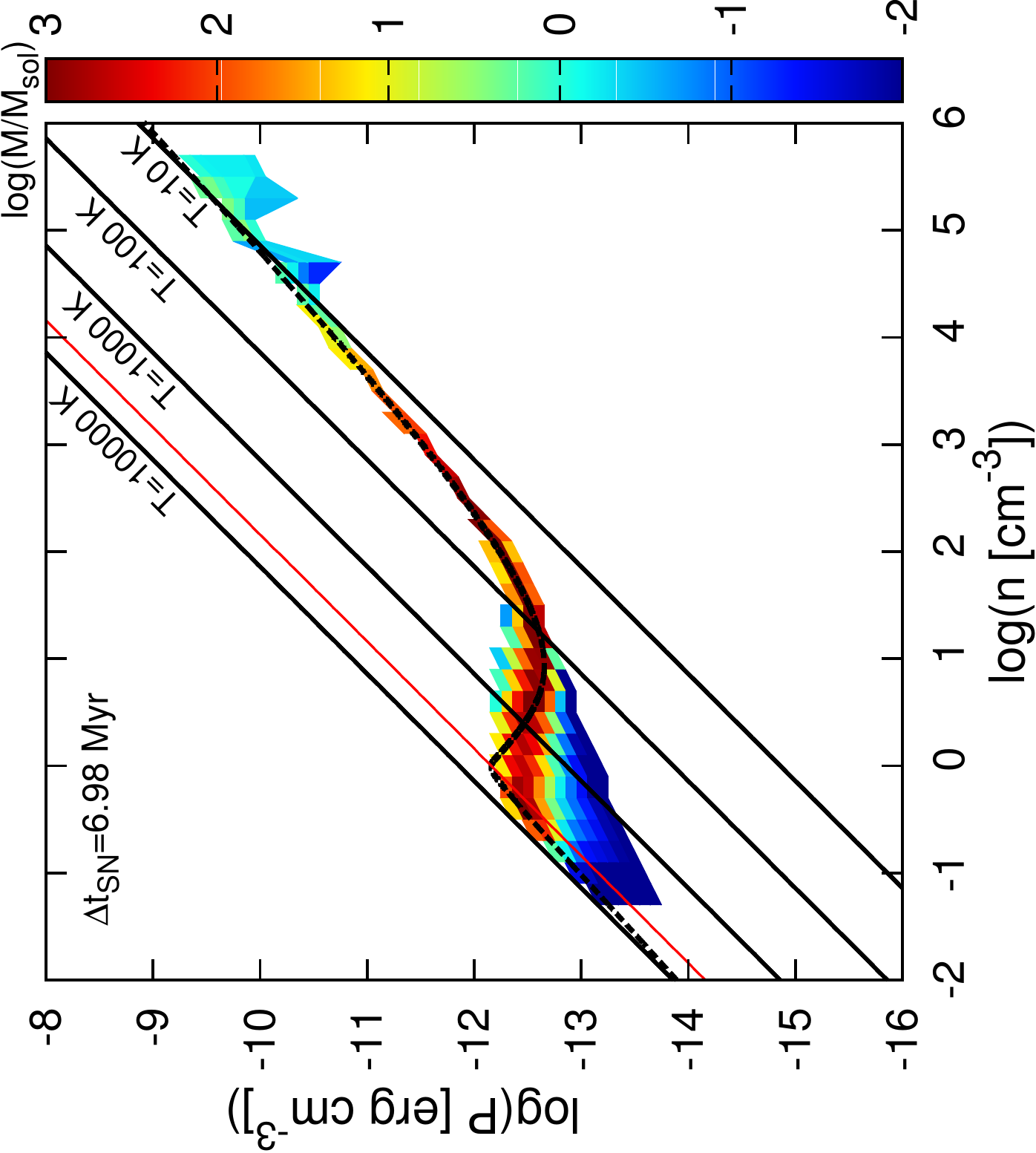}&\includegraphics[height=5.3cm,angle=-90]{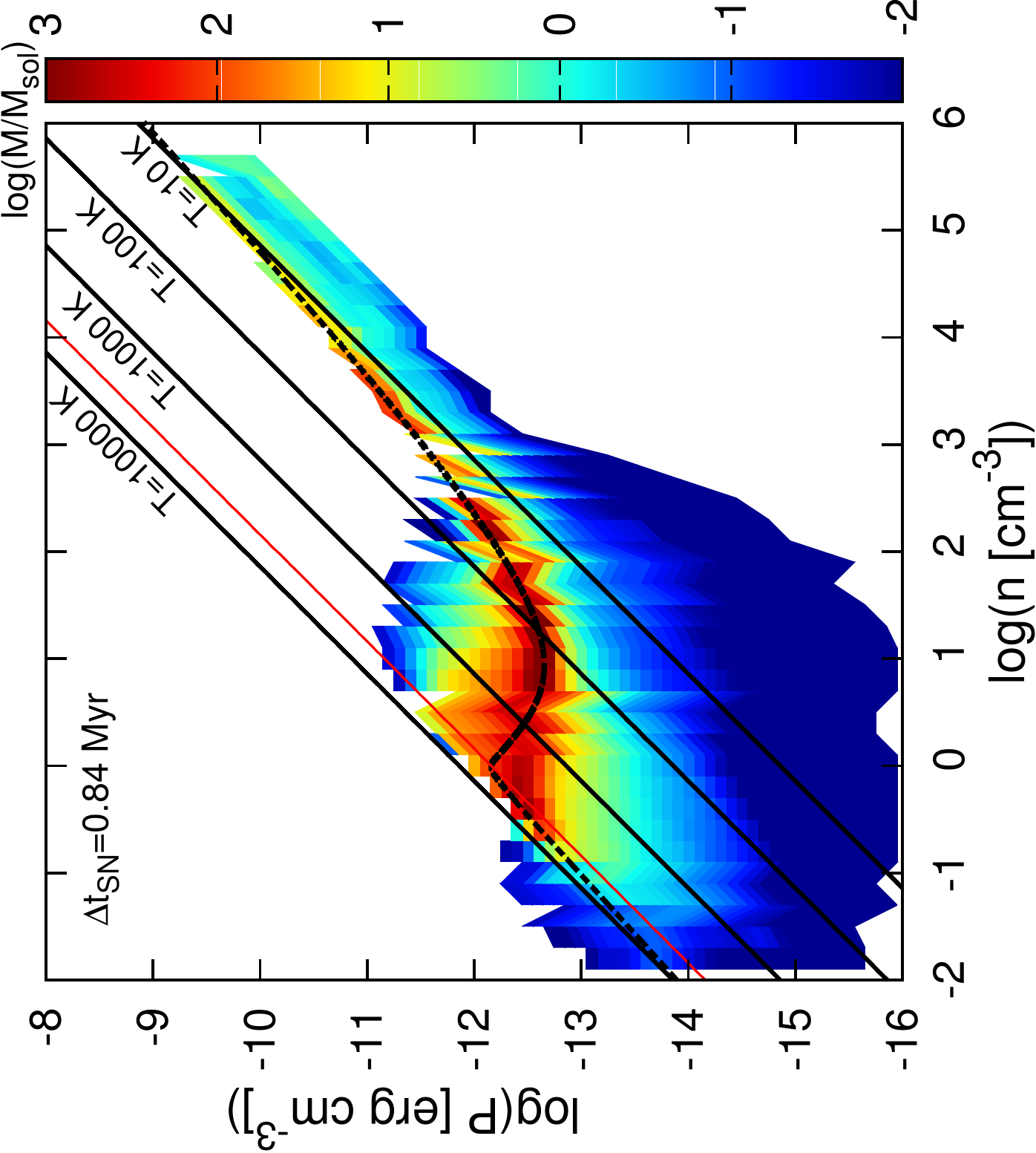}\\
%\hline
%\hline
%&\LARGE{\fat{$M_\mathrm{RMS}=1.0$}}	&\\%[0.1cm]
%\hline
&&\\
\Large{\fat{20.0 Myr}}&\Large{\fat{25.0 Myr}}&\Large{\fat{30.0 Myr}}\\
\includegraphics[height=5.3cm,angle=-90]{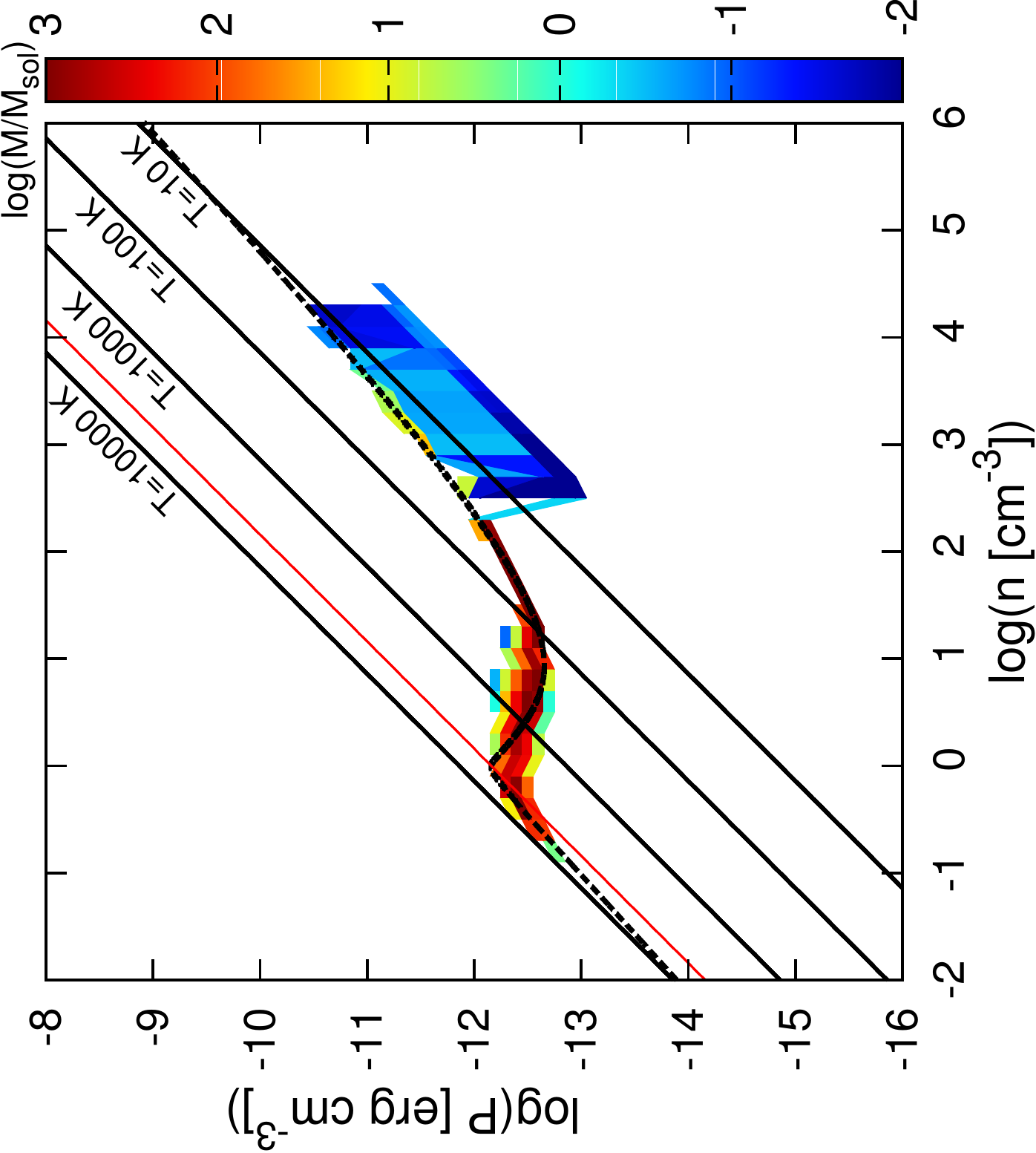}
&\includegraphics[height=5.3cm,angle=-90]{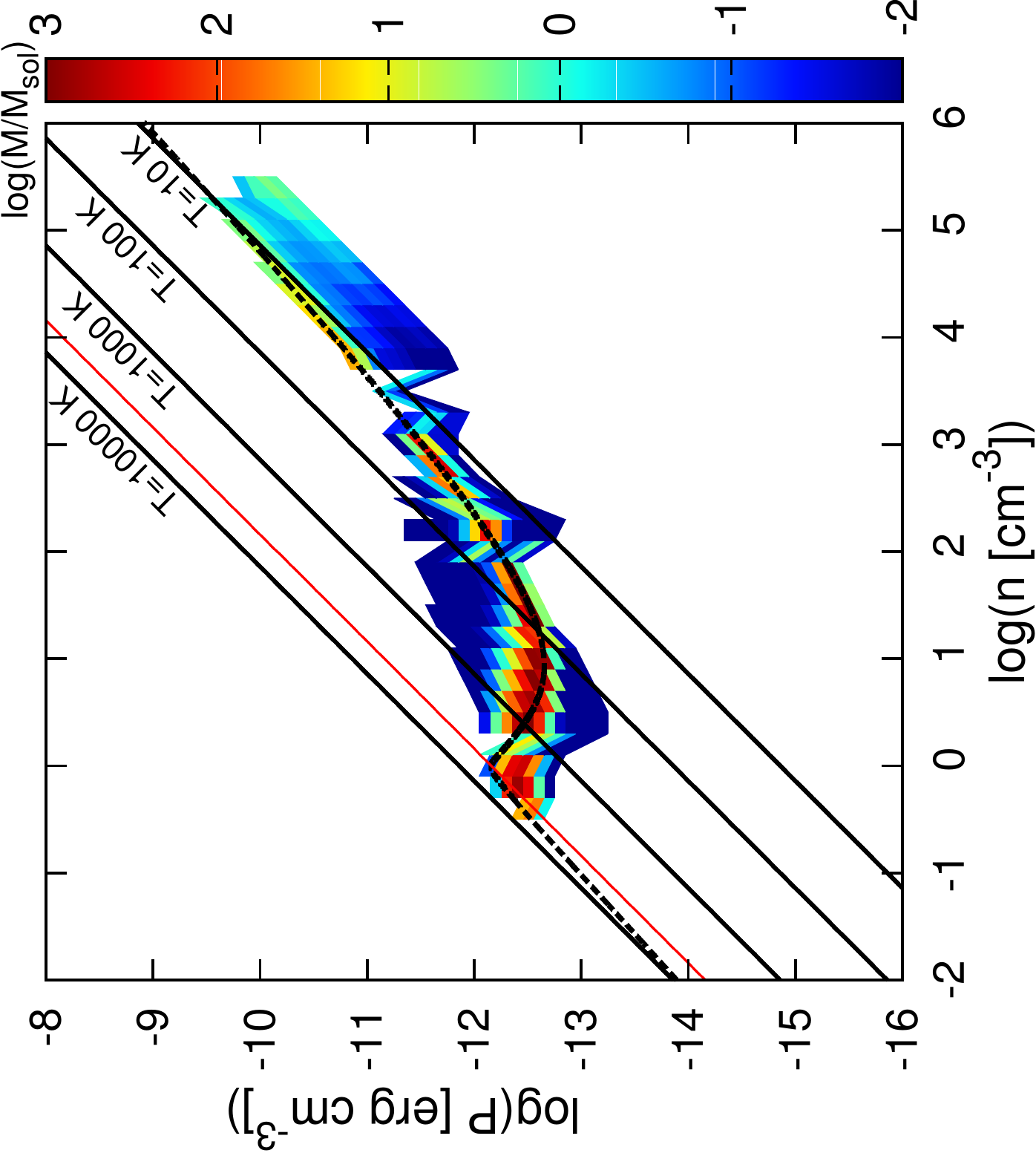}&\includegraphics[height=5.3cm,angle=-90]{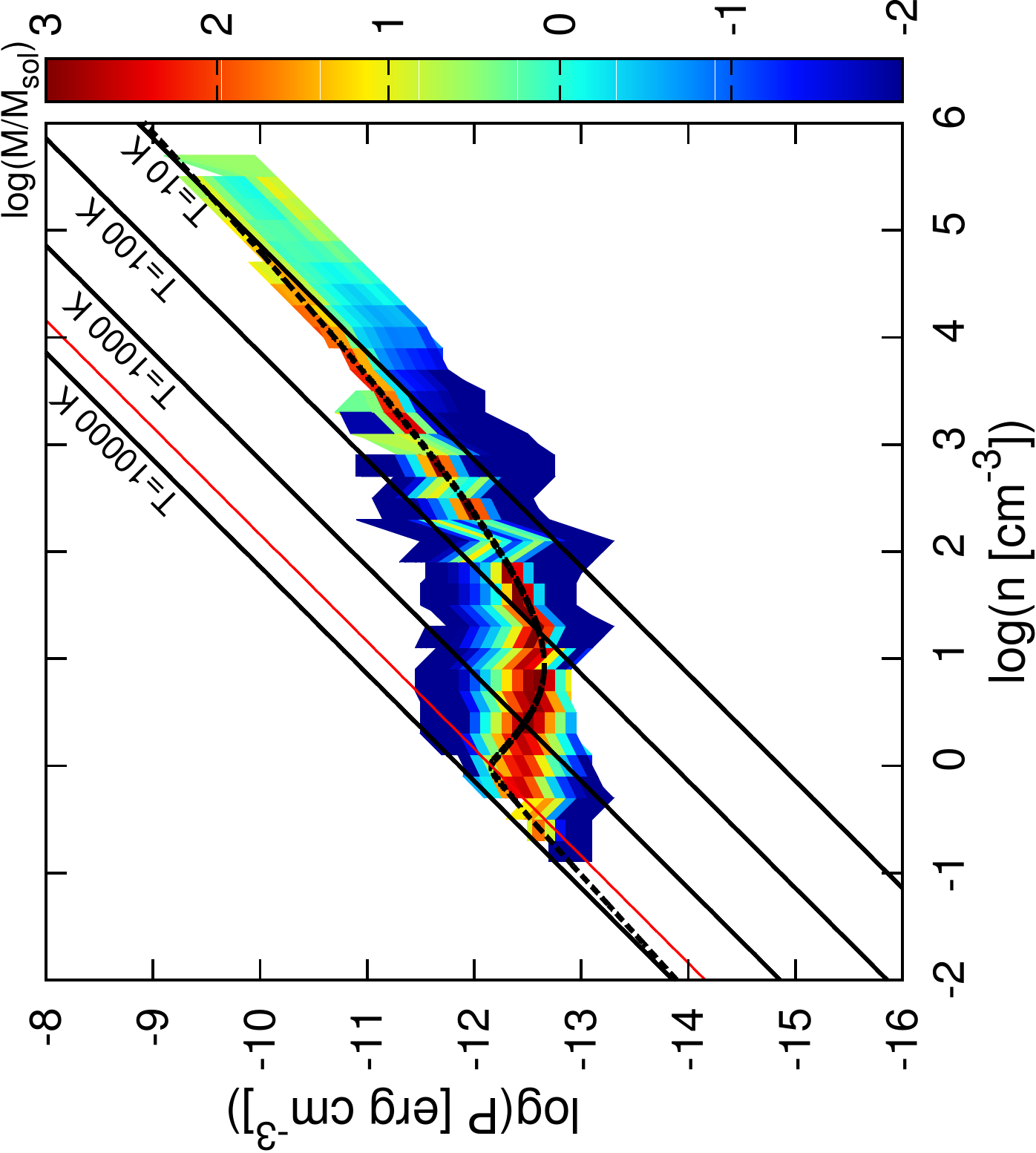}\\
&\includegraphics[height=5.3cm,angle=-90]{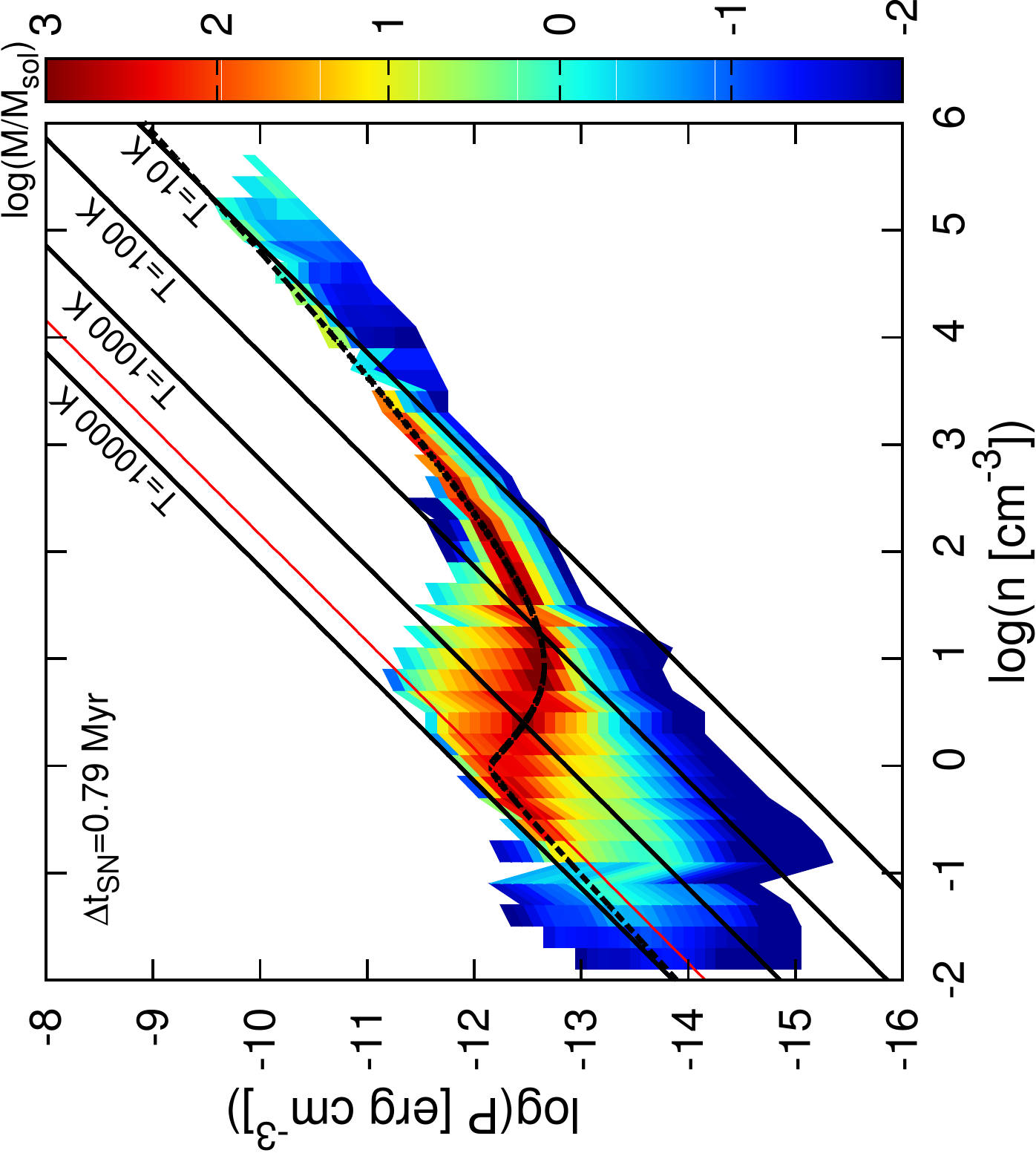}&\includegraphics[height=5.3cm,angle=-90]{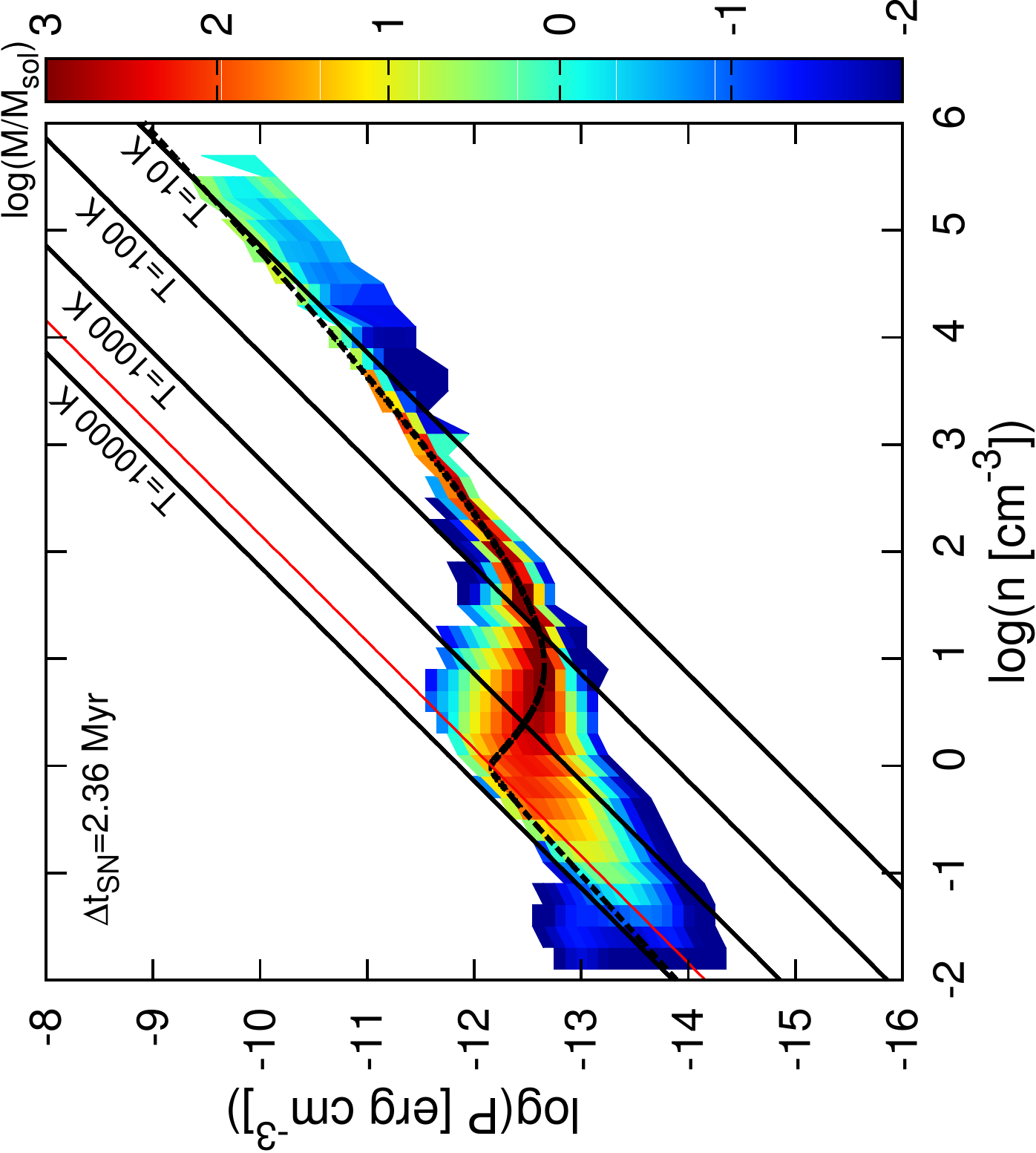}\\
\end{tabular}
\caption{Phase diagrams for the runs HR0.8N,HR0.8Y (top two rows) as well as HR1.0N and HR1.0Y 
(bottom two rows),
 respectively, without (upper row) and with (lower row) supernova feedback. Colour coded is the mass in each bin. 
Also shown are the isotherms for $T=10,100,1000,10000$\,K (solid black lines) and the isotherm for 
$T=5000$\,K (solid red line). The dashed line marks the equilibrium 
pressure. The times within some of the plots denote the elapsed time since the last SN. Most of the mass is in the cold phase with the scatter being produced by turbulence. Interestingly, supernovae only produce short--lived strong deviations 
from pressure equilibrium. After a while the gas approaches equilibrium conditions again. The major 
difference at later times is the occurence of a hot gas phase at low densities, which can be 
attributed to the cavity that has been created by the supernova explosion. Notice that HR1.0N and HR1.0Y are similar 
at $t=20\,$Myr. Hence, we do not show the plot for HR1.0Y.}
\label{fig2}
\end{figure*}

\begin{figure*}
\includegraphics[height=0.45\textwidth,angle=-90]{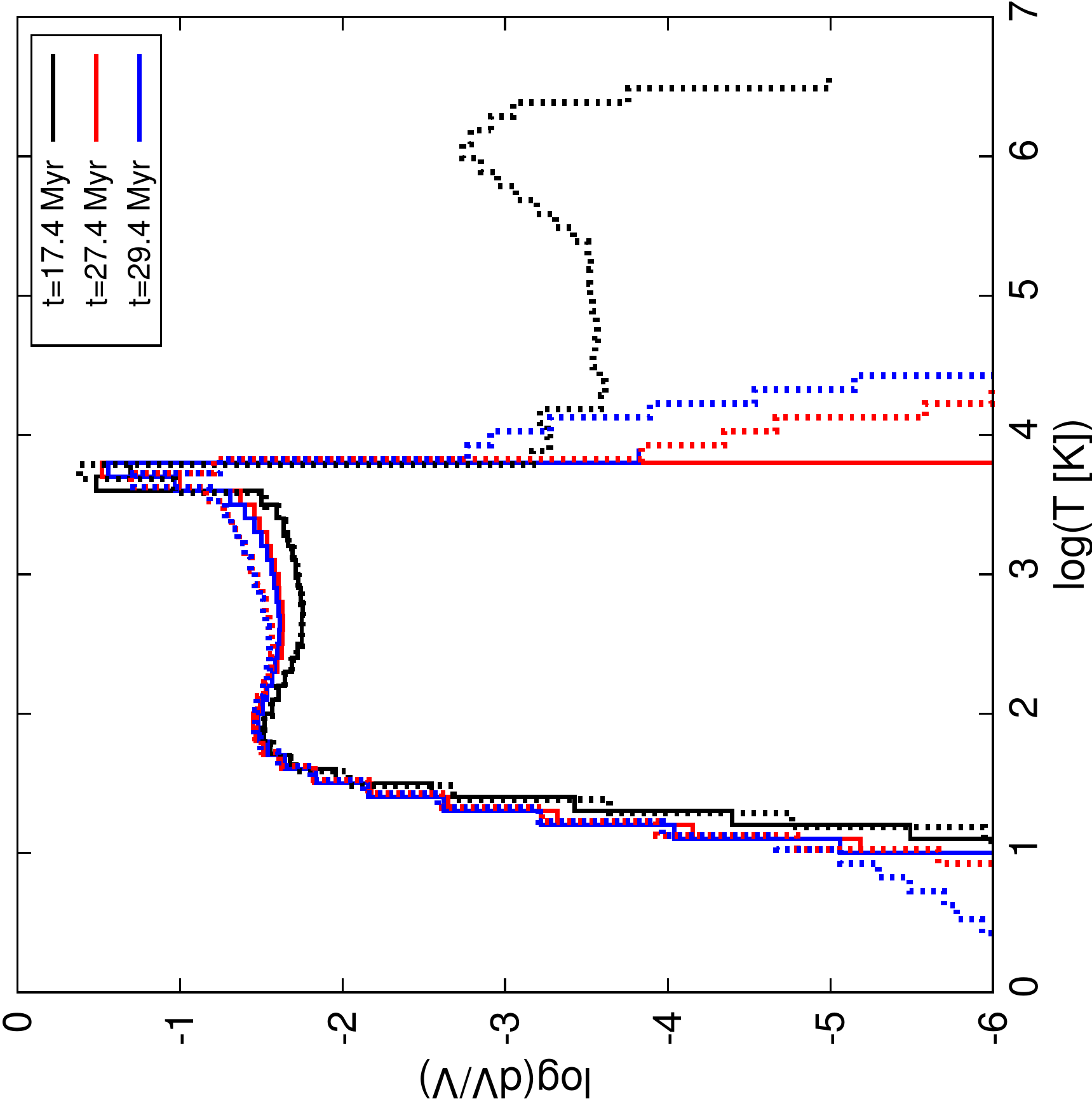}\qquad\includegraphics[height=0.45\textwidth,angle=-90]{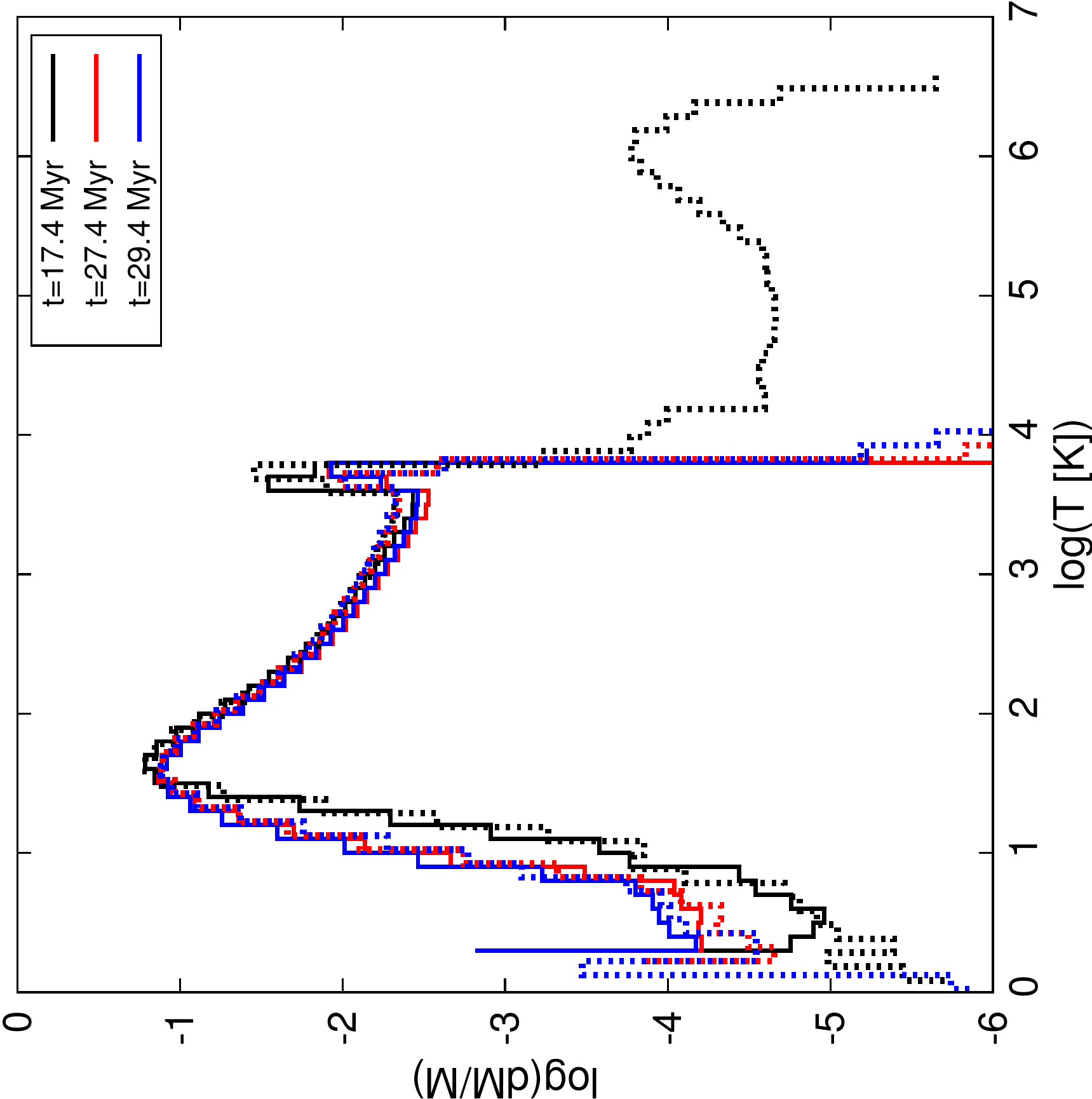}
\caption{Temperature histogram for different evolutionary stages with (dotted) and without (solid) feedback for runs 
HR0.8. 
Note the large increase in temperature due to a supernova explosion (which went off shortly 
before the shown time). SN feedback results in an increased amount of material in the thermally unstable regime.}
\label{figTempHist}
\end{figure*}

The redistribution of matter on large scales is accompanied by mixing of warm and cold gas on smaller scales. This is because of the 
turbulence generated by the multiple shock waves interacting with the substructures in the cloud as well as their 
interaction with each other.\\
The resulting increase of gas in the thermally unstable regime is seen in figure \ref{fig2}. 
We show the temporal evolution 
of phase diagrams for runs HR0.8N, HR0.8Y, HR1.0N, and HR1.0Y at three different times.The data shown refer
 to stages prior to, shortly after or long after supernova feedback.
In general, the gas 
evolves along the equilibrium curve, where most of the gas resides in the cold, stable regime. However, a significant part is also 
detected in the unstable regime, which is material from the halo surrounding the cloud. The scatter around the 
equilibrium curve is due to turbulent fluctuations in the gas that generate dilatational and compressive regions and drive the gas out of the cold and warm \ita{stable} phase into the unstable regime \citep[e.g.][]{Seifried11c}. 
%For the clouds without feedback there is a small trend for larger scatter as function of time.
Even in the case without feedback the pressure scatter increases with time (see the last column of the runs without feedback, i.e. rows one and three). 
This is a result of global collapse and 
conversion of gravitational into (turbulent) kinetic energy. \\
%For run HR1.0N, there seems to be a larger scatter at earlier times, due to the initially higher 
%turbulence. However, with time those fluctuations decay.
The individual supernovae have a large, but short--lived impact on the phase diagrams. The evolution of the supernova 
remnant creates over--pressured volumes with high temperatures, as well as under--pressured volumes with very 
low temperatures. Both phases are primarily seen in the low--density regime. However, the long--term evolution -- indicated by a large $\Delta t$ in the plots -- reveals that the gas is cooled faster than it is 
heated. This is best seen at $t=24.3\,\mathrm{Myr}$ in run HR0.8Y, where there is only some 
scatter observed in the under--pressured low--density regime.\\ % From $n\approx10\,\mathrm{cm}^{-3}$ the gas is in pressure equilibrium. For comparison, run HR0.8N reveals turbulent scatter in the same density regime.\\ 
The densest parts of the molecular cloud are barely affected. There only occurs a small decrease in mass, because 
the shock wave is not able to sufficiently disperse these regions. 
%Further temporal evolution shows that the hot (over--pressurised) gas is cooled down and the cold (under--pressurised) gas is heated up. 
%Furthermore the gas approaches equilibrium conditions \far{again} by either being heated (cold gas) or being cooled (hot gas). 
In the end of the simulation, the phase diagrams look similar in the intermediate density regime 
$\left(10<n/\mathrm{cm}^{-3}<100\right)$ for 
cases with and without feedback in that way that most of the gas evolves along the equilibrium curve. 
All clouds affected by SN feedback reveal the emergence of low--density $\left(n\leq0.1\,\mathrm{cm}^{-3}\right)$, warm material with temperatures 
of $10^3-10^4\,\mathrm{K}$, which can be attributed to the supernova remnants. These regions stay warm for 
 $\sim 7$\,Myr (see region between $0<y/\mathrm{pc}<20$ and $-40<z/\mathrm{pc}<0$ in the temperature slices of figure \ref{fig2d}).\\  
%\far{Since they are also very
 %teneous, those regions will be under--pressured with respect to the surrounding medium and hence they might be
 %subject to gravitational recollapse. This recollapse is indicated in figure \ref{fig2d} in the column density map in the range
 %$-20<z/\mathrm{pc}<0$ and $0<y/\mathrm{pc}<20$ where the column density in the region first affected by a SN increases over time. Note, however, that in addition further supernovae can push cloud material towards these 
%under--pressured regions.}\\
%\subsubsection{Temperature Histograms}
In figure \ref{figTempHist} we show volume weighted and mass
 weighted temperature histograms. Most of the mass is in the cold gas. The WNM instead contributes most to the volume fraction.The two major thermodynamic phases of the ISM are clearly identified with 
temperatures of about $T=30-50\,\mathrm{K}$ for the cold gas and $T\approx5500\,\mathrm{K}$ for the WNM.
A \ita{three--phase} medium is only being generated for a transient period of time, with the additional phase 
being the hot gas \citep[see also][]{McKee77}. A more persistent effect is that supernova feedback converts cold gas to gas with moderate temperatures of 
$2.5\,\leq\,\mathrm{log}(T/\mathrm{K})\,\leq\,3.5$ with a net increase of $\approx15\%$ in volume. The mass fraction, however, shows an increase of less than 1\,\% in this temperature regime.
We point out that an increase of gas in the 
thermally unstable regime can also be achieved via turbulent mixing alone \citep[e.g.][]{Seifried11c}. 
%However, in the case of supernova feedback it \far{might be} a combination of (enhanced) turbulent mixing due to turbulence generated behind the shock fronts and efficient cooling of \far{compressed} gas. The latter is a result of moderate temperature and density enhancements 
%in regions that are not directly affected by the hot stages of the SN, but rather by acoustic waves.\\
%Interestingly, the mass--weighted histogram looks very 
%similar to the one from \citet[][]{Hill12} in the range $\left|z\right|\leq20\,$pc, where $z$ is the height above/below the 
%disc midplane. 

%\subsubsection{Turbulent velocity \& temperature of the dense gas}
\subsubsection{Long--term dynamical evolution of the dense gas}
\label{secturbtemp}
\begin{figure*}
\includegraphics[height=0.45\textwidth,angle=-90]{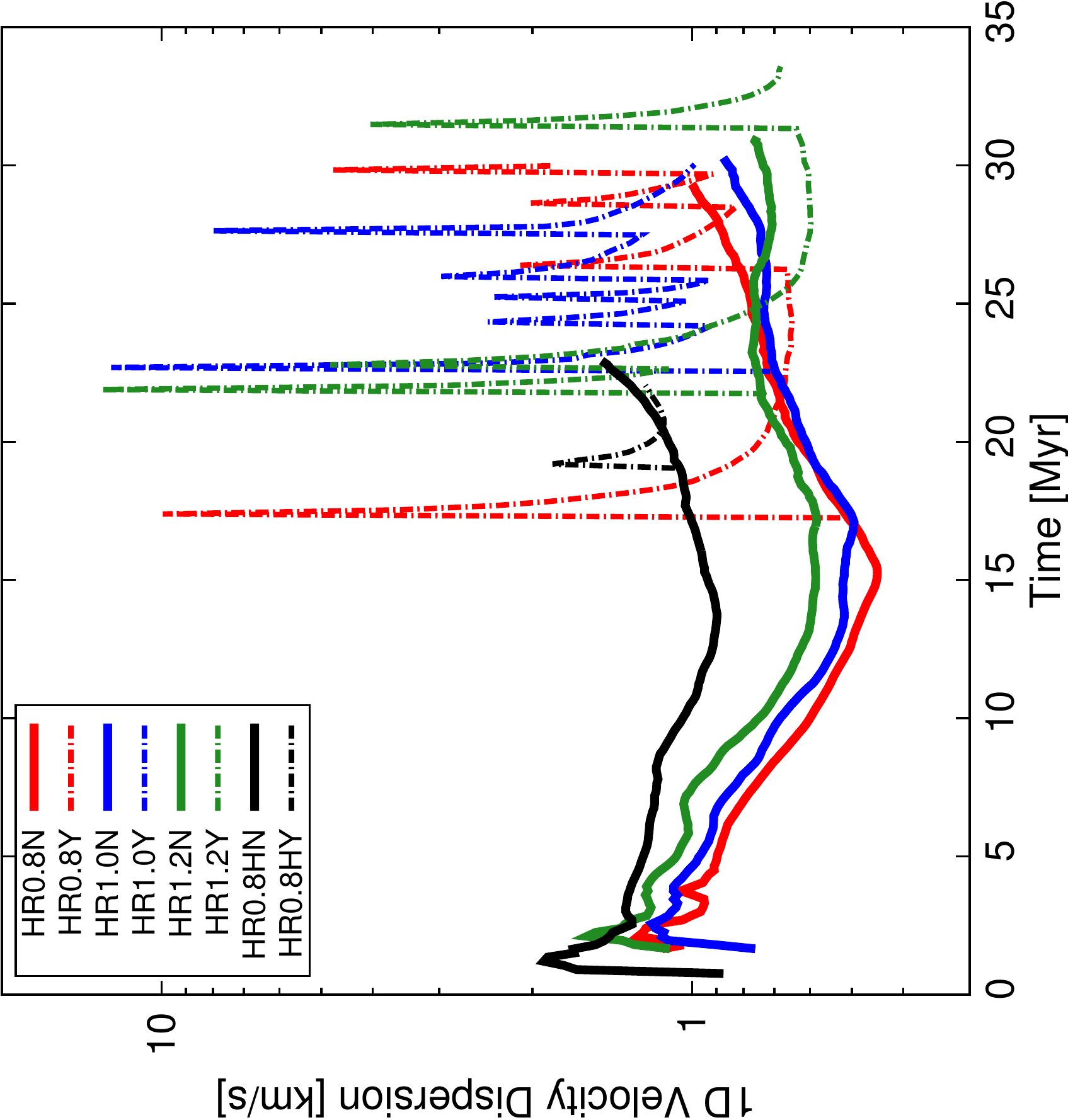}\qquad\includegraphics[height=0.45\textwidth,angle=-90]{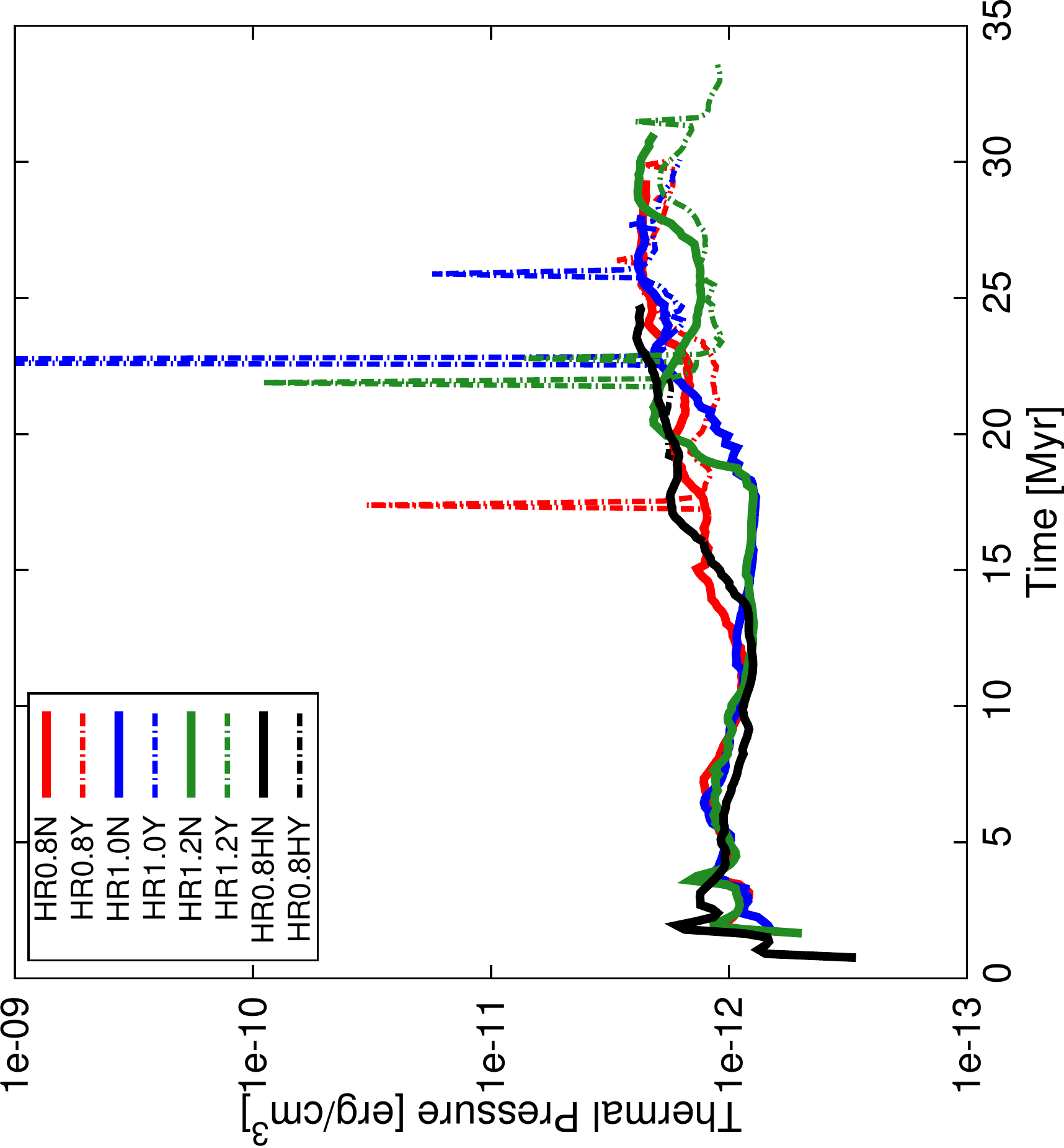}
\caption{\ita{Left:} Temporal evolution of the RMS velocity of the dense gas for all simulations. The 
supernovae only temporarily increase the velocity dispersion. This is due to the fact that the 
dense gas is even more compressed and most of the kinetic energy is hence converted into 
compressive work. Note that there occur stages where the RMS velocity in the SN runs 
falls below the no-SN values. Here, global collapse of dense regions towards the centre of the cloud is hampered. 
\ita{Right:} Evolution of the dense gas' thermal pressure. The densest parts of the molecular cloud show no significant
 increase of thermal pressure.}
\label{figRMSvel}
\end{figure*}

In figure \ref{figRMSvel} we show the one--dimensional velocity
dispersion (henceforth 1D--dispersion) and thermal pressure of the clouds. The former is calculated 
in accordance with \citet[][their eqs. (9) and (10)]{Gatto15}. \\
In general, the 1D--dispersion is higher for the hydro runs than for the MHD runs, roughly by a factor of 2--3.
Hence the ambient magnetic field suppresses velocity fluctuations by the influence of magnetic pressure and magnetic tension. The single supernova explosions are clearly identified by the sudden increase of the 1D--dispersion. Depending on the density of the region in which the supernovae go off,
 the 1D--dispersion reaches values of only  $12$\,km/s. However, there are also peaks of only a few km/s in case the SN goes off in regions with high densities. 
This stage of increased velocity does not last long (up to $\approx2\,\mathrm{Myr}$). Hence, momentum 
is not persistently transferred to the dense gas. The decrease in the 1D--dispersion is due to the conversion of kinetic 
energy into compressive work onto the gas. Additionally, the interaction of the shocks with the collapsing gas yield that the velocities fall below the values of the runs without supernova feedback (compare with the velocity pattern in figure \ref{fig2d}). %But this effect strongly depends on 
%the number of supernovae going off, on their individual position and on their temporal sequence. Their relative 
%position to each other is important because previous SN explosions clear out the stellar environment from dense gas.
 However, for run HR1.0Y there is an obvious net increase in  1D--dispersion by a factor of $\approx 2$. This is 
due to the formed supernova bubble, which is much more efficient in dispersing the dense gas and driving mixing motions within it\footnote{Please note that the formation of a supernova bubble is simply because of the 
clustered sink particles. In this sense, the efficiency of SN feedback in this simulation would change if no bubble was formed.}\citep[][]{Sharma14}.\\
The evolution of the thermal pressure of the dense gas, $P_\mathrm{th}\left(n>100\,\mathrm{cm}^{-3}\right)$, is quite
 similar. The initial thermal pressure is 
$P_\mathrm{th,init}\,\approx 7\times10^{-13}\,\mathrm{erg\,cm}^{-3}$. The compression by the flows and the turbulent fluctuations trigger thermal instability. The isobaric phase of this instability explains the occurence of 
dense gas at pressures near the initial value. This phase does not last long and thermal pressure is increased over time. After the flows have deceased ($t\approx10\,\mathrm{Myr}$), the pressure almost stays constant, indicating the negligible
 influence of the flows on the thermodynamical state of the cloud.\\
The individual SN events are clearly identified by the sudden increase. However, most of the thermal energy is 
radiated away very rapidly. In the end, thermal 
pressure in the clouds subject to SN feedback approaches the one in the clouds without feedback.\\

%\far{Mention H$\alpha$ line width and the lack to hold RMS velocity up at 10 km/s (which is observed in HI)}

\subsubsection{Evolution of energy ratios of the dense gas}
\begin{figure*}
\includegraphics[height=0.45\textwidth,angle=-90]{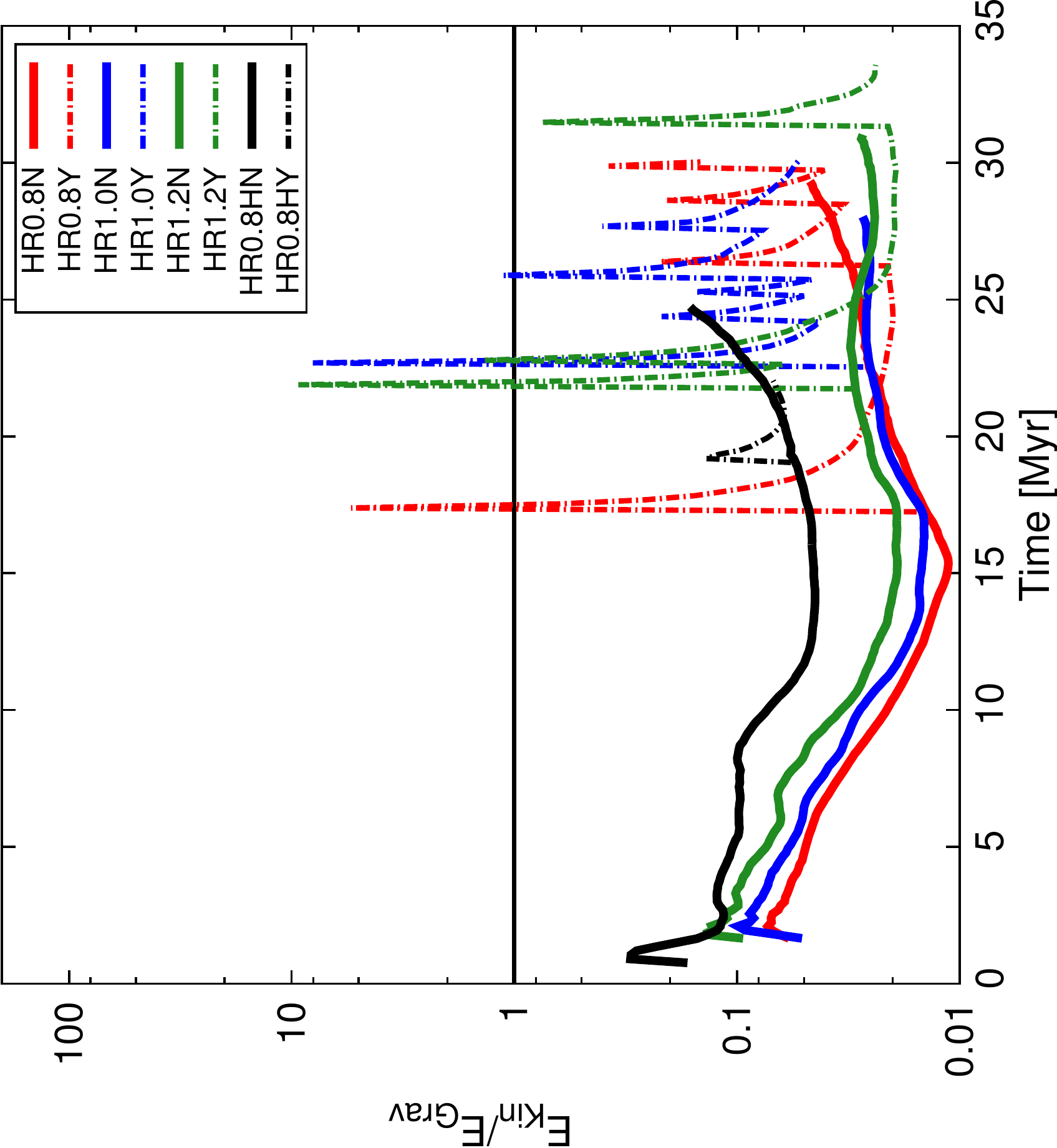}\includegraphics[height=0.45\textwidth,angle=-90]{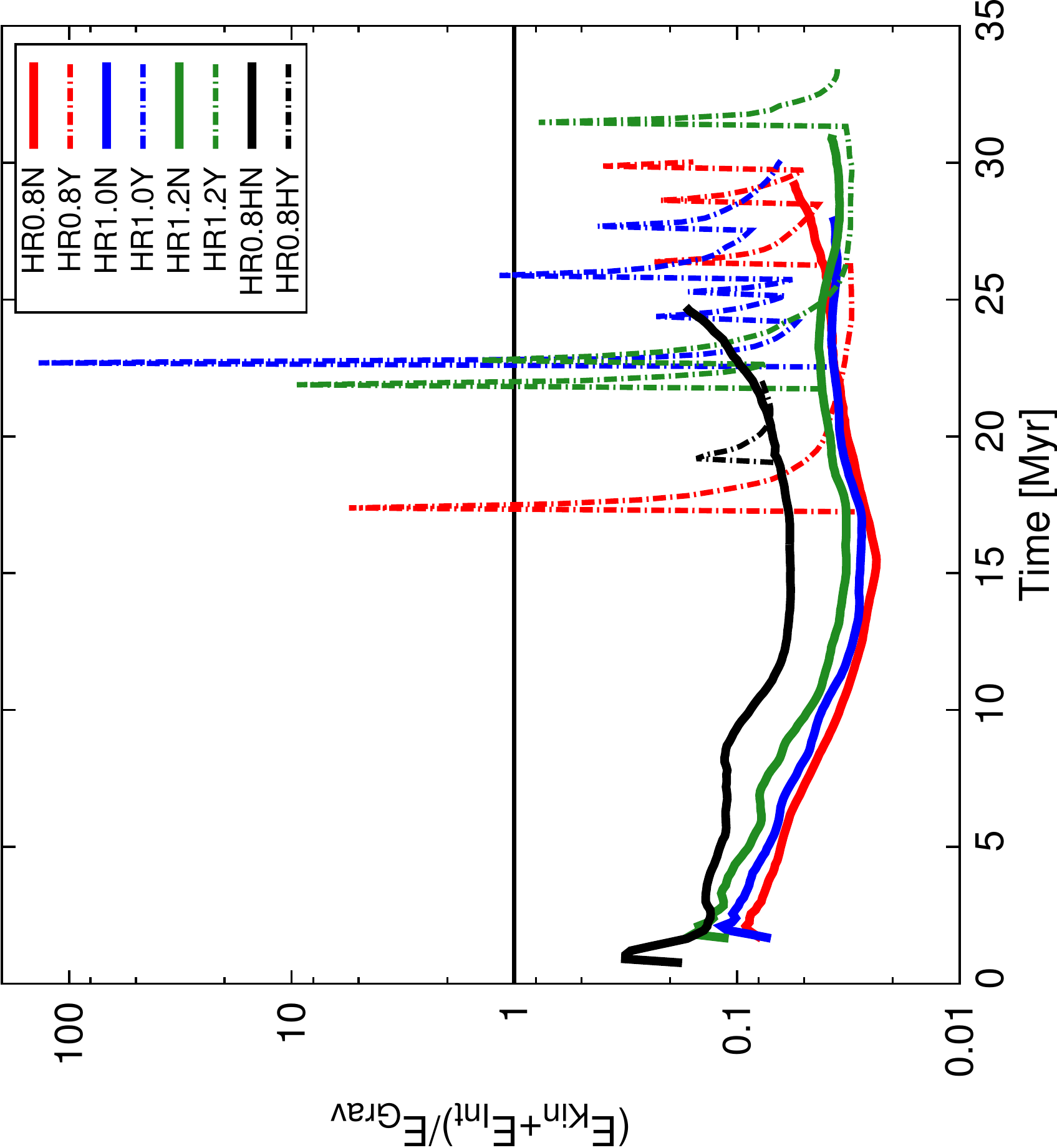}
\caption{\ita{Left:} Ratio of kinetic to gravitational energy. \ita{Right:} Ratio of kinetic plus thermal to gravitational energy. Only at the time when a supernova goes off, kinetic and thermal energy dominate. 
The ratio tends to increase with increasing number of supernovae, but it seems to strongly depend on the cloud morphology or the environment where the supernova goes off.}
\label{figekinegra}
\end{figure*}

Figure \ref{figekinegra} shows the evolution of the ratio of kinetic to gravitational energy as well as the ratio of total 
(thermal plus kinetic) to gravitational energy of the \ita{clouds}. The former ratio is being defined as  
\beq
\alpha=\frac{\frac{1}{2}\sum_i^{N}{\mathcal{V}_i\varrho_i\left|\textbf{\ita{v}}_i\right|^{2}}}{\sum_i^{N}{\mathcal{V}_i\varrho_i\Phi_i}}=
\frac{E_\mathrm{Kin}}{E_\mathrm{Grav}},
\eeq 
with $\mathcal{V}_i$ being the volume of the i--th cell and N being the number of cells with $n\geq100\,\mathrm{cm}^{-3}$.$\Phi_i$ is the gravitational potential in cell $i$.
The first stage between 
0 and 15 Myr is characterised by mass accretion. From 15 Myr on the ratio increases due to the conversion of 
gravitational energy into kinetic energy due to collapse \citep[][]{Vazquez07}. Feedback increases the ratio for 
a small amount of time. During the supernovae the energy budget of the dense gas is purely controlled by kinetic 
(and thermal) energy. In this time interval, \ita{the cloud seems to be rendered unbound} with $\alpha\,\geq\,2$.
However, gravitational energy immediately dominates again. If the time between succeeding 
supernovae is too long, the ratio drops below the ratio in the cases without feedback.
%This is due to the outward propagating shock wave interacting with the inward propagating molecular gas. 
If the time between individual 
explosions is short (as in case HR1.0Y), the energy input yields a net increase of the ratio. However, there need to be 
far more supernova explosions in order to achieve (virial) equilibrium stages. 
%Inspection of the density dependence of 
%the energy ratios (not shown) reveals that equipartition is attained at densities of about 
%$n\approx10\,\mathrm{cm}^{-3}$ for all runs, far below those that have been used to define the cloud.

%\far{Peaks show that whole energy budget is controlled by kinetic energy during the supernova event. -> Implications for 
%observations? Non--virialised entities can be a temporal stage due to stellar feedback.}

\subsection{Star Formation}
\subsubsection{Number and mass of sink particles}
\begin{figure*}
\includegraphics[height=0.45\textwidth,angle=-90]{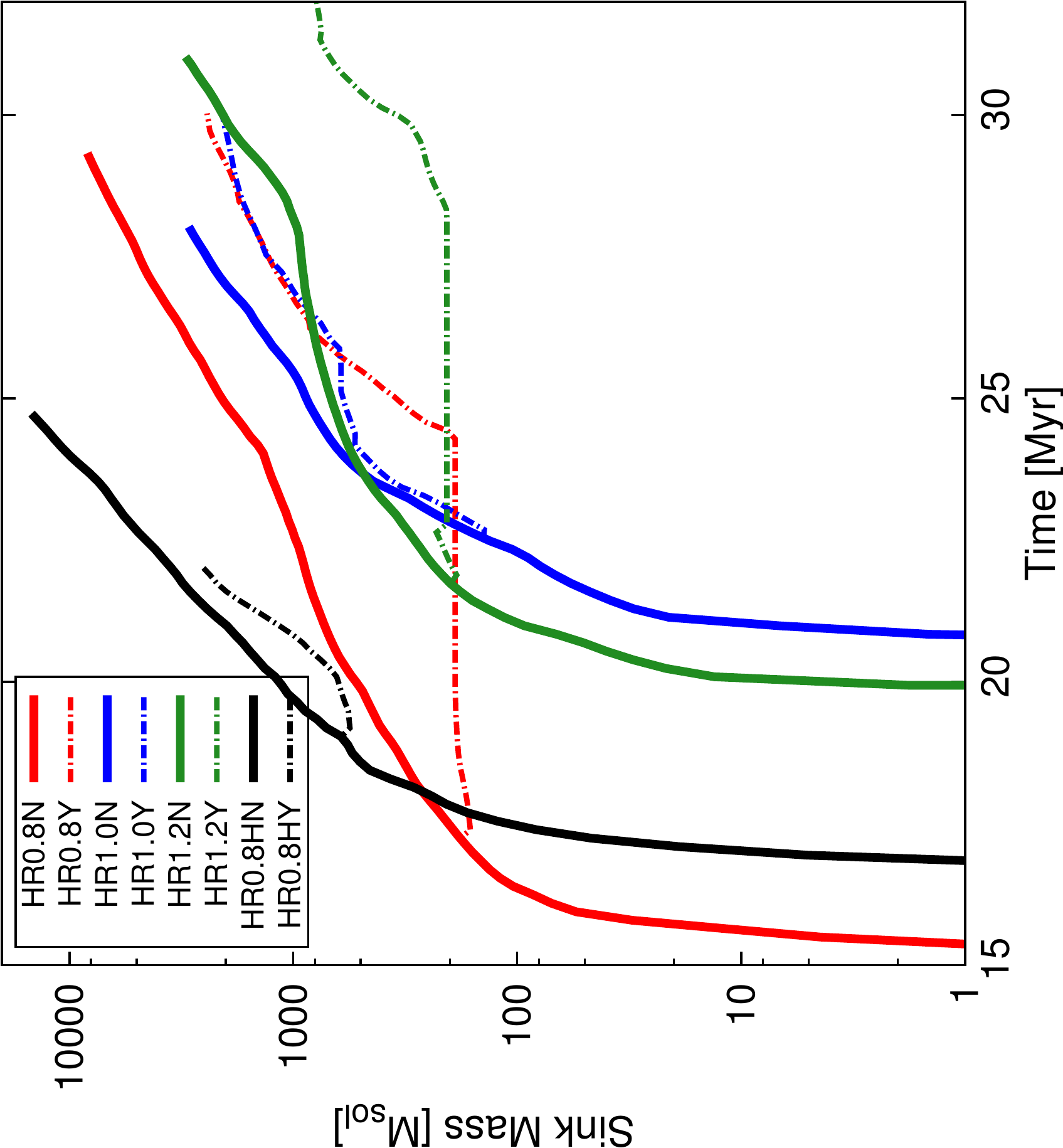}\qquad\includegraphics[height=0.43\textwidth,angle=-90]{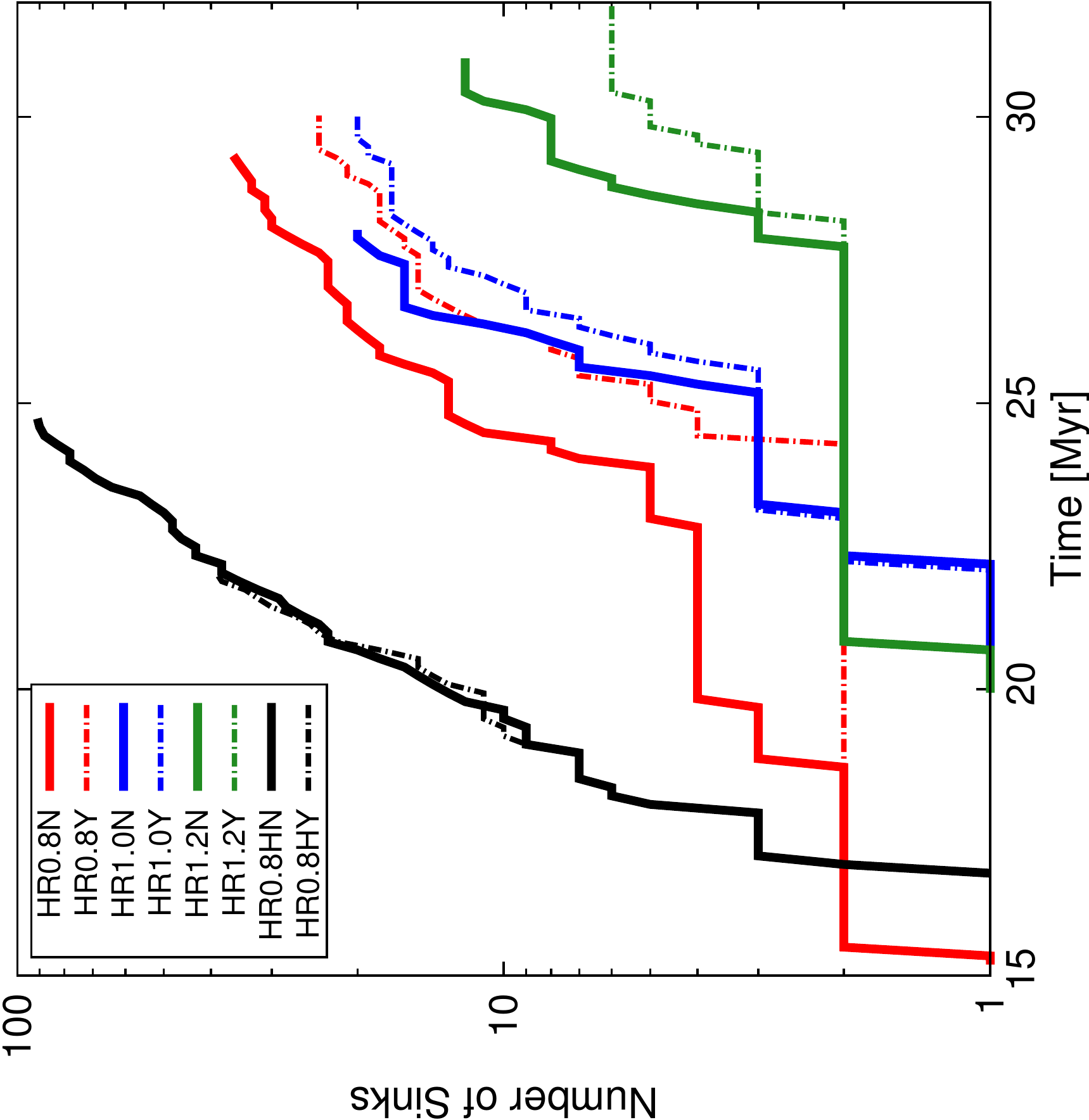}
\caption{\ita{Left:} Temporal evolution of sink particle mass (solid: without feedback, dash-dotted: with feedback).\ita{Right:} Number of particles (solid: without feedback, dash-dotted: with feedback).}
\label{figsinks}
\end{figure*}

%A reduction of the SFR can directly translate to a reduction in the total stellar mass. 
Figure \ref{figsinks} shows the temporal evolution of the total sink particle mass as well as of the number of particles. The sink particles accrete gas and the mass keeps on increasing over time. A decrease in total mass is seen as 
the initial turbulence is increased, because the accretion rates are influenced by the velocity fluctuations. Additionally, dense regions are more stable against collapse or they are dispersed very quickly. 
This also results in a smaller number of sink particles in the cloud.
Also note the large difference of the particle mass and number in the clouds of the hydrodynamic and MHD case with $\mathcal{M}_\mathrm{turb}=0.8$, indicating the balancing  
impact of the magnetic field \citep[][]{Vazquez11,Hennebelle13,Koertgen15}.\\
In turn, if supernova feedback is included, the number of sink particles is reduced. Dense regions are dispersed and hence the seeds for sink formation are missing.\\ 
%However, a closer look to the runs HR0.8H shows that there are more \far{sinks} in the clouds subject to feedback, than in the clouds without stellar feedback. This indicates the possible triggering of \far{sink particle} formation. However, triggered \far{sink particle} formation is not seen in the MHD runs, but analysis of this effect is beyond the scope of the underlying study.\\
The accretion rates are also being reduced by the supernova explosions. This leads to an overall reduction of the 
total mass, which is of about a 
factor of two for the MHD runs, but less for the hydro run. The latter is due to the limited simulation duration. Further evolution should show a greater decrease in mass. One interesting aspect 
concerning the total mass is seen in the evolution. For runs HR0.8Y and HR1.2Y there is a period of nearly constant particle mass, although there exist two sink particles in the cloud. 
Since only one sink has gone off 
as a supernova, this indicates that the shock wave swept over the second one. The second sink's mass supply is being dispersed, thus stopping its accretion either completely or reducing it 
to very low 
values. The increase of the mass at later times begins at roughly the same time as the formation of new sinks.

%\far{Hydrorun: Number of sinks in SN case larger than for noSN case -> triggered SF?}

\subsubsection{Star Formation Efficiency \& Rate}
\begin{figure*}
\includegraphics[height=0.45\textwidth,angle=-90]{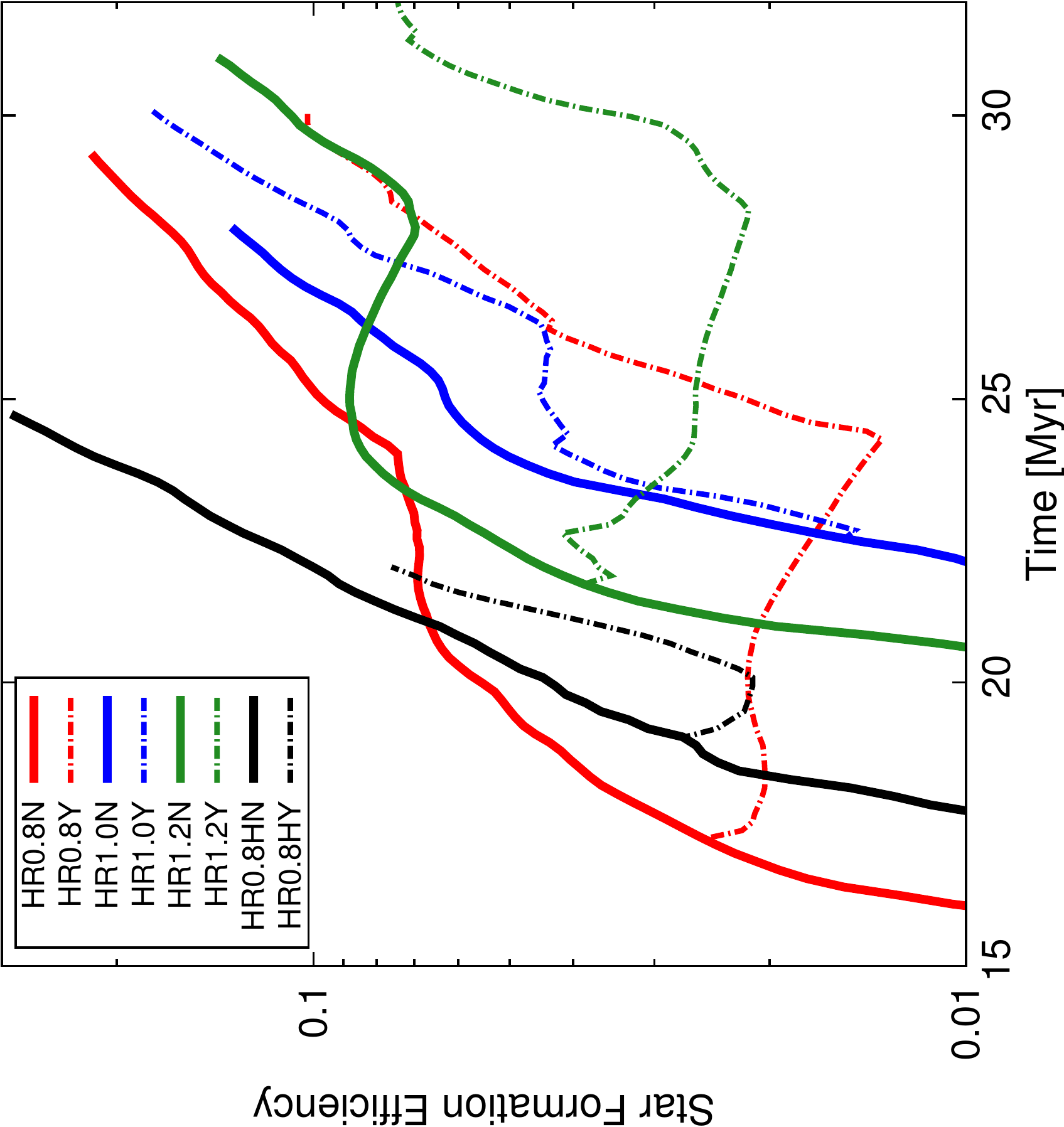}\qquad \includegraphics[height=0.46\textwidth,angle=-90]{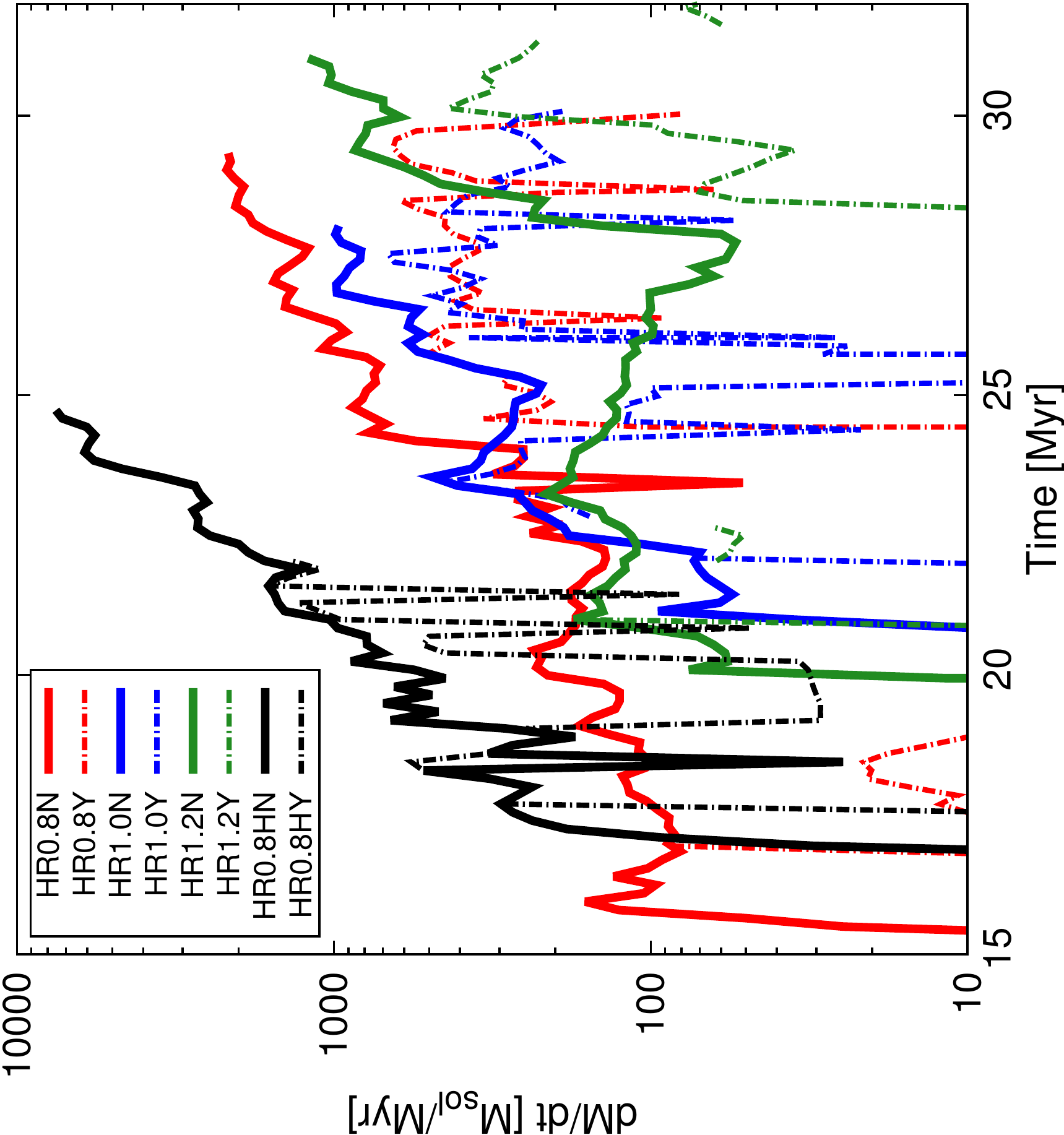}
\caption{Temporal evolution of the star formation efficiency (left) and corresponding star formation rate (right). Supernova 
feedback decreases the SFR by about a factor of 2--4. The SFE is only reduced by at most a factor of 2.}
\label{fig3}
\end{figure*}

Figure \ref{fig3} shows the star formation efficiency (SFE), defined as $SFE(t)=M_*(t)/(M_*(t)+M_\mathrm{cloud}(t))$, and the temporal derivative of the \ita{total} sink particle mass, which we refer to
 as star formation rate (SFR), as function of time. In general, both 
quantities are seen to increase with time. The non--magnetised clouds show a steeper increase of the SFE, as well as of the SFR, at least during the later 
evolutionary stages, due to the lack of additional magnetic support against gravity \citep[see e.g.][]{Koertgen15}. The SFE for the magnetised clouds shows a decrease with increasing initial turbulent Mach number.
The strong variation for run HR1.2N is due to the increase of the cloud's mass. This variation is also seen in the other two clouds (HR0.8 and HR1.0), but in a weaker fashion. In the end, 
values of 15--20\,\% are obtained.\\
 A similar trend is seen in the SFR. Here, the major difference compared to the hydrodynamic cases is the almost constant evolution for the first 
5 to 10 Myr after star formation has begun. The late increase of the SFR is due to global contraction of the cloud, where the magnetic field is not capable of counterbalancing gravity.\\
If feedback is included, both quantities are significantly decreased. Temporal variations in the accretion properties of both clouds and stars yield reduction efficiencies of factors 2--4. In the end of the simulations, the SFE is reduced by at most 
a factor of 2.
The SFR shows a more pronounced evolution. The supernovae are obviously seen by the sudden decrease in the SFR. The overall impact of supernova feedback 
is firstly a reduction and secondly a roughly constant SFR. The former is due to less efficient accretion of the existing sinks as well as suppressed formation of new sinks. The latter is due to the evacuation of dense gas from the centre of the cloud, 
where most of the sink particles reside. This, in turn, affects the accretion behaviour of the sinks. In the runs without feedback, global collapse increases the amount of gas that can (and will) be accreted by the individual sinks. 
The SFR is finally being reduced by roughly a factor 2--4.

%\far{Initial stellar mass as function of time....Are SN-sinks initially more massive?->massive star formation via feedback?}

\subsection{The one--dimensional velocity dispersion}
\begin{figure}
\includegraphics[height=0.45\textwidth,angle=-90]{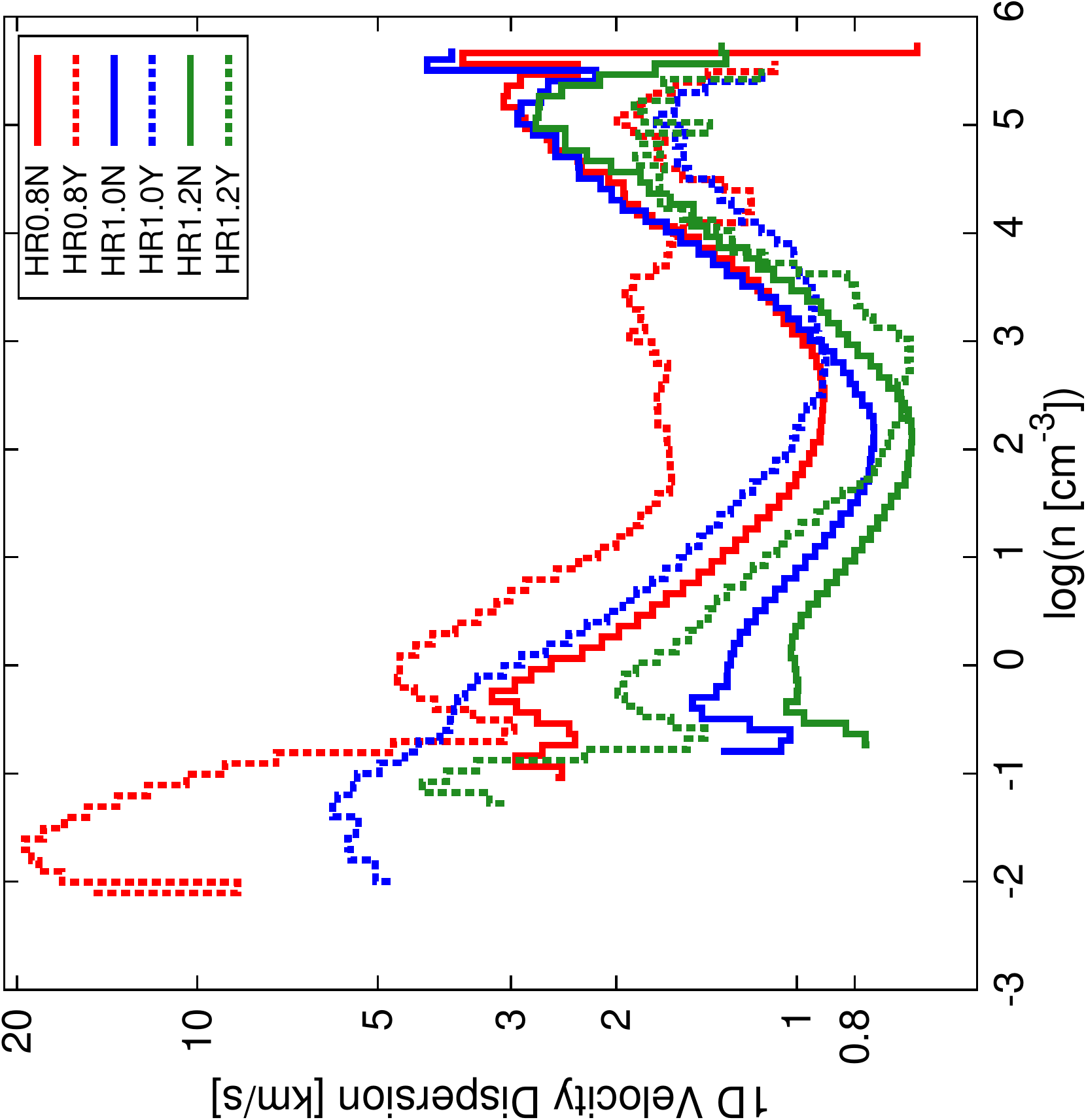}
\caption{One--dimensional velocity dispersion as function of number density. As is shown, the values fail to reproduce observational results, indicating that SN feedback alone may not be the source of the observed HI dispersion. Note the 
lower velocity dispersion in the densest parts of the cloud due to stalled gas motions.}
\label{fig1dveldis}
\end{figure}

Observations of HI in emission indicate that the one--dimensional velocity dispersion of the WNM is $\sigma_\mathrm{HI}\approx10\,\mathrm{km/s}$ \citep[e.g.][]{Heiles03,Tamburro09}. SN feedback is thought of driving such 
velocities and models including driven turbulence in the ISM often use these values as the typical turbulent velocity \citep[e.g.][and references therein]{Gatto15}. 
However, as \citet[][]{Gatto15} report, SN feedback seems to be not capable of driving such high velocity
 dispersions in HI for longer timescales. Figure \ref{fig1dveldis} shows the one--dimensional velocity dispersion as function of density at the end of each (MHD) simulation, using the recipe given in \citet[][their eqs. (9) and (10)]{Gatto15}\footnote{
Note that we do not include different chemical species. Thus the velocity dispersion is for one fluid and we compare our WNM regime with the HI emission results.}. 
As can be seen, velocity dispersions can be as high as $\approx20\,\mathrm{km/s}$, but only for the low--density gas. The WNM with densities of $0.5\leq n/\mathrm{cm}^{-3}\leq5$ reveals values of typically 1.5--5\,km/s in clouds subject to 
SN feedback, far lower than the one observed. The large spread (also for the clouds without feedback) is due to the different accretion properties of the clouds themselves. For run HR0.8Y, there occured a SN event shortly before the end of 
the simulation. That is why the velocity dispersion is higher compared to HR1.0Y and HR1.2Y. Even for the case of clustered supernovae, the one--dimensional velocity dispersion cannot reproduce observational results. 
\subsection{Lifetimes of individual regions within the clouds}
In order to evaluate the efficiency of SNe in disrupting small regions within molecular clouds, we give a comparative overview in table \ref{tabdest}. SN feedback is much 
more efficient in disrupting embedded structures like clumps and cores since the timescales (derived from simulation data: Estimate of cavity size at a timestep $t$ after a SN has gone off.) for the disruption 
are roughly an order of magnitude smaller than for ionisation feedback. 
Our results are in good agreement with the study by 
\citet[][]{Martizzi15}, who carried out simulations of individual SNe going off in an inhomogeneous medium. However, our achieved timescales are somewhat larger, because on the one hand, they did not include a 
magnetic field. On the other hand, the densities within our clumps, in which the massive stars explode, might be higher by up to two orders of magnitude. Hence, radiative cooling is much more efficient in our simulations. In contrast, 
\citet{Rogers13} give larger timescales for the disruption of a clump by SN feedback of approximately 1.5\,Myr, although the authors have included feedback mechanisms prior to the SN. 
%The reduction in the total cloud mass as well as of the SFE and SFR are also in good agreement with previous studies \citep[e.g][]{Hennebelle14,Iffrig15}. 
\begin{table*}
\caption{Estimated diameter and destruction timescale for different feedback mechanisms.}
\begin{tabular}{l c c l}
\hline
\hline
\fat{Source} &\fat{Diameter} &\fat{Timescale}	&\fat{Form of}\\
&(pc) &	&\fat{Feedback}\\
\hline
HR0.8Y$^a$	&4.8	&29\,kyr	&SN\\
HR1.0Y$^a$	&9.7	&50\,kyr	&SN\\
HR1.2Y$^a$	&6.5	&14\,kyr	&SN\\
\citet[][]{Rogers13}	&8	&$<$1.5\,Myr	&SN\\
\citet[][]{Colin13}		&10	&10--20\,Myr	&Ionisation\\
\citet[][]{Dale14}$^b$		&10	&10--20\,Myr	&Ionisation\\
\citet[][]{Martizzi15}$^c$	&10	&15\,kyr	&SN\\
\hline
\hline
\end{tabular}\\
\begin{flushleft}
$^a$\small{This study.}\\
$^b$\small{\citet{Dale14} give the size of the region, but do not give a concrete destruction timescale. However, they argue that their timescales are comparable to \citet{Vazquez10}.}\\
$^c$\small{The size of the region is read off by eye from their figure 1.}
\end{flushleft}
\label{tabdest}
\end{table*}

\section{Discussion}
\label{secdisc}
\subsection{Limitations}
We point out that our simulations lack the progenitor feedback mechanisms like stellar winds and the star's ionising radiation. In order to estimate the 
impact of progenitor feedback, one can calculate the cooling timescale
\beq
t_\mathrm{c}=\frac{3}{2}\frac{k_\mathrm{B}T}{n\Lambda\left(T\right)}.
\eeq 
Here, $k_\mathrm{B}$ is Boltzmann's constant, $T$ is temperature, $n$ is the number density of the heated gas, and $\Lambda\left(T\right)$ is the temperature dependent cooling function, respectively. $t_\mathrm{c}$ then gives the timescale 
when cooling starts to become dominant. For our simulations, typical densities in the stellar environment are in the range $n\,\in\,\left[10^2,10^5\right]\,\mathrm{cm}^{-3}$ and the SN temperatures are as high as $T=10^7-10^8\,\mathrm{K}$. 
The cooling function at these temperatures is roughly constant ($\Lambda\left(T\right)\approx5\times10^{-23}\,\mathrm{erg\,cm^{3}\,s^{-1}}$). These values give 
$t_\mathrm{c}\left(T=10^7\,\mathrm{K}\right)=1.3\times10^{-5}-1.3\times10^{-2}\,\mathrm{Myr}$ and $t_\mathrm{c}\left(T=10^8\,\mathrm{K}\right)=1.3\times10^{-4}-1.3\times10^{-1}\,\mathrm{Myr}$, which are still 
larger than the dynamical timescale, as stated in section \ref{secsubgrid}. Hence, the major part of the injected 
thermal energy is radiated away within only a few timesteps. The resulting heating of parts of the molecular cloud is then due to shock heating and the dispersion of gas clumps is driven by momentum input. In contrast, if the massive star 
generates a large HII region, the SN will go off in a region of teneous gas with densities of the order of $n_\mathrm{HII}=0.1-1\,\mathrm{cm}^{-3}$. The cooling timescale then increases to 
$t_\mathrm{c}\left(T=10^7\,\mathrm{K}\right)=1.3-13\,\mathrm{Myr}$ and $t_\mathrm{c}\left(T=10^8\,\mathrm{K}\right)=13-130\,\mathrm{Myr}$, respectively. The SN remnant is hence only subject to adiabatic cooling and should expand much 
further due to its pressure--driven evolution up to the point where the SN remnant hits the shell that was being swept--up by the HII region. The combined effects of ionising radiation and SN should then be able to disrupt entire molecular 
clouds on timescales less than that for pure ionising feedback, that is, of the order of a few Myr. This is also in agreement with the study by \citet[][]{Sharma14}, who showed that superbubbles can retain up to 40\,\% of their energy over longer 
timescales, in stark contrast to the failure of individual SNe. We here point out that recently \citet[][]{Walch15} concluded that the combined effects of ionisation and SNe are \ita{not} capable of disrupting molecular clouds. 
The authors note that the HII region will generate regions of reduced density as well as regions where the density is 
increased due to compression. This will, on the one hand, help the SNe to disrupt parts of the cloud, but on the 
other hand is able to generate obstacles (due to gas compression) for the remnant where its evolution is stalled. The whole process will enhance 
the impact of the SNe by only 50\,\%. Note, however, that their total cloud mass is approximately an order of magnitude
 larger than the cloud masses discussed in this study. It is therefore possible that their results change with varying 
cloud mass. The latter has recently been investigated by \citet[][]{Dale14}. The authors find that ionising radiation is
 capable of disrupting large parts of clouds with masses of $M_\mathrm{cloud}\approx 10^4\,\mathrm{M}_\odot$, but 
fails to disrupt clouds with masses of $M_\mathrm{cloud}\approx 10^6\,\mathrm{M}_\odot$. The 
influence of the massive stars is enhanced when stellar winds are included. Please note further that our failure to fully disrupt the clouds with SNe alone, and our proposal that possibly the combined action of ionising radiation and SNe may accomplish this task, should not be confused with the above mentioned results by 
\citet[][see also \cite{Dale12}]{Dale14} . In their case, it is possible that the difficulty in destroying such clouds arises by the initial
conditions considered by those authors (initially spherical clouds), since the spherical geometry causes the deepest possible potential wells, while real clouds are more likely sheetlike or
filamentary \citep[e.g.][]{Bally01a,Heiles03}
as is the case of the cloud in our simulations. In our case, the inability of the SNe alone to destroy the clouds is due more to its brief, impulsive nature, and the combination of this kind of feedback with ionising radiation
may well be capable of destroying even very massive clouds. We plan to address this question in a future study.
%\far{Walch\&Naab,2015: Ionisation reshapes cloud interior. SN disrupts more easily regions which have been dispersed 
%by the ionisation. On the other hand, SN is stalled at shells which have been compressed by the ionisation. The authors 
%argue that the SN impact is only slightly increased. The momentum input is only 50 per cent higher compared to 
%cases without prior ionisation.}

\subsection{Time duration of the simulations}
We ran the simulations for a maximum time of $t\sim35\,\mathrm{Myr}$. During this time the molecular clouds are not fully disrupted, hence star formation continues throughout. 
As reviewed by \citet[][]{Blitz07}, 
observations of molecular clouds indicate life times of 20--30\,Myr\footnote{Though having an uncertainty of $\sim\,50$\%.} in agreement with our simulations. We thus argue that this time span is enough 
to study star formation and stellar feedback effects in single, isolated clouds.

%For our simulations $n\approx10^2-10^3\,\mathrm{cm}^{-3}$, $T=10^7\,\mathrm{K}$ and thus $\Lambda\left(T\right)=5\times10^{-23}\,\mathrm{erg\,cm^{3}\,s^{-1}}$. Hence, $t_\mathrm{c}=1.3-13\,\mathrm{kyr}$. In typical HII region, densities are 
%as low as $n_\mathrm{HII}=0.1-1\,\mathrm{cm}^{-3}$. With the same temperature and cooling function, $t_\mathrm{c}=1.3-13\,\mathrm{Myr}$! Hence, the SN ejecta will only be subject to adiabatic cooling, since radiative cooling is 
%inefficient as long as the SNR does not interact with the shell swept--up by the HII region.
\section{Summary}
\label{secsum}
In this study we have presented results from numerical simulations on molecular cloud evolution including supernova feedback from massive stars.\\ The results suggest that supernova feedback alone is not sufficient to disrupt molecular clouds, consistent with previous studies. The dispersal is only restricted to some fraction of the parental cloud. 
Though the efficiency in disrupting the 
cloud is very low, supernovae still create regions of moderate temperature, which affects the thermodynamic behaviour of the gas. The efficiency also strongly depends on where the supernovae go off, on the number of supernova events, 
as well as on the clumpyness of the cloud. Single supernovae initially show signs of compression, which might lead to 
triggered star formation. With time, the shocks disperse those regions and the net effect is a 
negative feedback (disruption). If the supernovae are clustered, their combined energy and momentum input 
is sufficient to disrupt larger amounts of the parental cloud. However, even with clustered supernovae, the cloud is 
not fully destroyed. The supernovae are still able to remove up to 50\,\% of the total cloud mass. The inhomogeneity of the cloud due to initial turbulent fluctuations enables energy from the SN to escape through low--density
 channels. On the other hand, more turbulent clouds are also less compact and the substructures are hence dispersed more easily.\\
 The suppression of star formation, however, is quite effective with reduction of the SFE and SFR by factors of 2--4, again consistent with previous studies on SN feedback.
 This is due to the fact that there occurs a short--period, but sufficient 
momentum transfer to the dense gas, which leads to its dispersion. However, 
our results indicate that star formation is not halted and continues throughout the simulation ($t\sim35\,\mathrm{Myr}$) in 
all cases.\\

%We point out the lack of additional feedback mechanisms prior to the supernova. However, ionisation feedback as the main feedback process will disperse the gas surrounding the massive star, leaving behind a giant H\,II region. This will 
%enable the supernova to go off in a teneous, hot medium. Hence, the cooling will be less efficient and more energy is available to be converted to kinetic energy and thus momentum. We therefore propose the dispersal of molecular clouds 
%by the combined effects of ionising radiation and supernovae. The dispersal timescale will then be less than the dispersal timescale due to ionisation feedback alone.

%\subsection{Notes}
%Under--pressurised gas is created because of efficient cooling. $(n=0.1)^2\Lambda(T=10^8)>(n=1)^2\Lambda(T=10^3)$. The supernova remnants are at first over--pressurised volumes that become under--pressurised. This induces an 
%implosion--like process of recollapsing gas, although 'initial' gas in this volume is very thin. Hot bubbles are thus only created by interacting supernovae (?).

\section*{Acknowledgements}
We thank an anonymous referee for his/her comments, which helped to improve the quality of this study.
BK acknowledges hospitality at Centro de Radioastronom\'{i}a y Astrof\'{i}sica, Universidad Nacional Aut\'{o}noma M\'{e}xico, 
during the initial stages of this study and Gilberto C. G\'{o}mez for stimulating discussions. 
DS acknowledges funding by the Bonn-Cologne Graduate School as well as the Deutsche Forschungsgemeinschaft
(DFG) via the Sonderforschungsbereich SFB 956 Conditions and Impact of Star Formation. 
RB acknowledges funding by the DFG via the Emmy-Noether grant BA
3706/1-1, the ISM-SPP 1573 grants BA 3706/3-1
and BA 3706/3-2, as well as for the grant BA 3706/4-1.
The simulations were run on HLRN--III under project grant hhp00022. The authors also gratefully acknowledge the Gauss Centre for Supercomputing e.V. (www.gauss-centre.eu) for funding this project (project--id: pr85ga) by providing computing time on the GCS Supercomputer SuperMUC at Leibniz Supercomputing Centre (LRZ, www.lrz.de).
The authors furthermore gratefully acknowledge the Gauss Centre for Supercomputing (GCS) for providing computing time through the John von Neumann Institute for Computing (NIC) on the GCS share of the supercomputer JUQUEEN
\footnote{J\"ulich Supercomputing Centre. (2015). JUQUEEN: IBM Blue Gene/Q Supercomputer System at the J\"ulich Supercomputing Centre. Journal of large-scale research facilities, 1, A1. http://dx.doi.org/10.17815/jlsrf-1-18} at J\"ulich Supercomputing Centre (JSC). GCS is the alliance of the three national supercomputing centres HLRS (Universit\"at Stuttgart), JSC (Forschungszentrum J\"ulich), and LRZ (Bayerische Akademie der Wissenschaften), funded by the German Federal Ministry of Education and Research (BMBF) and the German State Ministries for Research of Baden-W\"urttemberg (MWK), Bayern (StMWFK) and Nordrhein-Westfalen (MIWF).
The software used in this work was in part developed by the DOE--supported ASC/Alliance Center for Astrophysical 
Thermonuclear Flashes at the University of Chicago.

\bibliographystyle{aa}
\bibliography{astro}

\label{lastpage}

\begin{appendix}

\section{Resolution study: Sink particles}
\begin{figure}
\centering
\includegraphics[height=0.5\textwidth,angle=-90]{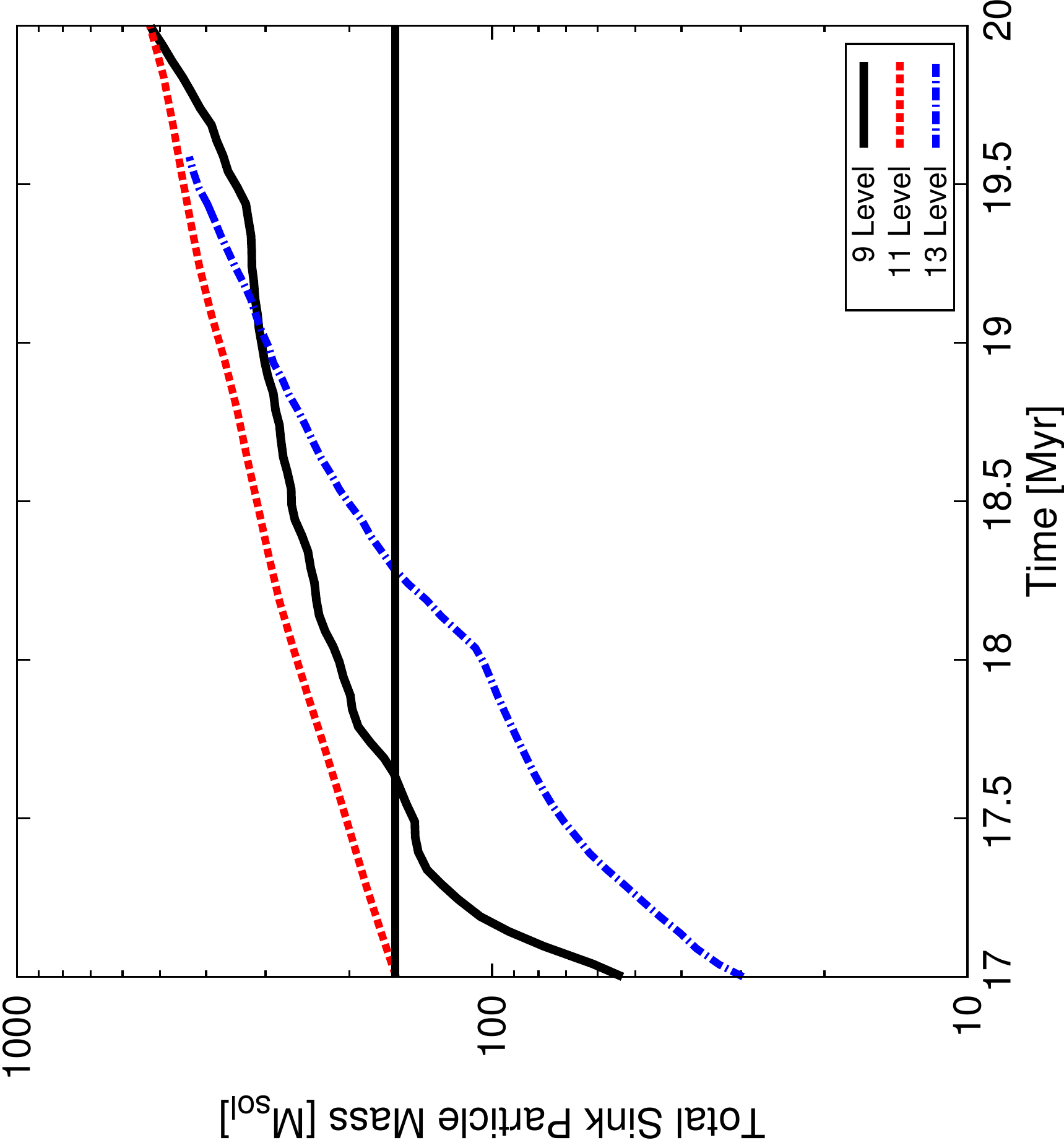}
\caption{Resolution study showing the mass of all sink particles as function of time for a time interval of $\Delta t\approx 3\,\mathrm{Myr}$. The horizontal black line denotes the critical cluster--mass for having at least one massive star. The temporal difference for reaching the critical mass is about $t_\mathrm{diff}\approx 1.2\,\mathrm{Myr}$. It is thus not significant for the long--term dynamical evolution of the cloud. The total mass in 
all sink particles converges after some Myr of evolution. }
\label{appfig1}
\end{figure}

As stated in section \ref{secsubgrid}, supernova feedback is only enabled if the total sink particle mass exceeds $M_\mathrm{Kroupa}=160\,\mathrm{M}_\odot$.
Figure \ref{appfig1} shows the total mass of stars for three simulations with varying numerical resolution. The Kroupa--mass (horizontal solid black line) is reached at different times. However, the temporal difference is not 
significant for the global evolution of the cloud since it is only about 1\,Myr. With time, the total stellar mass converges. It is thus independent of the numerical resolution. 
The usage of our IMF--fitting approach then gives the same supernova features for different resolutions. Note that the initial stages of the sink particle evolution differ due to different threshold densities. These densities influence the formation of 
sink particles as well as their accretion behaviour (gas is only accreted onto the sink particle, if the density in the respective cells exceeds the threshold density). 
\section{Resolution study: Supernova remnant}
\begin{figure}
\centering
\includegraphics[height=0.5\textwidth,angle=-90]{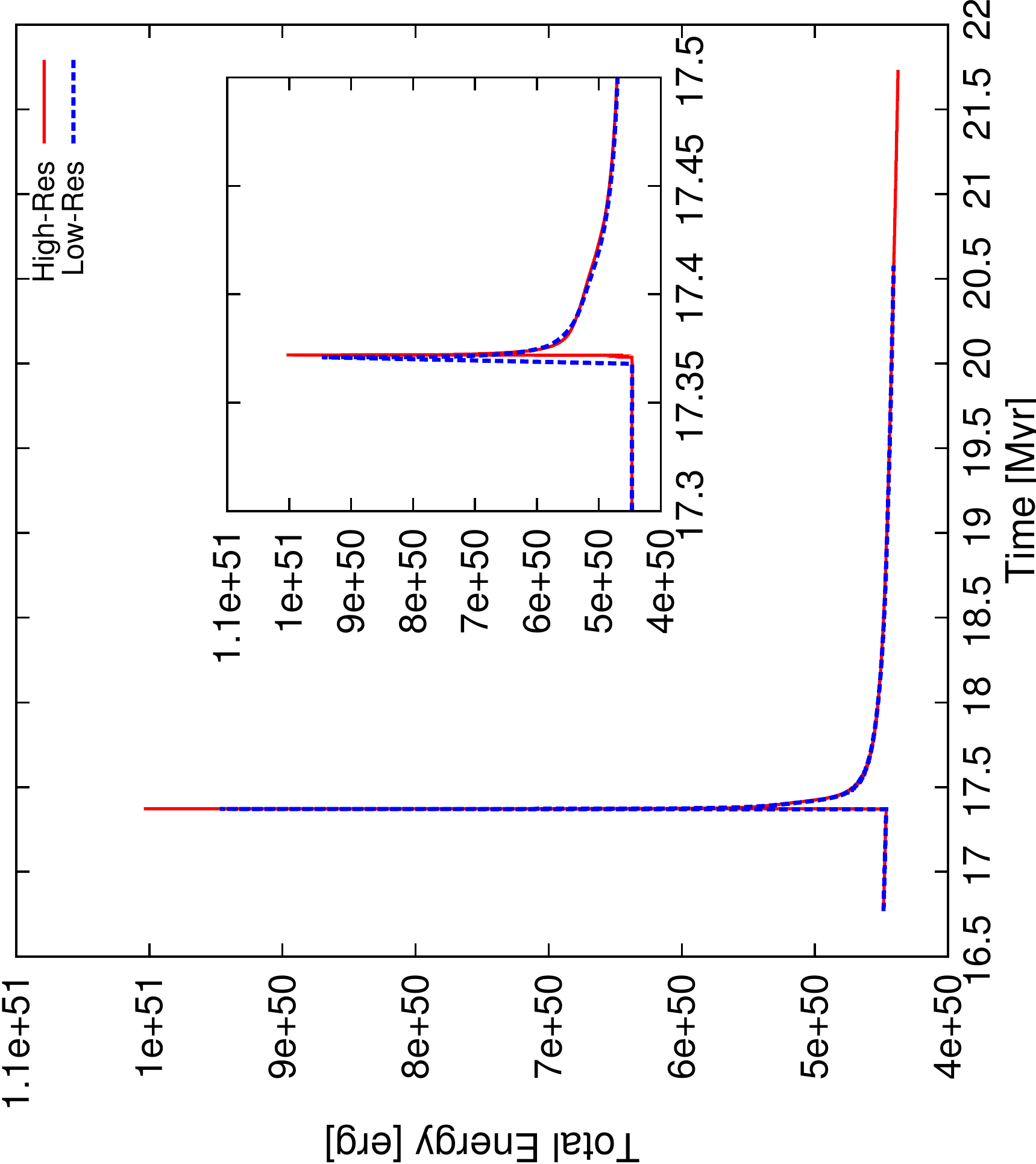}
\caption{Resolution study showing the temporal evolution of the total (thermal+kinetic) energy of run 
HR0.8Y. We compare two ways of resolving the radius of the supernova remnant. Our fiducial run uses 
two cells for resolving the SNR (termed \ita{Low-Res}) and the control simulation resolves the SNR with ten grid cells 
(termed \ita{High-Res}).
It is seen that the evolution is quite similar and that our fiducial value of 2 grid cells is sufficient.}
\label{appfig2}
\end{figure}

Sufficient numerical resolution of the supernova remnant is crucial for its further evolution. However, using too many cells 
may lead to undesired effects on the environment of the exploding sink particle. In figure \ref{appfig2} we compare 
two simulations in which the radius of the remnant is resolved with 2 grid cells (Low-Res in the figure; our fiducial value) and with 10 cells 
(High-Res). Shown is the evolution of the total (thermal plus kinetic) energy of the gas. For this resolution study, the simulation is stopped a few Myr after the first SN has gone off.  As can be seen, only minor deviations
 occur at the time of energy injection. The total energy injected in the fiducial run reaches $\sim 9.5\times10^{50}\,\mathrm{erg}$. These deviations from the model value ($10^{51}\,\mathrm{erg}$) are due to the small number of cells for resolving a spherical supernova remnant with cubic grid cells. However, the long--term evolution shows no significant difference (the deviations are $\leq 1\,\%$ ) and we conclude that the radius of the remnant is sufficiently resolved with 2 grid cells.

\section{Notes on the supernova rate}
We use a supernova rate (in terms of \ita{supernovae per solar mass}) as a combination of the observed supernova rate (in terms of \ita{supernovae per year}) and the Galactic star formation rate (in terms of 
\ita{solar mass per year}). This gives 
\beq
SNR_\mathrm{M} (\#/\mathrm{M}_\odot)=\frac{SNR_\mathrm{yr} (\#/\mathrm{yr})}{SFR (\mathrm{M}_\odot/\mathrm{yr})}.
\eeq
Using values for $SNR_\mathrm{yr}=(44\,\mathrm{yr})^{-1}$ from \cite[][]{Tammann94} and $SFR=1\,\mathrm{M}_\odot\,\mathrm{yr}^{-1}$ from \cite[][]{MacLow04}, the supernova rate becomes 
\beq
SNR_\mathrm{M}=(44\,\mathrm{M}_\odot)^{-1}.
\eeq
This is analogous to a calculation of the supernova rate directly from an IMF. Using an IMF, SNR$_\mathrm{M}$ is just the number of massive stars per unit solar mass. For a Kroupa--IMF 
$SNR_\mathrm{M}=(100\,\mathrm{M}_\odot)^{-1}$\,--\,$(160\,\mathrm{M}_\odot)^{-1}$, depending on the detailled numerical constants. The SNR in this study is hence three to four times higher than those 
from IMF estimates and thus gives an 
upper limit on the efficiency of cloud dispersion and disruption by supernovae.\\
For comparison, \citet[][]{Joung06} and \citet[][]{Gatto15} use the rate from \citet[][]{Tammann94} and scale it down to the respective size of the simulation box (compared to the area of the Galaxy). This gives 
\beq
SNR_{256\times256\,\mathrm{pc}^{2}}=2\,\mathrm{SNe}/\mathrm{Myr}.
\eeq
Knowing the number of SNe and the time interval in which they are going off, we are able to calculate a SN--rate in terms of \ita{supernovae per Myr}. The results give 
$SNR_\mathrm{Myr}\approx 1.7\,\mathrm{SNe}/\mathrm{Myr}-2.1\,\mathrm{SNe}/\mathrm{Myr}$ in very good aggreement with the above mentioned studies.
\end{appendix}

\end{document}